\documentclass[11pt]{article}

\usepackage[margin=1in]{geometry}  
\usepackage[english]{babel}
\usepackage[utf8x]{inputenc}
\usepackage[T1]{fontenc}
\usepackage{parskip}
\usepackage[dvipsnames]{xcolor}
\usepackage{graphicx}
\usepackage{subcaption}
\usepackage{longtable} 
\usepackage{multirow}
\usepackage{listings}
\usepackage{makecell}
\usepackage{array}
\usepackage{float}
\usepackage{dsfont}
\usepackage{rotating}
\usepackage{booktabs}
\usepackage{enumerate}
\usepackage{extarrows}
\usepackage{tikz}
\usetikzlibrary{chains}
\usetikzlibrary{arrows}
\usepackage{pgf}
\usepackage{longtable,booktabs}
\usepackage{makecell}
\usepackage{soul}
\usepackage[normalem]{ulem}
\usepackage{lscape}

\usepackage{amsmath}
\usepackage{amssymb}
\usepackage{amsthm}
\usepackage{bm}
\usepackage{mathtools}
\usepackage{mathrsfs}
\usepackage{nicefrac}       
\usepackage{algorithmic}
\usepackage[ruled,vlined]{algorithm2e}

\SetCommentSty{algocmtstyle}
\usepackage{wasysym}

\usepackage{natbib}
\usepackage[
  colorlinks,
  citecolor=blue,
  linkcolor=red,
  anchorcolor=red,
  urlcolor=blue
]{hyperref}
\mathtoolsset{showonlyrefs}
\usepackage{authblk}
\usepackage{todonotes}

\usepackage{xcolor}
\definecolor{yg}{RGB}{245, 75, 66}

\theoremstyle{plain}

\newtheorem{definition}{Definition}
\newtheorem{theorem}{Theorem}
\newtheorem{lemma}{Lemma}

\newtheorem{remark}{Remark}

\newtheorem{proposition}{Proposition}

\newtheorem{counterexample}{Counterexample}

\newcommand{\RR}{\mathbb{R}}
\newcommand{\ZZ}{\mathbb{Z}}

\newcommand{\EE}{\mathbb{E}}
\newcommand{\PP}{\mathbb{P}}
\newcommand{\ind}{\mathds{1}}

\newcommand{\indep}{\perp\!\!\!\perp}

\newcommand{\fdr}{\textnormal{FDR}}

\newcommand{\argmax}[1]{\underset{#1}{\arg\!\max}}
\newcommand{\argmin}[1]{\underset{#1}{\arg\!\min}}

\usepackage{xspace}

\newcommand{\stepa}[1]{\overset{\rm (a)}{#1}}
\newcommand{\stepb}[1]{\overset{\rm (b)}{#1}}

\newcommand{\floor}[1]{{\left\lfloor {#1} \right \rfloor}}
\newcommand{\ceil}[1]{{\left\lceil {#1} \right \rceil}}
\newcommand{\given}{{\,|\,}}

\newcommand{\Biggiven}{\,\Big{|}\,}
\newcommand{\bigggiven}{\,\bigg{|}\,}

\def\@#1\@{\begin{align}#1\end{align}}
\def\$#1\${\begin{align*}#1\end{align*}}

\definecolor{myblue}{rgb}{.8, .8, 1}
\definecolor{mathblue}{rgb}{0.2472, 0.24, 0.6} 
\definecolor{mathred}{rgb}{0.6, 0.24, 0.442893}
\definecolor{mathyellow}{rgb}{0.6, 0.547014, 0.24}

\newcommand{\cA}{{\mathcal{A}}}

\newcommand{\cC}{{\mathcal{C}}}

\newcommand{\cE}{{\mathcal{E}}}
\newcommand{\cF}{{\mathcal{F}}}
\newcommand{\cG}{{\mathcal{G}}}
\newcommand{\cH}{{\mathcal{H}}}
\newcommand{\cI}{{\mathcal{I}}}
\newcommand{\cJ}{{\mathcal{J}}}

\newcommand{\cM}{{\mathcal{M}}}
\newcommand{\cN}{{\mathcal{N}}}
\newcommand{\cO}{{\mathcal{O}}}

\newcommand{\cQ}{{\mathcal{Q}}}
\newcommand{\cR}{{\mathcal{R}}}
\newcommand{\cS}{{\mathcal{S}}}
\newcommand{\cT}{{\mathcal{T}}}

\newcommand{\cZ}{{\mathcal{Z}}}

\newcommand{\fg}{\mathfrak{g}}

\newcommand{\fullcirc}{\CIRCLE}
\newcommand{\halfcirc}{\LEFTcircle}
\newcommand{\bhonly}{\RIGHTcircle}
\newcommand{\notsel}{\Circle}

\newcommand{\sgn}{\textnormal{sgn}}

\newcommand{\bY}{\bm{Y}}

\newcommand{\bL}{\bm{L}}
\newcommand{\bV}{\bm{V}}
\newcommand{\pscr}{\mathsf{pscr}}
\newcommand{\obs}{\mathsf{obs}}
\newcommand{\rand}{\mathsf{rand}}
\newcommand{\early}{\mathsf{early}}
\newcommand{\cc}{\mathsf{cc}}

\graphicspath{figures-and-tables/}
\long\def\comment#1{}

\definecolor{LightGreen}{rgb}{0.0, 0.5, 0.0}

\title{Adaptive discovery of effect modification in matched observational studies}
\author{Yu Gui, Dylan S. Small, and Zhimei Ren\thanks{Corresponding author, \href{mailto:zren@wharton.upenn.edu}{zren@wharton.upenn.edu}.}}
\affil{Department of Statistics and Data Science,\\ The Wharton School, University of Pennsylvania}
\date{\today}

\begin{document}
\maketitle

\begin{abstract}

Understanding effect modification---how treatment effects vary across subpopulations---is 
practically important in observational studies, 
as it 
helps identify which subgroups are likely to benefit from a given treatment.
In this paper, we study the discovery of effect modification in matched observational 
studies, where each treated unit may be matched to 
multiple controls.
We develop a finite-sample valid procedure for identifying and selecting covariate-interpretable subgroups, 
with exact control of the subgroup-level false discovery rate (FDR). 
Our method explicitly accounts for unmeasured confounding via sensitivity models, 
and leverages multiple matched controls to improve statistical power.
We demonstrate the favorable performance of our method relative to baseline methods 
through extensive simulation studies and a real-world application 
to the economic returns to college education.

\end{abstract}

\section{Introduction}\label{sec:intro}

\subsection{Heterogeneous benefits of college education}\label{sec:example}
Is college worth it?
Understanding the economic returns to college education has long been a central topic in the social sciences \citep{baum2014higher,hout2012social}.
The question has become increasingly salient as college participation expands: 
the immediate college enrollment rate of recent high school completers rose from $51.7\%$ in 1970 to $62.0\%$ in 2022, while the annual number of undergraduate degrees (associate's plus bachelor's) grew from about one million in 1969--70 to about three million in 2021--22 \citep{andrews2024returns}.\footnote{\emph{The Digest of Education Statistics} Tables 302.10 and 318.10. \url{https://nces.ed.gov/programs/digest/current_tables.asp}.}
Although the young generation has long been encouraged to attend college by the conventional opinion that college education will lead to positive economic returns, some social scientists remain skeptical \citep{hout2012social}, and public skepticism has become increasingly visible in recent discussions \citep{blinder2025college}.

Whether college is worth it may depend on a person's characteristics, e.g., their interests, their high school grades, or their parents' education.  One approach to understanding the heterogeneity would be to directly model the conditional treatment effect as a function of a person's characteristics (covariates).  However, such a model may rely on parametric or structural assumptions and may not be that interpretable.  A more interpretable alternative is to study the heterogeneity through subgroups: rather than attempting to recover the full conditional treatment effect, one seeks to identify subpopulations defined by observed characteristics for which the treatment effect may differ and then identify the subpopulations for which the treatment has a positive effect, e.g., for which subpopulations college has positive returns \citep{perna2005benefits,brand2010benefits}.
Realizing this goal calls for a data-adaptive method that first identifies candidate subgroups from 
the data and then selects the significant ones to report with valid error-rate control, while 
relying on weak assumptions about the underlying distribution and remaining 
robust to possible unmeasured confounding.

\subsection{Adaptive discovery of effect modification with controlled error}
Effect modification arises when the treatment effect of interest varies across levels of observed covariates. 
Identifying such heterogeneity not only offers insights into effect heterogeneity but may also help define tailored inferential targets for follow-up investigations and decision-making.
In observational studies, treatment effect estimates may be biased by unmeasured confounding, e.g., people who are more likely to attend college may also have parents with higher socioeconomic status (SES) that is not fully measured, and this unmeasured SES may also affect a person's earnings in ways other than making a person more likely to go to college.  Sensitivity analysis can be used to study how sensitive a treatment effect estimate is to unmeasured confounding \citep{cornfield1959smoking,rosenbaum1987sensitivity,rosenbaum2005sensitivity}.  
Discovering effect modification can make findings less sensitive to bias from unmeasured confounding \citep{hsu2013effect,lee2018discovering}.

\paragraph{Detecting effect modification via matching.}
As alluded to above, in observational studies, 
directly defining subgroups and performing inference based on 
a model for the conditional treatment effect can be 
unreliable when the model is misspecified.
An alternative is \emph{matching} \citep{cochran1973controlling,rubin1973matching,rubin1973use,rosenbaum2020modern,stuart2010matching},
which constructs matched sets of treated and control units that are comparable in observed covariates.  The simplest type of matching is pair matching, which constructs matched pairs of treated and control units.
By balancing covariate distributions prior to outcome analysis, 
matching reduces overt bias and limits extrapolation to regions with poor overlap, thereby making subsequent subgroup comparisons more interpretable without imposing a fully specified model for the conditional treatment effect.

While it is natural to use the same matched dataset to choose subgroups and make inferences about treatment effects within those subgroups, this reuse of data could invalidate
inference without careful adjustment.
Data splitting is a classical solution:
subgroups are defined with a pilot dataset and 
inference is conducted on the identified subgroups using the remaining subset \citep{cox1975note,chen2019heterogeneous}.
But this comes with a price: by halving the effective sample size, 
sample splitting can lead to a loss in power, especially in small-sample settings. 
This motivates the development of a disciplined and fine-grained inferential procedure 
that allows selection over data-driven subgroups while retaining valid error control conditional on the identified subgroups.
\noindent Due to space constraints, discussion of additional related work on subgroup partition and selection is deferred to Appendix~\ref{app:related}, including CATE-based subgroup selection \citep{wager2018estimation,muller2025isotonic} and interactive methods with type-I error control in randomized experiments \citep{cheng2025chiseling}.

\paragraph{Quantification of selection error.}
Once the candidate subgroups have been specified, 
our goal is to select the subgroups in which the treatment has an effect.
To quantify the error rate of subgroup selection,~\citet{hsu2015strong} considered  
{\em family-wise error rate (FWER)}, i.e., the probability of making at least one false discovery, and 
proposed an algorithm that controls the FWER at any given level 
$\alpha \in (0,1)$ with a finite sample.
However, FWER control can be too stringent when the number of subgroups is large, since guarding against even a single false discovery can substantially reduce power 
and lead to few selected subgroups.
An alternative error rate quantity that adapts to the scale of the problem is the 
{\em false discovery rate (FDR)}, which focuses on the proportion of false discoveries
among the selected subgroups.
Concretely, the goal is to control
\begin{align}
\text{FDR} = \EE\left[\frac{\sum_{\fg}\ind\{\text{subgroup }\fg \text{ is selected but has zero effect}\}}{\text{number of selected subgroups} \vee 1}\right] \leq \alpha,
\end{align}
where $a\vee b = \max(a,b)$, for $a,b \in \RR$.
To achieve finite-sample FDR control, 
\cite{karmakar2018false} proposed applying the Benjamini-Hochberg (BH) 
procedure \citep{benjamini1995controlling} to subgroup-level p-values,
e.g., Wilcoxon signed rank p-values.
However, this approach has several limitations. 
First, it does not support interactive/adaptive data analysis, in which data analysts 
adapt their decisions after viewing (part of) the data. 
Second, because these subgroup-level tests are often discrete, the resulting 
p-values can be coarse, especially for small subgroups, which in turn can substantially reduce selection power. 
More recently,~\citet{duan2024interactive} proposed an adaptive multiple testing procedure 
for subgroup selection with FDR control that addresses these issues.
Their method, however, is tailored primarily to settings without unobserved confounders,
and can become quite conservative when such confounding is taken into account.
A detailed comparison with~\citet{duan2024interactive} is provided in Section~\ref{sec:baseline}.



\paragraph{Enhancing power by leveraging multiple matched control units.}
Beyond valid error control, it is also crucial to design powerful statistics 
that effectively use the observational data. 
One potentially valuable source of information is the presence of 
multiple matched control units, which commonly arises in relatively large cohorts 
with limited treated units \citep{smith19976,rubin2000combining,rosenbaum1987role}. 
\cite{rosenbaum2013impact} demonstrates that the use of multiple controls often yields a nontrivial reduction in sensitivity to unmeasured biases. 
However, in the context of effect modification with data-driven subgroups, 
it remains unclear how to effectively leverage multiple control units to improve statistical power while maintaining rigorous error control (see discussion in Section~\ref{sec:ada-test} under ``Multiple controls'').


\subsection{Adaptive multiple testing}\label{sec:ada-test}
Our subgroup selection method adopts the adaptive multiple testing 
framework, which permits sequential, data-driven decisions based on a masked view of the test statistics, while the remaining information is reserved for rigorous error rate control
(see e.g.,~\citet{barber2015controlling,candes2018panning,lei2018adapt,ren2023knockoffs,lei2021general,chao2021adapt,duan2020familywise,
duan2022interactive,duan2024interactive,freestone2024semi}, among others). 
The multiple testing problem induced by our main research question, however,  
introduces new challenges that are not addressed by existing methods. 
In developing efficient subgroup selection procedures, we 
also expand the methodological toolbox of adaptive multiple testing itself.
We outline these challenges below.

\begin{figure}[ht]
    \centering
    \begin{subfigure}[t]{0.31\textwidth}
        \centering
        \includegraphics[height=4.2cm]{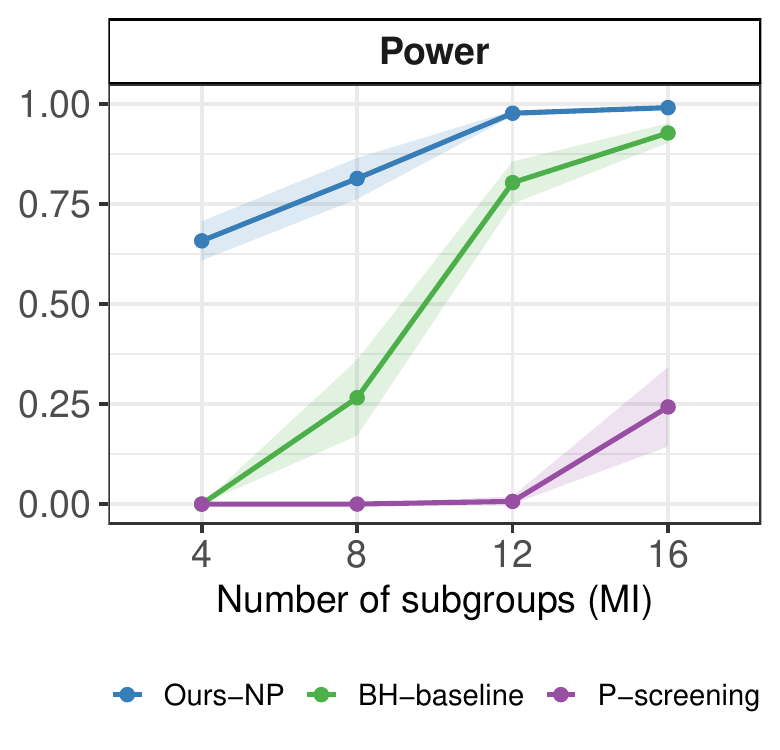}
        \caption{}
        \label{fig:groupsize}
    \end{subfigure}
    \begin{subfigure}[t]{0.31\textwidth}
        \centering
        \includegraphics[height=4.2cm]{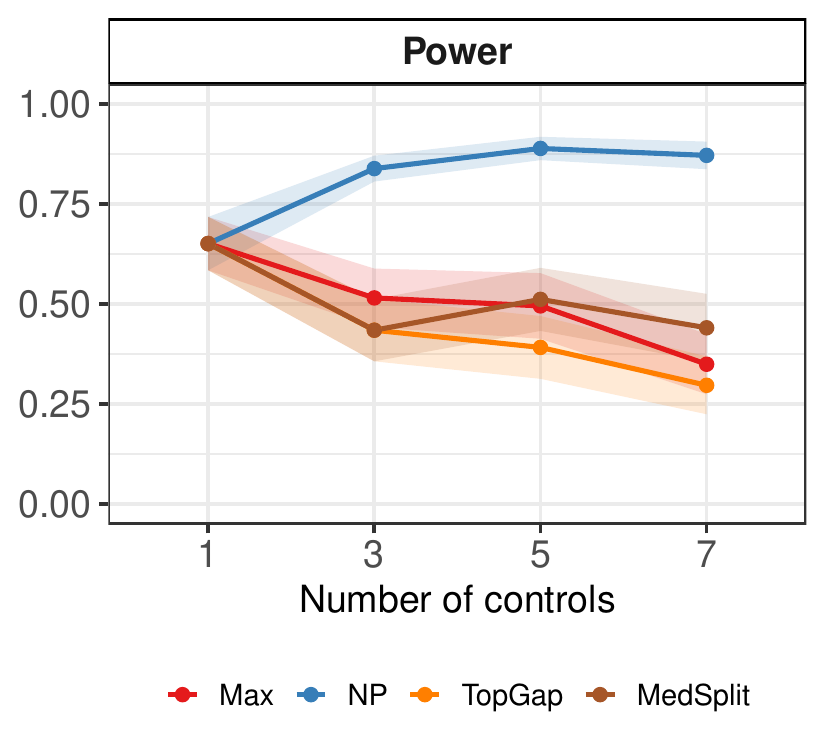}
        \caption{}
        \label{fig:compare_power}
    \end{subfigure}
    \begin{subfigure}[t]{0.35\textwidth}
        \centering
        \includegraphics[height=4.2cm]{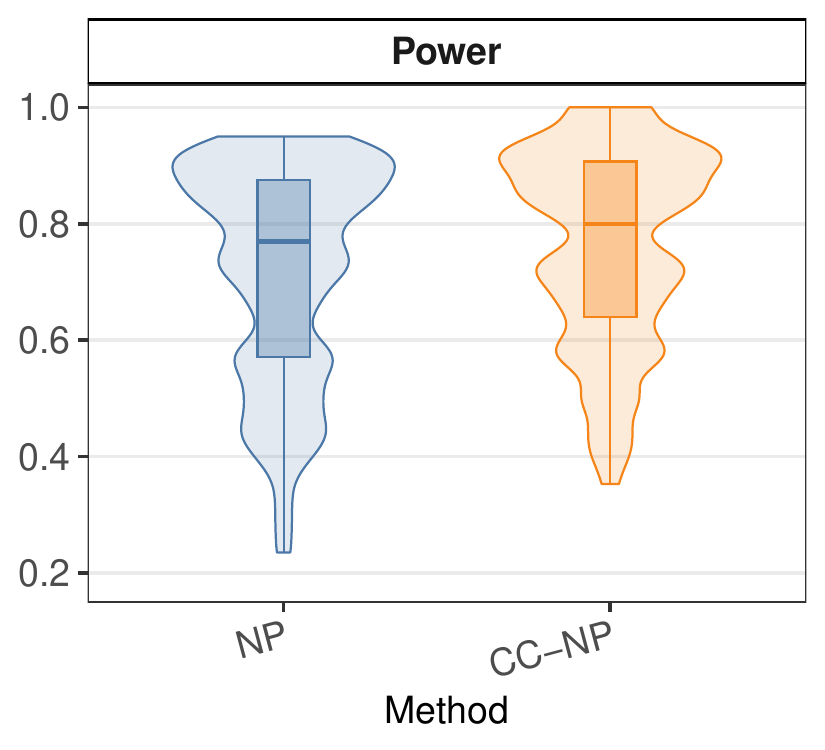}
        \caption{}
        \label{fig:compare_cc}
    \end{subfigure}
    \caption{(a) Power comparison of our method (\texttt{Ours-NP}) with 
    the baselines \texttt{BH-baseline}~\citep{karmakar2018false} and \texttt{P-screening}~\citep{duan2024interactive} as subgroup size varies. 
    (b) Power of our proposed method under different choices of masking methods 
    as a function of the number of control units. The power curves
\texttt{Max}, \texttt{TopGap}, \texttt{MedSplit} correspond to existing 
methods, while \texttt{NP} denotes our proposal.
(c) Power comparison between our method with and without conditional calibration.
See details in Section~\ref{sec:simu}.
}
    \label{fig:plot-intro}
\end{figure}

\paragraph{Composite null.}
In observational studies, valid inference on treatment effects must 
account for unmeasured confounding. Under such confounding,  
the null hypothesis of no treatment effect is inherently composite.
Existing adaptive multiple testing methods are largely designed for 
testing sharp nulls, or for the composite nulls under the additional requirement that  
each hypothesis admits a p-value with a non-decreasing 
density~\citep{duan2020familywise,chao2021adapt}. However, as we show
in Section~\ref{sec:warmup}, the commonly used Wilcoxon signed-rank p-values 
{\em do not} satisfy the 
non-decreasing condition, making existing composite null-oriented 
methods not generally applicable. Meanwhile, applying methods tailored to sharp nulls
often leads to suboptimal statistical power. 
Our proposed method directly targets composite nulls without imposing 
density conditions on the p-values, achieving substantially improved power 
and rigorously controlled FDR; see Figure~\ref{fig:groupsize} 
for a preview of the power comparison between our proposed method, \texttt{Ours-NP},
and the sharp null-oriented method from~\citet{duan2024interactive}, \texttt{P-screening};
the performance of the other baseline~\citep{karmakar2018false}, \texttt{BH-baseline}, is also included.

\paragraph{Multiple controls.}
A central question in this paper is how to 
leverage multiple controls to improve statistical efficiency. 
When applying existing multiple testing methods that incorporate 
multiple controls~\citep{gimenez2019improving,he2021identification} to 
the subgroup selection task, we observe a counter-intuitive phenomenon: 
power {\em decreases} as the number of 
controls grows (see the \texttt{Max}, \texttt{TopGap}, \texttt{MedSplit} curves in Figure~\ref{fig:compare_power}).
Upon closer inspection, these methods primarily differ in how the test statistics are masked, 
yet all their masking schemes ``dilute'' the signal as the number of controls increases. 
Our approach identifies and retains a key piece of information that is lost under 
existing masking strategies; by incorporating it into the masking step, 
we achieve increasing power as the number of controls grows, as shown 
by the \texttt{NP} curve in Figure~\ref{fig:compare_power}.

\paragraph{Offset effect.}
In the adaptive multiple testing framework, FDR control 
is typically achieved by forming an FDR estimate 
from the unmasked information and choosing the selection 
set for which this estimate falls below the target level. 
Due to an artifact of this construction---specifically, the granularity of the estimator---these 
procedures become effectively powerless unless the number of nonnulls exceeds a certain 
threshold. This phenomenon, referred to as the 
{\em threshold effect}~\citep{barber2015controlling,gimenez2019improving,luo2022improving,duan2024interactive},  
is common across existing adaptive multiple testing methods.
In the context of subgroup selection, it manifests when only a few 
(potentially large) subgroups are under consideration, 
making reliable selection difficult.
To alleviate this issue, we adopt the {\em conditional calibration} technique~\citep{luo2022improving,lee2024boosting}, which 
boosts power in such regimes while maintaining rigorous 
FDR control. As illustrated in Figure~\ref{fig:compare_cc}, the conditionally calibrated 
version (\texttt{-cc}) indeed improves power in this regime. 
Meanwhile, our instantiation of conditional calibration 
for composite nulls of this type appears to be new in the literature.

\subsection{Our contributions}
Our main contributions are summarized below.
\begin{itemize}
    \item[(1)] We develop an adaptive framework for selecting data-driven, covariate-interpretable subgroups
    with significant treatment effects. Our approach controls the subgroup-level FDR under Rosenbaum's $\Gamma$-sensitivity model~\citep{rosenbaum1987sensitivity} and accommodates flexible modeling choices.
    \item[(2)] Our procedure generalizes beyond matched pairs and applies to 
    the setting with multiple controls, where a treated unit can be matched with multiple control units. 
    By leveraging the masked rank of the treated outcome, 
    it substantially improves the discovery power relative to existing multi-control testing approaches. 
    The proposed test statistic is based on the likelihood ratio, and we also investigate 
    its optimality theoretically. 
    \item[(3)] We develop several techniques to further enhance the power of the subgroup selection procedure,
    which may be of independent interest within the broader framework of adaptive multiple testing.
    \item[(4)] The empirical advantages of our method 
    are supported by extensive simulations. 
    In a real-data study on the economic return to college education, 
    our method selects strictly more subgroups than the comparison methods across varying values of sensitivity parameters.
    The selected subgroups provide insights into the heterogeneity of benefits 
    from college and corroborate the negative selection theory in sociology that the groups who are least likely to attend college are the ones that would most benefit from it.
\end{itemize}

\section{Problem setup}\label{sec:setup}

Consider an observational study with $I$ matched sets. 
For each $i \in [I]:=\{1,2,\ldots,I\}$, set $i$ contains 
$n_i \geq 2$ units that share the same covariates $x_i \in \RR^d$; 
for any $j\in[n_i]$, let $Z_{ij} \in \{0,1\}$ denote the assigned treatment, where 
$Z_{ij}= 1$ indicates treatment and $Z_{ij}=0$ indicates control. 
The $Z_{ij}$'s are assumed to be independent across $i$, and 
we impose the constraint that $\sum^{n_i}_{j=1} Z_{ij}=1$, i.e., each matched set
contains exactly one treated unit. 
Let $t_{ij}$ and $c_{ij}$ denote 
the potential outcomes of unit $(i,j)$ with and without treatment, respectively, and let $u_{ij}$ represent all the unmeasured confounding factors. 
We adopt the {\em stable unit treatment value assumption (SUTVA)}, 
under which the observed outcome can be written as
\begin{align}\label{eq:sutva}
    R_{ij} = Z_{ij}\cdot t_{ij} + (1-Z_{ij})\cdot c_{ij}.
\tag{SUTVA}
\end{align}
Following the convention in the literature of observational studies, we denote 
\[
\cF = \left\{x_i, u_{ij}, t_{ij}, c_{ij}\right\}_{i\in[I], j \in [n_i]} \qquad \text{and} \qquad \cZ = \left\{Z_i:\; i \in [I],\; \sum_{j=1}^{n_i} Z_{ij} = 1\right\}.
\]
Inference is conditional on 
$\cF$ and $\cZ$ throughout, and we do not explicitly write down the conditional events when the context is clear.
Within each matched set, units $(i,j)$ and $(i,j')$ can differ in the odds of being treated due to unmeasured confounding. 
In this paper, we adopt Rosenbaum's $\Gamma$-sensitivity model \citep{rosenbaum1987sensitivity} to quantify the difference in the odds: for some $\Gamma\ge 1$, 
\begin{align}\label{eq:sens}
\Gamma^{-1} \le \frac{\pi_{ij}/(1-\pi_{ij})}{\pi_{ij'}/(1-\pi_{ij'})} \le \Gamma,
\quad \forall i\in [n] \text{ and } \forall j,j' \in [n_i],
\end{align}
where we write $\pi_{ij} = \PP(Z_{ij} = 1 \mid \cF)$ for notational simplicity.
Under~\eqref{eq:sens}, there is
\begin{align}\label{eq:sens_deriv}
\Gamma^{-1} \le \frac{\PP(Z_{ij}=1 \mid \cF, \cZ)}{\PP(Z_{ij'}=1 \given \cF, \cZ) } \le \Gamma,
\quad \forall i\in [n] \text{ and } \forall j,j' \in [n_i].
\end{align}

To identify effect modification, we partition the full cohort into non-overlapping 
subgroups using the information in $\cF$. The guiding principle 
is to construct groups that are as homogeneous as possible with respect 
to the treatment effect---so that units within a group share the same 
treatment effect whereas effects vary across groups.
Formally, denoting $\cA_{G}$ as an algorithm of subgroup partition, we obtain data-driven subgroups
$\cG = \cA_{G}\left(\cF\right) = (\fg_1, \cdots, \fg_K)$,
which is deterministic given $\cF$.

\paragraph{Inferential target.}
At the unit level, we consider the null hypothesis: $\forall i\in [I]$, $\forall j\in[n_i]$, 
\begin{align} 
H^{(0)}_{ij}: t_{ij} = c_{ij} \text{ versus }
H^{(1)}_{ij}: t_{ij} \neq c_{ij}.
\end{align}
In concrete contexts, we may also be interested in positive effects, in which case we consider $H^{(0)}_{ij}: t_{ij} = c_{ij}$ versus $H^{(1)}_{ij}: t_{ij} > c_{ij}$.
At the matched-set level, the null hypothesis for each $i \in [I]$ is defined by 
the intersection of the individual-level nulls, where $H_i^{(0)}: \cap_{j \in [n_i]}\;H_{ij}^{(0)}$.
Given subgroups $\cG = (\fg_1, \cdots, \fg_K)$,
we further define the subgroup-level null hypotheses:
\begin{align}
    H^{(0)}_{\fg} = \bigcap_{i \in \fg}\; H_i^{(0)}, \qquad \fg \in \cG.
\end{align}
Formulating subgroup analysis as a multiple testing problem for $\{H^{(0)}_{\fg}$, $\fg \in \cG\}$, 
our goal is to identify a set of subgroups $\cS_{\cG} \subseteq \cG$ that are likely to 
have significant treatment effect while controlling the FDR: 
\begin{align}
    \mathrm{FDR}(\cS_{\cG}; \cG) = \EE\left[\frac{\sum_{\fg \in \cG} \ind\left\{\fg \in \cS_{\cG},\;H^{(0)}_{\fg}\;\text{true}\right\}}{|\cS_{\cG}| \vee 1} ~\bigg|~ \cA_{G}\left(\cF\right) = \cG\right] \leq \alpha,
\end{align}
for a given nominal level $\alpha \in (0,1)$. 
Here, the expectation is conditional on the identified subgroups as the output of $\cA_G$, 
and post-selection bias should be taken into consideration.

We emphasize that our inferential target is at the subgroup level. 
While one could instead conduct inference at the matched-set level (e.g., in \cite{duan2024interactive}), 
the resulting discoveries---indexed by $i \in [I]$---can be hard to interpret 
and to generalize beyond the observed samples.
For example, in the college education dataset, if a matched pair of graduates from Wisconsin with covariates \texttt{rural residence=Yes}, \texttt{mother's education=12 years}, and \texttt{number of siblings=3} is selected as a ``subgroup'', it is not clear how this result translates into our understanding of the 
benefit of college education. 
Moreover, for inference at the matched set level, within-subgroup variability cannot be estimated, making the interpretability, replicability, and statistical significance of such findings unclear.
\section{Subgroup selection with matched pairs}\label{sec:warmup}

We begin with the matched-pair setting, i.e., $n_i = 2$ for all $i \in [I]$. 
To fix ideas, we focus on the one-sided unit-level alternative 
$H_{ij}^{(1)}: t_{ij}>c_{ij}$, together with the induced matched-set- and group-level hypotheses.
Extension to two-sided alternatives is provided to Appendix~\ref{sec:two-sided}.
In the college education example, this formulation corresponds to identifying subgroups in which college education exerts a positive effect on individuals' income.
\subsection{Construction of test statistics}\label{sec:L-agg}
Suppose the partition $\cG$ is given. For each matched pair $i\in[I]$, define 
the treated-control difference in outcomes $Y_i = (R_{i1}- R_{i2})(Z_{i1}-Z_{i2})$
and decompose it as
\$
L_i = \sgn(Y_i), 
\quad 
W_i = |Y_i|.
\$
Above, the {\em sign} statistic $L_i$ captures the direction of the effect,
while the {\em magnitude} statistic $W_i$ estimates its size. 
Under $H_i^{(0)}$, we have 
\begin{align}\label{eq:sens1}
    \PP(L_i = 1 \mid W_i, \cF,\cZ) = \PP(Z_{i1} = \ind\{c_{i1} - c_{i2} > 0\} \mid \cF, \cZ) \in \left[\frac{1}{1+\Gamma}, \frac{\Gamma}{1+\Gamma}\right],
\end{align}
where the equality follows from
$W_i = |c_{i1}-c_{i2}|$ under $H_i^{(0)}$ and 
the final inclusion is a direct consequence of Equation~\eqref{eq:sens_deriv}
under the $\Gamma$-sensitivity model.

We now pass from the matched-pair level to the group level.
For each $\fg \in \cG$, denote $\cQ_{\fg} \subseteq \fg$ a subset of indices 
determined solely by $\{W_i:\;i\in \fg\}$.
For example, we may define $\cQ_{\fg} = \{i \in \fg: W_i \geq W_{(k)}\}$, 
where $W_{(k)}$ is the $k$-th largest value among $\{W_i:\;i \in \fg\}$.
Using the units in $\cQ_\fg$ as the group representative, we define:
\@ \label{eq:agg}
L_\fg \;=\; 2 \cdot \ind\!\left\{\,\sum_{i \in \cQ_{\fg}}\, L_i \;>\; 2\,\eta_\fg - |\cQ_\fg|\,\right\} - 1,
\qquad
V_\fg \;=\; \bigl(L_{j}\bigr)_{j\in \fg,\; j \notin \cQ_{\fg}},
\@
where the threshold $\eta_\fg$ is the upper $\Gamma/(1+\Gamma)$-quantile of
the worst-case binomial distribution,
\begin{equation}
\eta_\fg \;:=\; \min\!\left\{\, t \in \ZZ_{\ge 0} \;:\;
  \PP\!\left(\mathrm{Bin}\bigl(|\cQ_\fg|,\, \tfrac{\Gamma}{1+\Gamma}\bigr) > t\right)
  \;\le\; \tfrac{\Gamma}{1+\Gamma} \,\right\}.
\label{eq:eta-def}
\end{equation}
One special case is when $\cQ_\fg$ contains only 
$i^*(\fg) = \argmax{i \in \fg} ~W_i$, i.e., the unit in the group with the largest magnitude outcome difference. 
In this case, $|\cQ_\fg|=1$,
the group-level sign statistic reduces to $L_\fg = L_{i^*(\fg)}$,
and $V_\fg = (L_{j})_{j\in \fg, j\neq i^*(\fg)}$.

In words, $L_\fg$ serves as a proxy for the overall direction of the treatment effect 
within group $\fg$, while $V_\fg$ collects the remaining sign information in the group. 
Writing $W_\fg = (W_i)_{i\in\fg}$, we define the  
bounded skewness condition for $L_\fg$ conditional on $(V_\fg,W_\fg)$.

\begin{definition}[$\kappa$-bounded skewness]
\label{def:bdd-skewness}
For any $\kappa > 0$, the group-level statistics $(L_\fg, V_\fg, W_\fg)$ are 
said to satisfy the $\kappa$-bounded
skewness condition if, $\forall \fg \in \cG$ such that $H_{\fg}^{(0)}$ is true,
\begin{align} \label{eq:set-bounded-skewness}
\PP\big(L_{\fg} = 1 \mid V_\fg,W_\fg,\cF,\cZ\big) 
\le \kappa \cdot \PP\big(L_{\fg} = -1 \mid V_\fg, W_\fg,\cF,\cZ\big).
\end{align}
\end{definition}
The following lemma characterizes the bounded skewness of the design in~\eqref{eq:agg}.
\begin{lemma}
\label{lem:skewness}
For each $\fg \in \cG$ such that $H_{\fg}^{(0)}$ is true, the group-level statistics $(L_\fg, V_\fg, W_\fg)$ defined in~\eqref{eq:agg} 
satisfy the $\kappa$-bounded skewness condition with $\kappa = \PP(\mathrm{Bin}(|\cQ_\fg|,\tfrac{\Gamma}{1+\Gamma}) > \eta_\fg)/\PP(\mathrm{Bin}(|\cQ_\fg|,\tfrac{\Gamma}{1+\Gamma}) \leq \eta_\fg)$,
which is further bounded by $\Gamma$.
\end{lemma}

The proof of Lemma~\ref{lem:skewness} is provided in Appendix~\ref{sec:proof-lemma1}.
The bounded skewness condition enforces a form of partial ``orthogonality''
between $L_\fg$ and $(V_\fg,W_\fg)$. In particular, it allows us 
to treat $(V_\fg,W_\fg)$ as fixed---that is, as the masked statistics introduced earlier---and 
to estimate the number of null groups 
with $L_{\fg} = 1$ by counting the number of subgroups with $L_{\fg} = -1$ as a proxy.

\subsection{The subgroup selection procedure}\label{sec:alg}
Our proposed method performs subgroup selection through a sequential screening procedure. 
In this process, subgroups exhibiting weak evidence of positive treatment effects are progressively 
filtered out, while the FDP among the remaining subgroups is monitored. Once the estimated 
FDP falls below the target level $\alpha$, the procedure stops, and the remaining subgroups 
are declared discoveries.

The key to FDR control is that the sign statistics $L_\fg$ are (partially)
masked during screening. Instead of directly using their values, 
we rely on the remaining information to predict or ``guess'' them. 
Error control follows from keeping the true signs hidden, 
while the power of the procedure depends on how accurately we can predict them.
The procedure consists of four main components: (i) initialization, (ii) the screening order, (iii) the FDP estimator, and (iv) the stopping rule. We detail each component in turn.

\paragraph{Initialization.}
For each group $\fg \in \cG$, we generate a variable 
$\xi_\fg \overset{\text{iid}}{\sim}\text{Bern}(\gamma)$, 
independent of all other quantities. 
These auxiliary binary variables are used to randomly split the groups with
$L_\fg=-1$, where we define: 
\@\label{eq:split} 
\cI^-_1 =  \{\fg \in \cG: L_\fg = -1, \xi_\fg = 0\},
\quad
\cI^-_2 =  \{\fg \in \cG: L_\fg = -1, \xi_\fg = 1\},
\@
and $\cI^+ =  \{\fg \in \cG: L_\fg = 1\}$.
That is, the groups with negative representative signs are randomly divided into two folds, 
while the positive groups are kept intact. By construction, $\cI^-_1$ 
consists of groups that appear unlikely to exhibit a positive treatment effect 
and will serve as a labeled set of ``negative'' examples. In contrast, 
$\cI^-_2 \cup \cI^+$ forms a mixed pool containing both pessimistic 
and potentially promising groups.

We then treat $\cI^-_1$ as (negatively) labeled data and $\cI^-_2 \cup \cI^+$ 
as unlabeled data to build an initial model $\hat \mu_0$ for $L_\fg$. 
The model is trained to predict $L_\fg$ from the group-specific features 
$(x_i)_{i \in \fg}$, together with $(V_\fg,W_\fg)$.
This setup naturally fits into the semi-supervised learning framework
(e.g.,~\citet{zhu2002learning,du2014analysis}), and in principle,  
any semi-supervised learning algorithm may be used to fit $\hat \mu_0$, 
provided that the true signs $\{L_\fg\}_{\fg \in \cI^-_2 \cup \cI^+}$ 
remain masked throughout the training process.

\paragraph{Screening order.}
Equipped with the initial model $\hat \mu_0$, we proceed to
examine the groups in $\cI := \cI^-_2 \cup \cI^+$.
Let $\cO_t \subseteq \cG$ denote the set of subgroups that have been screened out at step
$t \in \{0,1,\ldots,|\cI|\}$, initialized at $\cO_0 = \cI^-_1$.
At time $t\ge 1$, we use the prediction model $\hat \mu_{t-1}$ to score 
each remaining group: 
\$
\hat L_\fg := \hat \mu_{t-1}\big((x_i)_{i\in\fg}, V_\fg, W_\fg\big), ~\forall \fg\in \cO_{t-1},
\$
with the convention that a larger value of $\hat L_\fg$ indicates a higher 
likelihood that $L_\fg = 1$.
We then remove the group with the weakest predicted evidence for a positive effect, updating
\begin{align}\label{eq:update_rule}
\cO_{t} = \cO_{t-1} \cup \bigg\{\pi(t):= 
\argmin{\fg\notin \cO_{t-1}}~\hat L_\fg\bigg\}.
\end{align}
The prediction model is then updated to $\hat \mu_t$.
Crucially, its training may only use the information revealed at step $t$, namely, $\cM_t: = \{\{x_i\}_{i\in [I]}, \{V_\fg,W_\fg\}_{\fg \in \cG}, \{L_i: i \in \fg\}_{\fg \in \cO_{t}}\}$.
That is, while the covariates, treatment-difference magnitudes $W_\fg$, and residual sign information 
$V_\fg$ are always available, the sign statistics $L_\fg$ 
are only used for groups that have already been screened out. The signs of the unscreened groups remain masked.
The updated prediction model $\hat \mu_t$ is then used to re-score the groups $\fg \notin \cO_t$,
and the screened-out set is updated according to~\eqref{eq:update_rule}.

\paragraph{FDP estimate and stopping rule.}
At each step $t \ge 0$, we estimate the FDP by 
\begin{align}
    \widehat{\mathrm{FDP}}(t) = \frac{\kappa}{\gamma}
    \cdot \frac{(1 + N_t)}{1 \vee P_t},\quad \text{where } P_t \coloneqq \sum_{\fg \notin \cO_t} \ind\{L_{\fg} = 1\},\;\;N_t \coloneqq \sum_{\fg \notin \cO_t} \ind\{L_{\fg} = -1\}.
\end{align}
We define the stopping time as $\tau \coloneqq \inf\{t \in \{0\} \cup [|\cI|]:\;\widehat{\mathrm{FDP}}(t) \leq \alpha\}$
with the convention that $\inf \varnothing = \infty$ and $\cO_{\infty} = \cG$.
We obtain a selected subset of subgroups
\begin{align}\label{eq:SS}
\hat\cS \coloneqq \left\{\fg \in \cG:\;\fg \notin \cO_{\tau}, L_{\fg} > 0\right\}.
\end{align}
We conclude with a remark on the role of $\xi_\fg$'s.

\begin{remark}[The role of $\xi_\fg$'s]
Our proposed method incorporates external randomness 
through the $\xi_\fg$'s, which are introduced for sample splitting 
to improve the initialization of the prediction model.
This sample splitting strategy takes a special form that follows~\citet{freestone2024semi}.
Unlike standard sample-splitting procedures,  it only ``sacrifices''
the groups with $L_\fg = -1$ that were not to be selected otherwise.
We emphasize that this randomization is optional: when randomness is undesirable,  
it can be ``turned off'' by setting $\gamma = 1$. 
\end{remark}

\subsection{Finite-sample FDR control}
To understand why our proposed procedure leads to finite-sample FDR control, we 
first examine the intuition behind the FDP estimator. 
Recall that $L_{\fg}$ satisfies the $\kappa$-bounded skewness condition (Definition~\ref{def:bdd-skewness}). 
Consequently, among the subgroups still under consideration (unscreened) at step $t$, 
the number of nulls with 
$L_{\fg}=1$ can be upper bounded (heuristically) by those with $L_{\fg}=1$: 
\$
\sum_{\fg \notin O_t} \ind\{L_{\fg} = 1, H_{\fg}^{(0)} \text {is true}\}
& \approx
\sum_{\fg \notin O_t} \frac{1}{\gamma}\ind\{L_{\fg} = 1, \xi_\fg = 1, H_{\fg}^{(0)} \text {is true}\}\\
& \lesssim 
\sum_{\fg \notin O_t} \frac{\kappa}{\gamma} \ind\{L_{\fg} = -1, \xi_\fg = 1, H_{\fg}^{(0)} \text {is true}\} \le \frac{\kappa}{\gamma}(1 + N_t). 
\$
It follows that the true false discovery proportion among the unscreened subgroups,
\[
\text{FDP}(t) =  \frac{\sum_{\fg \notin O_t} \ind\{L_{\fg} = 1, H_{\fg}^{(0)}\text{ is true}\}}{1 \vee P_t}
\lesssim \widehat{\text{FDP}}(t).
\]
In words, $\widehat{\text{FDP}}(t)$ serves as a conservative estimate of FDP$(t)$,
and the choice of the stopping time $\tau$ ensures $\widehat{\text{FDP}}(\tau)$ 
to be bounded by $\alpha$, which in turn yields control of $\text{FDP}(\tau)$.
Of course, the derivation above is only heuristic, as indicated by the use of the 
``$\lesssim$'' notation and by informally treating $\tau$ as a constant.
The rigorous argument requires a more careful conditioning and leave-one-out analysis,
which is provided in Theorem~\ref{thm:fdr-general}. In what follows, 
we write $\cH_{0,\cG} = \{\fg \in \cG:\;H^{(0)}_{\fg}\;\text{is true}\}$
for the collection of null subgroups.


\begin{samepage}
\begin{theorem}\label{thm:fdr-general}
Assume~\eqref{eq:sutva} holds. Given the groups $\cG = \cA_G(\cF)$, suppose the 
group-level sign statistics $(L_{\fg},V_\fg,W_\fg)_{\fg \in \cG}$ satisfy the $\kappa$-boundedness
condition in Definition~\ref{def:bdd-skewness}. The selection set $\hat \cS$
produced by \eqref{eq:SS} satisfies
\[
\mathrm{FDR}(\hat\cS;\cG) = \EE\left[\frac{\sum_{\fg\in \cH_{0,\cG}} \ind\{\fg \in \hat \cS \}}{1 \vee |\hat \cS|} ~\bigg|~\cG\right] \leq \alpha.
\]
\end{theorem}
\end{samepage}
\noindent The proof of Theorem~\ref{thm:fdr-general} is provided in Appendix~\ref{app:proof-fdr-general}. 

\subsection{Further power improvement via conditional calibration}
\label{sec:cc}
When the number of tested groups or the number of 
groups with treatment effects is small, our 
proposed method in Section~\ref{sec:alg} may be underpowered due
to the ``+1'' term in the numerator. Indeed, we need $P_t$ to be 
at least $\kappa/(\gamma \alpha)$ for $\widehat{\text{FDP}}(t)$ to (possibly) drop 
below the required level. We alleviate this issue by applying the 
{\em conditional calibration} technique~\citep{luo2022improving,lee2024boosting} 
to our procedure in Section~\ref{sec:alg} as a post hoc adjustment.

To instantiate the conditional calibration framework, we consider
an auxiliary statistic $A_\fg = h(L_\fg, V_\fg)$ for some 
nonnegative function $h$ and define a modified stopping time 
\$ 
T^\early = \inf\bigg\{t \in \{0\}\cup [|\cI|]: \widehat{\text{FDP}}(t) \le \alpha \text{ or }
P_t < \frac{\kappa}{\gamma \alpha}\bigg\},
\$
where the sequential procedure is stopped earlier in the ``hopeless'' scenario, i.e., 
when $P_t < \kappa / (\gamma \alpha)$.
By construction, $T^\early$ only differs from $T$ when $\hat \cS = \varnothing$. We 
augment $\hat \cS$ as $\hat \cS^{\cc} = \hat \cS \cup \{\fg\in \cG \backslash \hat \cS: 
E_\fg(A_\fg) \le 0\}$,
where, with $L_i \overset{\textnormal{ind}}{\sim} \text{Bern}(p_i)$,
\$
& E_\fg(c) := \sup_{ p_i \in [l_\Gamma, u_\Gamma], i\in\fg}
\EE\bigg[\frac{\ind\{\fg \in \hat \cS \text{ or } A_\fg \ge c\}}
{|\hat \cS \cup \{\fg\}|}  - {{\frac{\alpha \gamma}{\kappa} \cdot}} \frac{\ind\{\fg \notin \cO_{T^\early}, L_\fg = 1\}}{1+N_{T^\early}} \Biggiven \{L_i\}_{i\notin \fg}, \{W_{\fg'},\xi_{\fg'}\}_{\fg'\in \cG}\bigg]
\$
and $(l_\Gamma,u_\Gamma)$ refer to the lower and upper bounds on 
$p_i$ under the $\Gamma$-sensitivity model. 

By construction, $\hat \cS^\cc \supseteq \hat\cS$, and therefore uniformly improves  
the power over the baseline.
The following theorem shows that $\hat \cS^\cc$ also controls the FDR.
\begin{theorem}\label{thm:cc}
Suppose the same assumptions of Theorem~\ref{thm:fdr-general} hold.
The augmented selection set satisfies $\text{FDR}(\hat \cS^\cc;\cG)\le \alpha$;  
\end{theorem}
\noindent We provide the proof in Appendix~\ref{sec:proof-cc} 
and discuss the implementation details in Section~\ref{sec:app-cc}.

\subsection{Comparison with existing work}\label{sec:baseline}
We now take a moment to compare our proposal with the closely related method of~\citet{duan2024interactive},\footnote{While~\citet{duan2024interactive} primarily
focus on discovery at the set level, we consider 
the version for group-level hypotheses introduced in Section 6 of their manuscript.} 
which we  refer to as {\em P-screening}.
Under the matched-pair setting in Section~\ref{sec:warmup} and 
with our notation, 
their procedure omits the initialization step (equivalently, it sets $\gamma \equiv 1$) 
and the conditional calibration step, 
constructing a p-value $p_{\fg}$ for any $\fg \in \cG$. 
These p-values may be obtained, for example, from a Wilcoxon signed-rank test or from a permutation test.
The p-values $p_{\fg}$ are then decomposed into two components, 
which in our notation are defined as 
$L^\pscr_{\fg} = 2\cdot\ind\{p_\fg<0.5\} - 1, V_{\fg}^\pscr = \min\{p_\fg, 1-p_\fg\}$.
Given these statistics, P-screening proceeds in a similar fashion: it sequentially 
screens out subgroups using information available in $\cM_t$, 
maintains an FDP estimate based on $L_{\fg}^{\pscr}$'s,
and stops once the estimated FDP drops below $\alpha$. To ensure FDR control, P-screening 
requires $(L_\fg^\pscr,V_\fg^\pscr, W_\fg)_{\fg \in \cG}$ to satisfy the bounded skewness condition with 
$\kappa = 1$.

In the presence of unmeasured confounders ($\Gamma>1$), however, the p-value-based 
construction presents two issues:
(1) the resulting FDP estimate may be overly conservative and (2) $(L_{\fg}^{\pscr}, V_{\fg}^{\pscr},W_\fg)$  
no longer satisfies the $1$-bounded skewness assumption, thereby invalidating the FDR control guarantee.
We discuss the two issues in detail below.

\begin{figure}[ht]
  \centering
  \begin{subfigure}[t]{0.53\textwidth}
    \centering
    \includegraphics[height=0.45\textwidth]{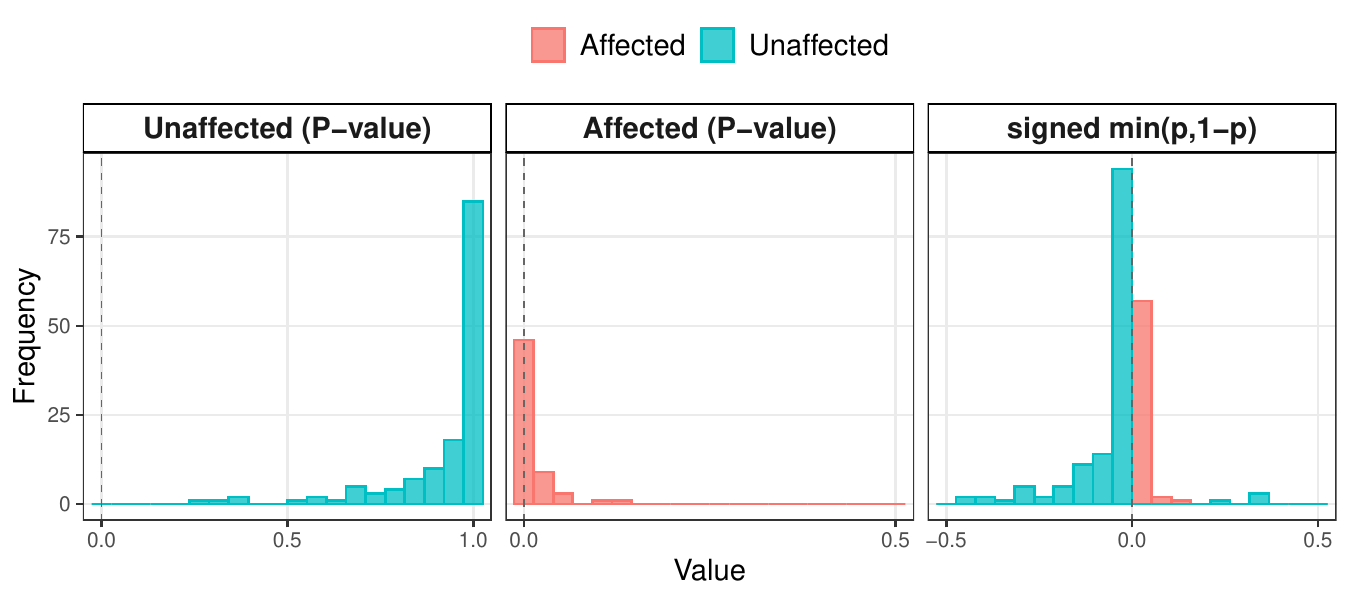}
    \caption{Left: p-values of null groups. Middle: p-values of non-null groups.
    Right: signed p-values $L_\fg^\pscr\times V_{\fg}^\pscr$.}
    \label{fig:pval-1.5-20-og}
  \end{subfigure}
  \hfill
  \begin{subfigure}[t]{0.45\textwidth}
    \centering
    \includegraphics[height=0.45\textwidth]{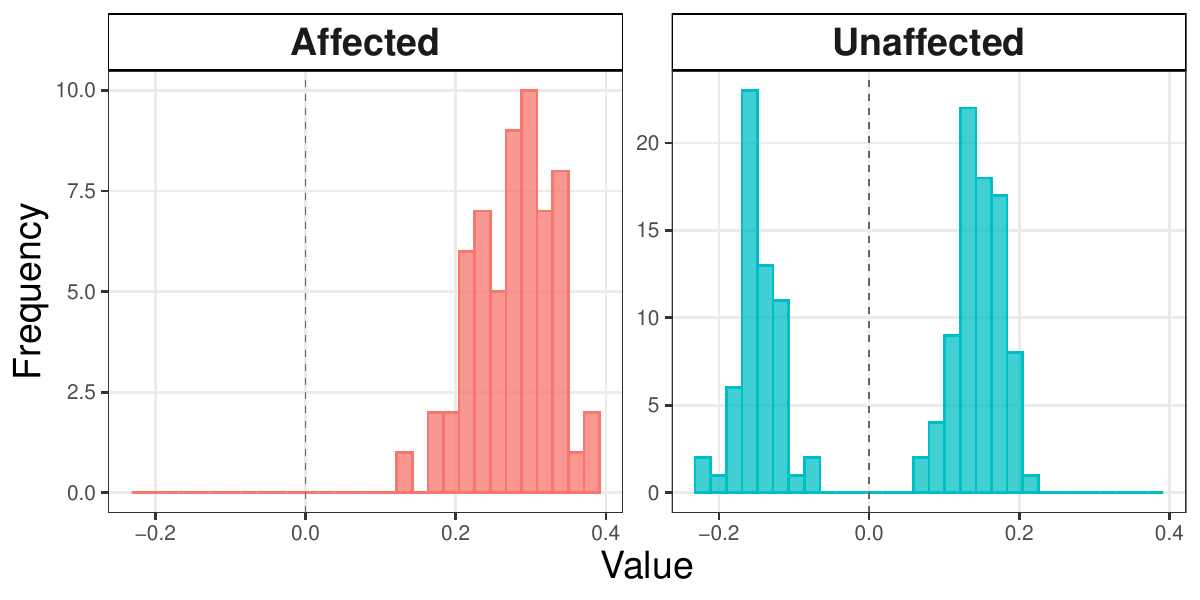}
    \caption{Distribution of $\hat L_\fg \cdot L_{\fg}$ under the alternative (left) and the null (right).}
    \label{fig:dist-WL-oneside}
  \end{subfigure}
  \caption{Histograms of subgroup-level p-values and $\hat L_\fg \cdot L_{\fg}$ ($\Gamma=3.0$).}
  \label{fig:pval-diagnostics}
\end{figure}

\paragraph{Overly conservative FDP estimate.}
The FDP estimate based on $L_\fg^{\pscr}$ relies on the heuristic that,
for any $t \ge 0$,
$\# \{\fg \in \cH_{0,\cG} \cap \cO_t: p_{\fg} < 0.5\}
\lesssim \# \{\fg \in \cO_t: p_{\fg} \ge 0.5\}$.
However, under the $\Gamma$-sensitivity model with $\Gamma > 1$,
null p-values can concentrate near~$1$, making this bound extremely loose.
To formalize this, consider the Wilcoxon signed-rank p-value $p_\fg$
computed under the least-favorable null $L_i^* \overset{\text{iid}}{\sim} \text{Bern}(\tfrac{\Gamma}{1+\Gamma})$,
and let $\Delta_\fg$ measure the standardized gap between the true
treatment-assignment probabilities $p_i$
and the least-favorable boundary $p^*=\tfrac{\Gamma}{1+\Gamma}$
(formal definitions are given in Appendix~\ref{app:proof-conservative-pval}).
\begin{proposition}\label{prop:conservative-pval}
For any $\fg \in \cH_{0,\cG}$, if $\Delta_{\fg}\rightarrow -\infty$ as $|\fg|\rightarrow \infty$,
then $p_{\fg} \rightarrow 1$ in probability.
\end{proposition}
\noindent The proof is in Appendix~\ref{app:proof-conservative-pval}.
As a consequence, even if a small fraction of units---on the order of
$\Omega(|\fg|^{3/4+\epsilon})$---have $p_i$ bounded away from the
least-favorable boundary, the group-level p-value $p_\fg \to 1$,
and the heuristic above becomes vacuous.
Figure~\ref{fig:pval-1.5-20-og} illustrates this empirically
(simulation details in Section~\ref{sec:simu}):
null p-values pile up near~$1$, making it impossible to distinguish
affected from unaffected groups via $V^\pscr_\fg = \min(p_\fg, 1-p_\fg)$.
In contrast, Figure~\ref{fig:dist-WL-oneside} shows that the statistic
$\hat L_\fg \times L_\fg$ remains approximately symmetric under the null
yet clearly shifted for nonnull groups.



\paragraph{Violation of FDR-controlling assumptions.}
The FDR control of P-screening requires 
\@\label{eq:cond_sym}
\PP(L^\pscr_\fg = -1 \given V^\pscr_\fg) \le \PP(L^\pscr_\fg = 1 \given V^\pscr_\fg), 
\text{ for all }\fg \in \cH_{0,\cG}.
\@ 
This condition holds when the p-values are uniformly distributed---or,  
more generally, when they are {\em mirror conservative} in the sense of~\citet{lei2018adapt}.
However, under the sensitivity model, such conditional symmetry is no longer guaranteed;
see the counterexample in Appendix~\ref{app:example}.
Since the condition in~\eqref{eq:cond_sym} may be violated under the sensitivity model, P-screening does not, 
in general, guarantee FDR control unless the screening order is independent of $V_\fg^\pscr$. 
In Appendix~\ref{app:split-method}, we further explore alternative splitting strategies within the 
P-screening framework---such as the decomposition proposed by~\citet{chao2021adapt}---for constructing
$(L_\fg^\pscr,V_\fg^\pscr)$, where we find that similar issues persist.


\section{Subgroup selection with multiple controls}\label{sec:multictrls}
We previously focused on the case $n_i=2$ for all $i \in [I]$, i.e., 
each matched set contains one treated and one control unit. 
We now extend our proposed method to full matching, where multiple controls can be 
matched to a treated unit.  As before, we restrict our attention to testing for a positive 
treatment effect for exposition.

Under the full-matching setting, the main difference lies in 
constructing the matched set-level statistics $(L_i,W_i)$. 
With multiple controls, each matched set contains richer information about the null distribution, 
which can---in principle---enable more powerful procedures. 
The key challenge, however, is how to effectively leverage the additional 
information provided by multiple controls to improve the statistical power 
of subgroup selection. As we will see, a na\"ive 
generalization of the $n_i\equiv 2$ case can in fact lead to power loss, 
while our proposed full-matching variant overcomes the issue.
We detail the construction of $(W_i,L_i)$ below; the remainder of the subgroup selection 
procedure follows as in the $n_i=2$ case.




\subsection{Matched set-level statistics}
Recall that in the $n_i=2$ case, we define the matched set-level sign statistic as
the sign of the treatment-control difference $L_i = \sgn((Z_{i1} - Z_{i2})(R_{i1} - R_{i2}))$.
Equivalently, we may write $L_i = \ind\{r_i=1\}-\ind\{r_i=2\}$ in the absence of ties, 
where $r_i$ is the rank of 
the treated outcome among $\{R_{i1},R_{i2}\}$ in decreasing order. 
This reformulation admits a natural generalization to the 
full-matching setting: for each $i\in[I]$, define
\begin{align}\label{eq:L-multi-ctrl}
L_i = \ind\left\{r_i \le \lfloor n_i / 2 \rfloor\right\} - \ind\left\{r_i \ge \lceil n_i / 2 \rceil + 1\right\},
\end{align}
where $r_i$ is now the rank of the treated outcome among $\{R_{ij}\}_{j\in [n_i]}$
in decreasing order, with the ties broken randomly.

Under the $\Gamma$-sensitivity model, if the null hypothesis $H_i^{(0)}$ is true, 
then for each $r \in [n_i]$, 
\begin{align}\label{eq:prob-sens}
    \PP\left(r_i = r \mid \{R_{i(j)}:\;j \in [n_i]\}\right) & = \PP\left(Z_{i(r)} = 1 \mid \{R_{i(j)}:\;j \in [n_i]\}\right)  
     \in \left[\frac{1}{1 + (n_i - 1)\Gamma}, \frac{\Gamma}{n_i - 1 + \Gamma}\right],
\end{align}
where $R_{i(r)}$ denotes the element in $\{R_{ij}:\;j \in [n_i]\}$ ranked $r$ and 
$Z_{i(r)}$ is its corresponding treatment assignment (conditioning on the tie-breaking randomness if there are ties). 
This characterization implies that $L_i$ obeys a (marginal) bounded skewness condition, 
formalized in Lemma~\ref{lem:marg-bdd-skewness};
its proof is provided 
in Appendix~\ref{app:proof-marg-bdd-skewness}.
\begin{lemma}\label{lem:marg-bdd-skewness}
For any $i\in[I]$, if $H_i^{(0)}$is true, then $L_i$ defined in \eqref{eq:L-multi-ctrl} satisfies 
\$ 
\PP(L_i = 1 \given \cF,\cZ) \le \kappa(n_i,\Gamma) \cdot \PP(L_i = -1 \given \cF,\cZ),
\text{ where }
\kappa(n_i,\Gamma) = \frac{(n_i-1)(\Gamma-1) + \floor{n_i/2}}{\lfloor n_i/2 \rfloor}.
\$
\end{lemma}

We now turn to the construction of magnitude statistics.
Recall that when $n_i \equiv 2$,  $W_i = |Y_i|$ serves as a proxy for the 
matched set-level effect magnitude. Directly extending the idea to
multiple controls leads to several natural choices: 
\@\label{eq:magstat-max}
(\text{Max})~W_i = R_{i(1)}; \quad
(\text{TopGap}) ~W_i = R_{i(1)} - R_{i(2)};\quad
(\text{MedSplit})~W_i = R_{i(1)} - R_{i(\lfloor n_i / 2\rfloor )}.
\@
Here, option (TopGap) is from~\citet{gimenez2019improving}, 
while option (MedSplit) is from~\citet{he2021identification}.
However, these direct generalizations may 
lead to power loss when $n_i$ increases, as illustrated in Figure~\ref{fig:compare_power}
in the introduction.

To understand this seemingly counterintuitive phenomenon, consider the following 
thought experiment. For each $i\in[I]$, suppose $c_{ij} \overset{\text{iid}}{\sim} \cN(0,1)$
and $t_{ij} \overset{\text{iid}}{\sim} \cN(\tau,1)$, for some constant $\tau>0$;
the treatment assignment is uniform in $[n_i]$.
As $n_i \rightarrow \infty$, we have $W_i = O_p(\sqrt{2\log n_i})$
for (Max) and (MedSplit) and $W_i = o_p(1)$ for (TopGap), regardless 
of whether hypothesis $H_i^{(0)}$ is null or non-null. Consequently, $W_i$ constructed in~\eqref{eq:magstat-max}
fails to effectively distinguish non-nulls from nulls.
This gap suggests that the current construction of the magnitude statistics
fails to exploit all available information. To fill in the gap, we introduce the 
``masked'' rank:
\$ 
g(r_i) = \min(r_i, n_i + 1 - r_i).
\$
By construction, $g(r_i)$ does not reveal $L_i = 1$ or $L_i = -1$ whenever it is nonzero.    
Crucially, the sign statistic $L_i$ continues to satisfy a bounded skewness condition even after
conditioning on the masked rank. We formalize this property below, 
and provide its proof in Appendix~\ref{sec:proof-skewness}.


\begin{lemma}\label{lem:multic-L}
For any $i\in[I]$, under $H_i^{(0)}$,  
the statistic $L_i$ defined in \eqref{eq:L-multi-ctrl} satisfies
\$ 
& \PP(L_i = 1\given g(r_i), \cF, \cZ) \le \Gamma \cdot \PP(L_i = -1\given g(r_i), \cF,\cZ).
\$
\end{lemma}
Lemma~\ref{lem:multic-L} shows that $L_i$ retains $\Gamma$-bounded skewness 
{\em conditional } on $g(r_i)$, revealing a form of partial orthogonality between $L_i$ and $g(r_i)$.
This allows us to leverage the additional information in $g(r_i)$ when constructing the magnitude statistic.

\begin{figure}[t]
    \centering
    \includegraphics[width=0.75\linewidth]{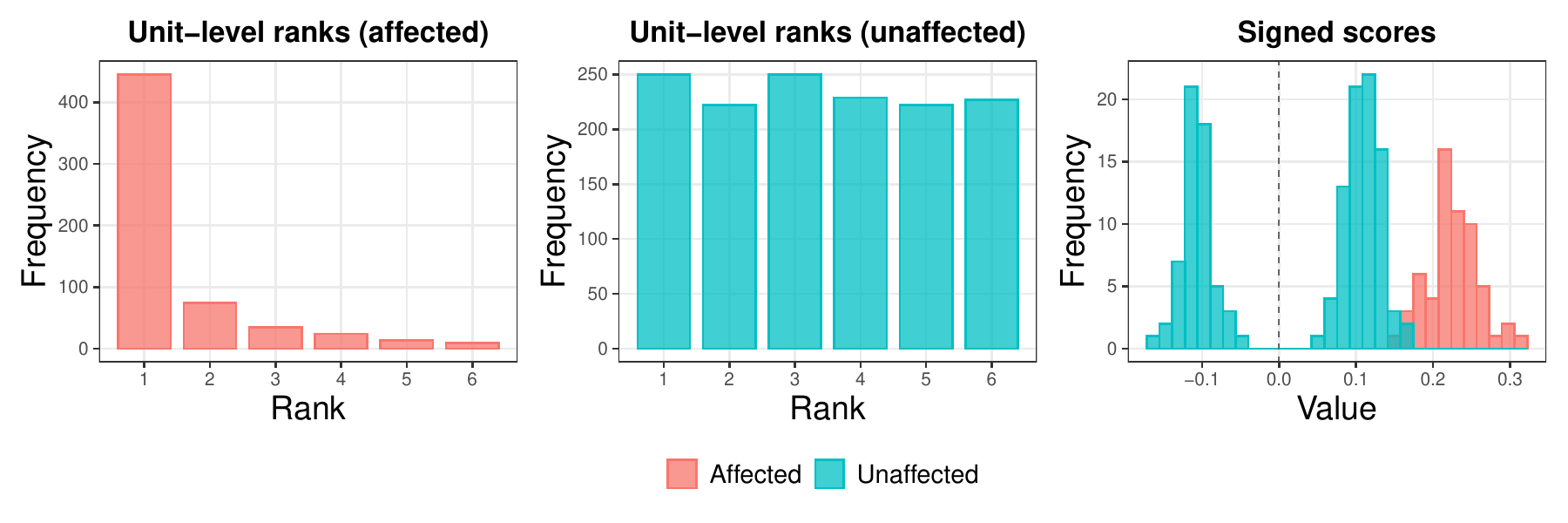}
    \caption{Histograms of unit-level ranks $r_i$ and subgroup-level statistics $L_{\fg} \cdot W^{\rm NP}_{\fg}$ ($5$ controls).}
    \label{fig:hist-multi-ctrl}
\end{figure}

Moreover, as $n_i$ increases, the rank $r_i$ becomes more granular, and $g(r_i)$ 
correspondingly more informative: under the alternative, small values of $g(r_i)$
occur more frequently than under the null (see Figure~\ref{fig:hist-multi-ctrl} for 
illustration). This separation makes it easier for 
us to distinguish nulls from non-nulls.

We note that~\citet{emery2020multiple}---in a different context---also considers the use of ranks in 
testing with multiple negative controls. Our result is more general by accommodating
non-exchangeability among treated and control units due to confounding, 
unequal group sizes, and interactive data analysis. Furthermore, we characterize
the optimal test statistic as a function of the rank and side information under the 
Gaussian model (see Section~\ref{sec:opt}).
The use of masking in our approach also  echoes ideas in~\citet{lei2018adapt,lei2021general,chao2021adapt}, 
among others, albeit in a substantially different context.


\subsection{Desiderata for the optimal magnitude statistic}\label{sec:opt}

We now turn to the design of $W_i$ using $g(r_i)$ together with $\cF$ and $\cZ$.
Given a masked rank $g(r_i)$, let $\cJ$ denote its preimage.
We note that $|\cJ| \in \{1,2\}$, with $|\cJ|=1$ if and only if $n_i$ is odd and $r_i = \lfloor n_i / 2 \rfloor + 1$;
in this case, $L_i=0$ and this matched set effectively does not affect our procedure.
Hence, without loss of generality, we assume $|\cJ|=2$.
We study the optimal form of $W_i$ under a Gaussian model: 
\begin{align}\label{eq:gaussian}
R_{ij} \given X_i \stackrel{\text{ind}}{\sim} \cN(\mu(X_i) + \tau \ind\{j=q\}, \sigma^2), \text{ for }j \in [n_i],
\end{align}
where $q \in [n_i]$ being the unknown index corresponding to the 
treated unit; $\mu(\cdot)$ denotes the baseline heterogeneous outcome, 
and $\tau>0$ corresponds to the positive treatment effect; 
$\sigma^2$ is the noise variance. They are all \emph{unknown}. 
Marginally, the covariate $X_i$ is generated from some unknown covariate distribution $P_X$. 
We denote the preimage of the masked rank as $\cJ_i = \{r_q, r_{q'}\}$, 
where $r_q$ is the rank of the treated unit and $r_{q'}$ is that of the control unit that is masked together with $r_q$.

Recalling the structure of our subgroup selection procedure, 
we see that its selection power is driven by the {\em order} of the groups:
more discoveries can be made if 
the groups with negative sign statistics are identified and revealed early in the 
screening procedure. Motivated by this, we seek to construct $W_i$ that 
optimally distinguishes between $L_i = -1$ and $L_i = 1$.

Concretely, we seek a predictor $\hat L_i \in \{\pm 1\}$, based on 
the masked index set $\cJ_i$, the order statistic $\tilde R_i = (R_{i(j)})_{j \in [n_i]}$,
and the covariate $X_i$, 
that minimizes the following classification error $\ell(\hat L_i) = \PP(L_i = 1, \hat L_i = -1) + \PP(L_i = -1, \hat L_i = 1)$.
Theorem~\ref{thm:optim} establishes that the optimal predictor is the sign of the log-likelihood ratio: 
\begin{align}\label{eq:rejection}
\hat L_i =\sgn\big(\beta(\cJ_i, \tilde R_i,X_i)\big), \text{ where }
\beta(\cJ_i, \tilde R_i,X_i) = \log\frac{\PP(L_i = 1 \mid \cJ_i, \tilde R_i,X_i)}{\PP(L_i = -1 \mid \cJ_i, \tilde R_i,X_i)},
\end{align}
and moreover, $\beta(\cJ_i, \tilde R_i,X_i)$ is a strictly increasing function of $W_i^{{\rm NP}} = | R_{iq} - R_{iq'} |$.
\begin{samepage}
\begin{theorem}\label{thm:optim}
Under the Gaussian model~\eqref{eq:gaussian}, for any fixed $\tau>0$, $q$, $\mu(\cdot)$, and $\sigma^2$, there is 
\begin{enumerate}
\item[\textup{(1)}] 
the predictor $\hat L_i$ defined in~\eqref{eq:rejection} minimizes the classification error $\ell(\hat L_i)$;
\item[\textup{(2)}] the log-likelihood ratio $\beta(\cJ_i,\tilde R_i,X_i)$ is strictly 
increasing in $W^{\rm NP}_i$.
\end{enumerate}
\end{theorem}
\end{samepage}
\noindent See proof in Appendix~\ref{sec:proof-optim}. Theorem~\ref{thm:optim} suggests 
that $\sgn(\beta(\cJ_i,\tilde R_i,X_i))$ provides the best prediction for $L_i$ while
$W_i^{\text{NP}}$ preserves its order; we therefore adopt $W_i = W_i^{\rm NP}$ as the 
default individual-level magnitude statistic, which forms the basis for the ordering. 

\section{Numerical simulations}\label{sec:simu}

This section presents simulation studies that empirically evaluate the proposed method and its variants against existing baselines. 

\subsection{Simulation setup}\label{sec:simu_setup}
For each matched set $i\in[I]$ with $n_i\ge 2$ units, we draw a set-level covariate $X_i\sim\cN(0,I_d)$ with $d=5$ (shared by all units in the set) and assign an ``importance'' label $S_i\in\{0,1\}$ with $\PP(S_i=1)=p_{\rm imp}=0.3$, indicating whether set $i$ is a nonnull subgroup. Each unit has a latent confounder $u_{ij}\sim{\rm Unif}(0,1)$, and the potential outcome is
$R_{ij} = \beta^\top X_i + \alpha_u\, u_{ij} +\tau(X_i) \cdot S_i \cdot \ind\{Z_{ij}=1\} + \varepsilon_{ij}$, where $\alpha_u=0.2$ controls the strength of the latent confounder and $\tau(X_i)$ is a heterogeneous treatment effect.
Exactly one unit per set is treated, with the treated index drawn under Rosenbaum's sensitivity model~\citep{rosenbaum2020design}, under which within-set treatment-probability ratios lie in $[\Gamma^{-1},\Gamma]$ (we use $\Gamma=3.0$).
Due to space constraints, we defer further details of the data-generating process to Appendix~\ref{app:simu_setup}.

\paragraph{Baseline methods.} 
We consider the following methods as baselines: (1) \texttt{BH-baseline} using subgroup-level p-values (see Section~\ref{sec:baseline});
(2) P-value screening (\texttt{P-screening}): we apply \cite{duan2024interactive} but with the masking function proposed in \cite{chao2021adapt} to alleviate the conservativeness under the composite null.\footnote{The sign statistic is defined as $L^\pscr_{\fg} = 1$ if $p_{\fg} \leq \underline{\alpha}$, $L^\pscr_{\fg} = -1$ if $p_{\fg} \in [\lambda, \nu]$, and $L^\pscr_{\fg} = 0$ otherwise. The magnitude statistic is set to $W^\pscr_{\fg} = (\varepsilon + \underline{\alpha}(\nu - p_{\fg})/(\nu - \lambda))^{-1}$ if $p_{\fg} \in [\lambda, \nu]$, and $W^\pscr_{\fg} = (\varepsilon + p_{\fg})^{-1}$ otherwise. 
We adopt the exact adaptive choice of masking thresholds in \cite{chao2021adapt}.}
All the p-values are subgroup-level Wilcoxon signed rank p-values.
We implement our method with $W_{\fg}^{\texttt{NP}}$, as well as the variants 
\texttt{Max}, \texttt{MedSplit}, and \texttt{TopGap} introduced in Section~\ref{sec:multictrls}
for comparison. {Throughout this section, when implementing the statistics aggregation in Section~\ref{sec:L-agg}, we adopt $|\cQ_{\fg}|=\min\{4, \min_{\fg' \in \cG}|\fg'|/2\}$ and $W_{\fg}=\sqrt{|\fg|^{-1}\sum_{i \in \fg} W_i^2}$. Results with different values of $|\cQ_{\fg}|$ can be found in Appendix~\ref{sec:ablate-agg}.}

\paragraph{Conditional calibration.}
We leverage conditional calibration to further boost the power of our methods with both Max-based and NP-based $W_{\fg}$'s. 
Implementation details are deferred to Appendix~\ref{sec:app-cc}.
We use \texttt{-cc} to indicate the use of conditional calibration.

\paragraph{Subgroup partition.}
In simulation settings, we consider two types of subgroup partitions:
\begin{itemize}
    \item \emph{Random partition}: given a targeted number of subgroups $K$, we randomly partition $\{i\in [I]:\;S_i = 1\}$ and $\{i\in [I]:\;S_i = 0\}$ into $\lfloor K \cdot p_{\rm imp} \rfloor$ and $K - \lfloor K \cdot p_{\rm imp} \rfloor$ subgroups. 
    \item \emph{Tree-based partition}: in each setting, with a treatment-independent statistic as the outcome (e.g., $|R_{i1} - R_{i2}|$ in Section~\ref{sec:warmup}), we fit a conditional inference tree \citep{hothorn2006unbiased} (implemented in the R package \texttt{partykit}) for the setting-specific outcome versus covariates and obtain subgroups as a partition of the covariate space. More implementation details can be found in Appendix~\ref{app:simu_setup}.
\end{itemize}
Here, for the random partition, the group size is denoted as $m_{\fg}$, and we accordingly set $I = \lfloor\max\{1200,\;m_{\fg} \cdot 200\}\rfloor$.
For the tree-based partition, from now on, we use group size to refer to \texttt{minsplit} in \texttt{partykit::ctree}, which controls the smallest sample size in each subgroup, and we set $I = 2\lfloor\max\{1200,\;m_{\fg} \cdot 200\}\rfloor$.
For a fair comparison, all methods share the same subgroup learning step in each realization; 
FDR and power are evaluated at the subgroup level based on these data-driven subgroups.

\subsection{Effect modifications with varying group sizes}\label{sec:simu-positive}
In this section, we consider the case with $n_i-1 =3$ control units for each matched set and study the performance of each method with one-sided and two-sided effects, respectively.
We first consider the case where the alternative has positive effects (Section~\ref{sec:warmup}), and vary the group sizes in the random partition in $\{5,10,15,20\}$.
To further incorporate two-sided effects, we flip the sign of treated $R_{ij}$ according to $B_{\fg} \sim {\rm Bern}(0.5)$ for each $i \in \fg$.

\begin{figure}[ht]
  \centering
  \begin{subfigure}[b]{0.49\textwidth}
    \centering
    \includegraphics[height=0.5\textwidth]{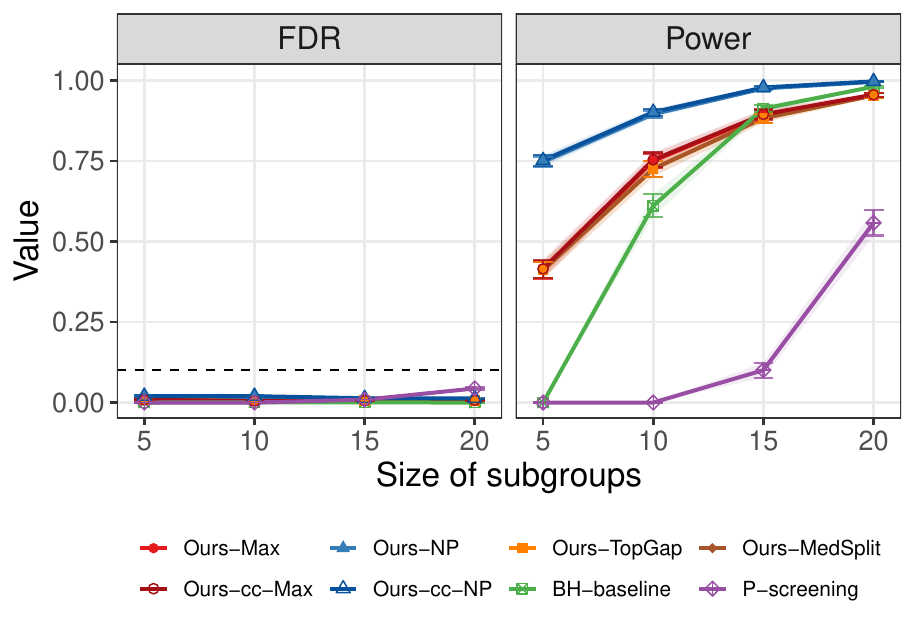}
    \caption{One-sided effects.}
    \label{fig:1side-1out-random}
  \end{subfigure}
  \hfill
  \begin{subfigure}[b]{0.49\textwidth}
    \centering
    \includegraphics[height=0.5\textwidth]{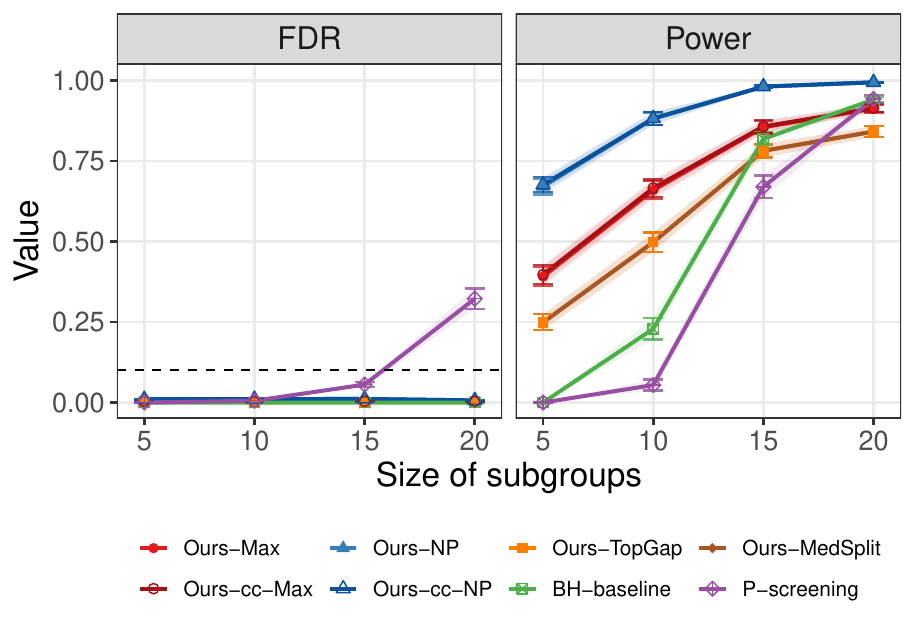}
    \caption{Two-sided effects.}
    \label{fig:2side-1out-random}
  \end{subfigure}
  \caption{FDR and power versus group size.}
  \label{fig:1out-random}
\end{figure}

Figure~\ref{fig:1out-random} reports the averaged FDP and power over 100 simulated datasets against varying group sizes.
All methods control averaged FDP below the nominal level $\alpha = 0.1$ in both panels.
Regarding power, \texttt{Ours-NP}---our method with NP-based $W_{\fg}$--- maintains consistently high discovery power across all group sizes, achieving power around $0.7$ even at the smallest group size of $5$. 
Notably, the performance of \texttt{Ours-NP} dominates other choices of $W_{\fg}$'s, especially when the subgroup size is small.
By contrast, the \texttt{BH-baseline} and \texttt{P-screening} are nearly powerless with the smallest group size and only become competitive once subgroup sizes exceed $15$, since subgroup-level p-values carry limited information about effect sizes in small samples.
The gap is especially pronounced under two-sided effects (Figure~\ref{fig:2side-1out-random}), where \texttt{Ours-NP} achieves power above $0.7$ at all group sizes while the two baselines remain near zero at group sizes $5,10$.
Results with tree-based partitions are deferred to Appendix~\ref{sec:add-simu-group}.

\subsection{Effect modifications with multiple controls}\label{sec:simu-ctrl}
\begin{figure}[ht]
    \centering
    \begin{subfigure}[b]{0.49\textwidth}
        \centering
        \includegraphics[height=0.5\textwidth]{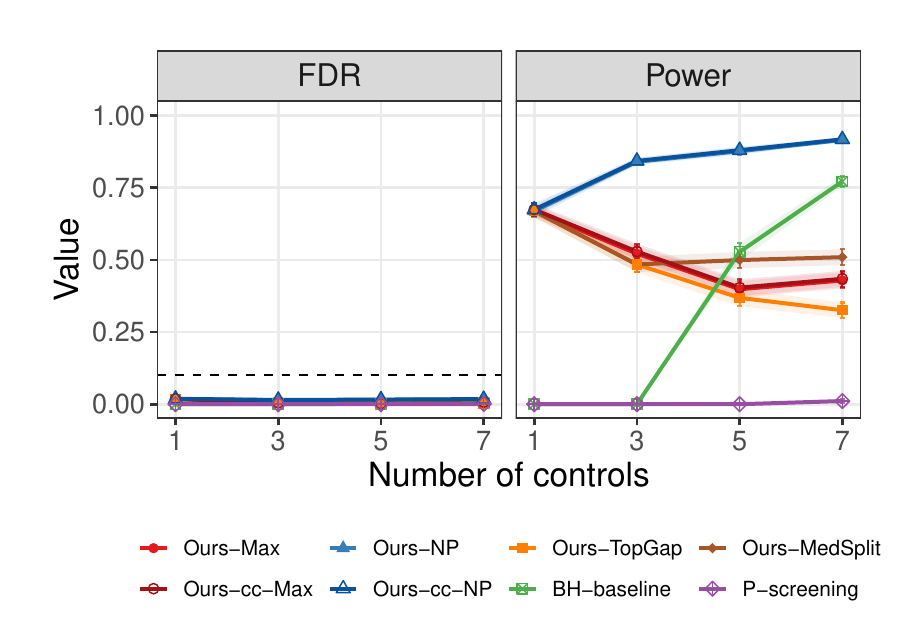}
        \caption{Group size $= 6$.}
        \label{fig:1side-1out-random-ctrl6}
    \end{subfigure}
    \hfill
    \begin{subfigure}[b]{0.49\textwidth}
        \centering
        \includegraphics[height=0.5\textwidth]{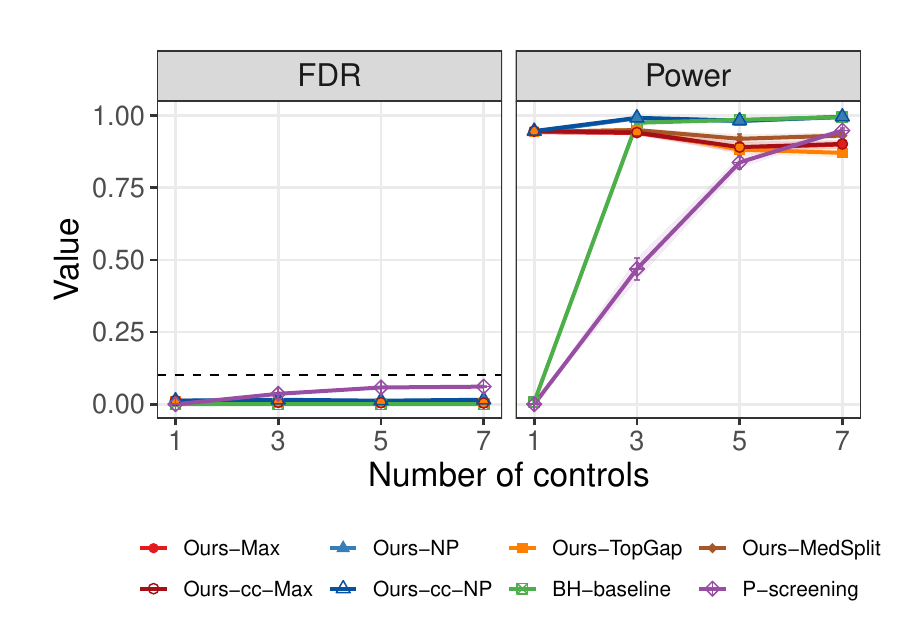}
        \caption{Group size $= 18$.}
        \label{fig:1side-1out-random-ctrl18}
    \end{subfigure}
    \caption{FDR and Power comparison with random subgroup partition. 
    }
    \label{fig:1side-1out-random-ctrl}
\end{figure}

\begin{figure}[ht]
    \centering
    \begin{subfigure}[t]{0.49\linewidth}
        \centering
        \includegraphics[height=0.5\textwidth]{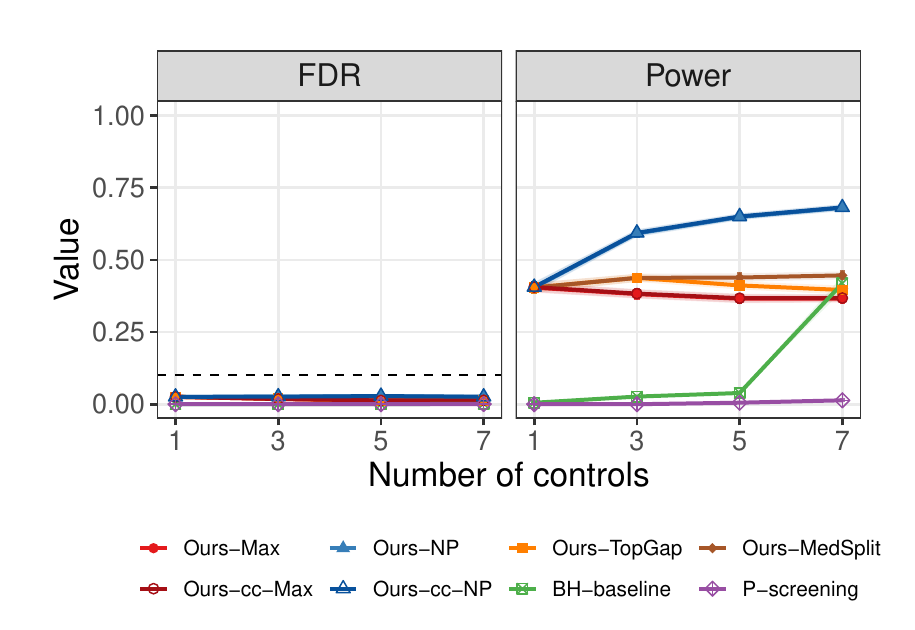}
        \caption{$\texttt{minsplit}=\texttt{minbucket}=2$.}
        \label{fig:1side-1out-tree-ctrl6}
    \end{subfigure}
    \hfill
    \begin{subfigure}[t]{0.49\linewidth}
        \centering
        \includegraphics[height=0.5\textwidth]{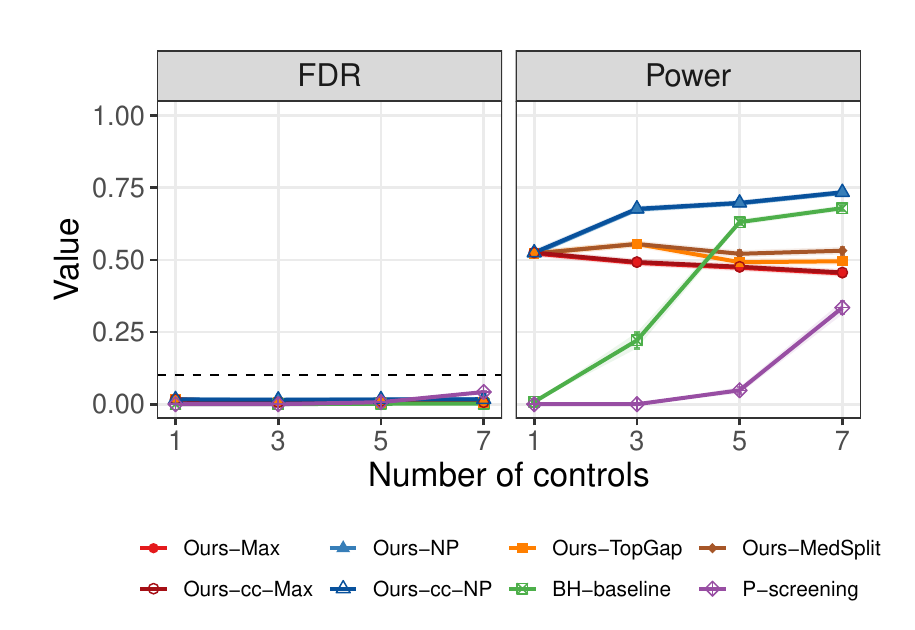}
        \caption{$\texttt{minsplit}=\texttt{minbucket}=4$.}
        \label{fig:1side-1out-tree-ctrl10}
    \end{subfigure}
    \caption{FDR and Power comparison with tree-based subgroup partition. 
    }
    \label{fig:1side-1out-tree-ctrl}
\end{figure}

We adopt the same data-generating process as in Section~\ref{sec:simu-positive} 
but vary the number of units per matched set $n_i$ in $\{2, 3,4,5,6\}$ 
(equivalently, $1, 2, 3, 4, 5$ controls) to investigate how additional controls affect each method. We consider the one-sided effect setting here; results with two-sided effects are deferred to Appendix~\ref{sec:add-simu-ctrl}.
Figures~\ref{fig:1side-1out-random-ctrl} and~\ref{fig:1side-1out-tree-ctrl} report results under random partitions (subgroup sizes $6$ and $18$) and tree-based partitions (with \texttt{minsplit} and \texttt{minbucket} in \texttt{partykit} set to $2$ and $4$), respectively.
The two partition schemes yield qualitatively similar trends: both baselines, \texttt{BH-baseline} and \texttt{P-screening}, are nearly powerless at small subgroups regardless of the number of controls, and \texttt{BH-baseline} becomes competitive only when subgroups are larger while still remaining below or comparable with our approach. \texttt{Ours-Max} and \texttt{Ours-TopGap} lose power when more controls are included, whereas \texttt{Ours-NP} maintains or consistently improves discovery power as the number of controls increases, with the gain more pronounced when the group sizes are small.

\section{Application to benefits of college education}\label{sec:real}

We use the Wisconsin Longitudinal Study (WLS) dataset to examine the heterogeneous effect of college education on earnings. 
The WLS follows $10,317$ men and women who graduated from Wisconsin high schools in 1957, 
with seven rounds of follow-up studies of this cohort designed 
to assess their educational and occupational attainment, well-being, 
and other aspects of life course.
Focusing on participants' earnings, we follow the setup in \cite{brand2010benefits} and 
take the outcome $Y$ to be hourly wages at the age of $53$, measured in 1993.
Following \cite{brand2010benefits}, we include the following $13$ covariates from four aspects:\footnote{More details on the definition of variables are presented in Table~\ref{tab:covariate_name_map} in Appendix~\ref{sec:app-college}.}
(1) {\em family background:} parents' income, father's education, mother's education, location of residence (rural or not), number of siblings, distance to the closest college, and an indicator for whether the graduate lived with both parents up to age 16;
(2) {\em academic performance:} class rank in high school and college preparation;
(3) {\em innate ability:} the Henmon-Nelson Test of Mental Ability scores (IQ);
(4) {\em social support:} teachers' encouragement to attend college, parents' encouragement, and friends' plan for college.

\paragraph{Matching and data-driven subgroups partition.}
We use propensity score nearest neighbor matching. Implementation details on matching are in Appendix~\ref{sec:app-college}. We match two controls to each treated unit (i.e., $n_i=3$), yielding $670$ matched sets, with a $1:2$ matching ratio adopted to retain sufficient matched sets for subgroup partition while maintaining adequate covariate balance.
To determine subgroups for later-stage inference, we fit a conditional inference tree \citep{hothorn2006unbiased} using the magnitude statistic as the response (implemented by \texttt{ctree} included in the R package \texttt{partykit}).
Since the validity guarantee is free of exploration using magnitude statistics, to enhance the interpretability of the defined subgroups, we screen the base $13$ covariates using a regression with magnitude statistics as the outcome, and use a subset of $9$ covariates ranked by exploratory p-values for tree fitting (more implementation details are in Appendix~\ref{sec:app-subgroup}).

\begin{figure}[!t]
  \centering

  \includegraphics[width=0.8\textwidth]{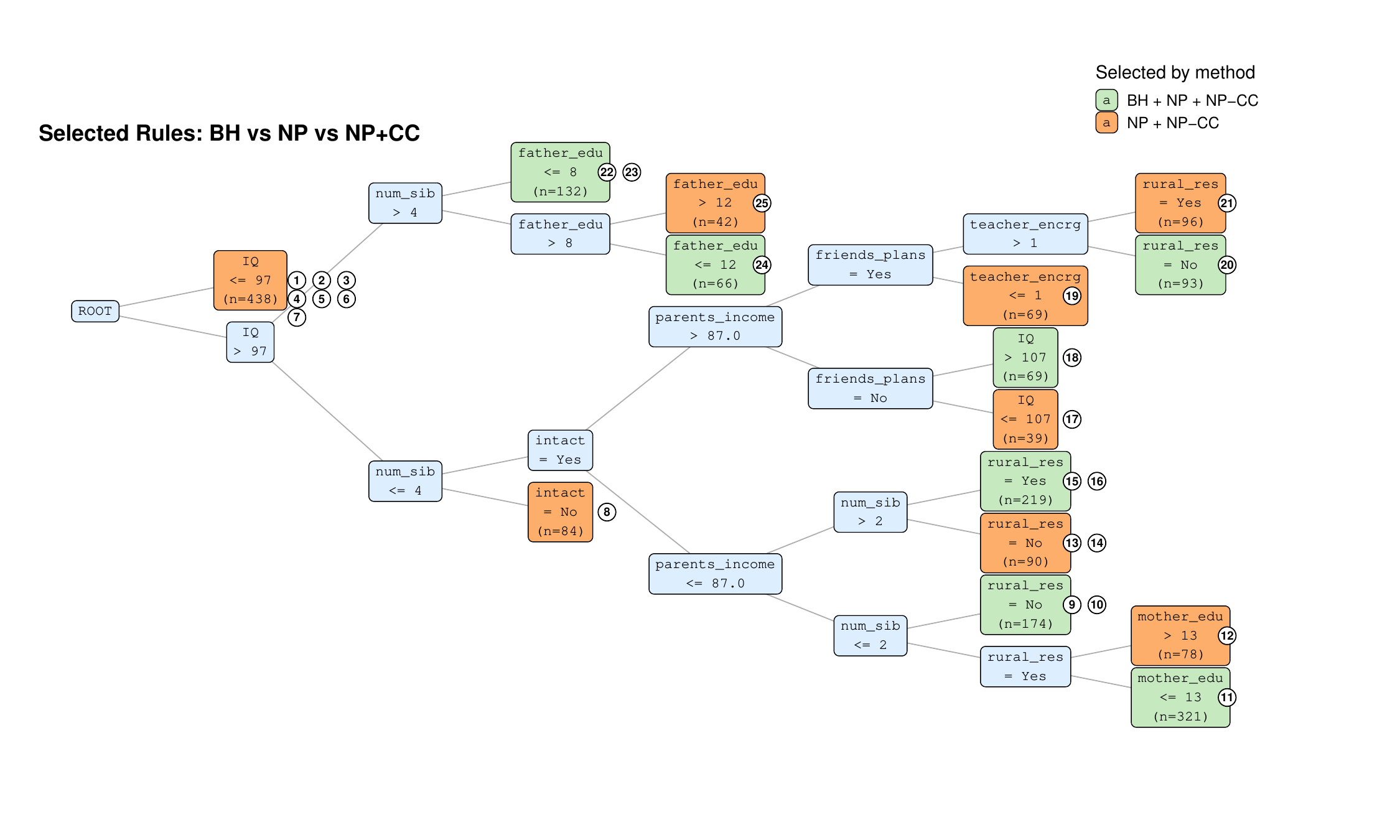}
  \caption{Visualization of subgroups partition and selection: $\Gamma=2$.}
  \label{fig:tree_plots_12}
\end{figure}

\paragraph{Results.}
We compare the performance of our method \texttt{NP}, as well as its conditionally calibrated counterpart \texttt{NP-CC}, with that of the \texttt{BH-baseline} in this section.
Since the P-screening method requires further conditions on the shape of null distributions 
for FDR control, we exclude it from our comparison.
Figure~\ref{fig:tree_plots_12} visualizes all the subgroups under consideration, 
with the colored ones corresponding to those declared significant at $\Gamma=2$.
Table~\ref{tab:sensitivity-fixed-1to2} collects subgroups with top $W_{\fg}$'s among all subgroups (where the magnitude statistic $W_{\fg}$ provides a measure of the extent of benefits of college education) and reports their selection status across $\Gamma\in\{1,1.5,2,2.5,3\}$, where the number of selections at the bottom of Table~\ref{tab:sensitivity-fixed-1to2} are calculated over all subgroups.\footnote{Due to space constraints, when plotting the tree in Figure~\ref{fig:tree_plots_12}, we collapse leaves to their parent node if all leaves of this node are selected by the same method. To present all subgroups, a full Table~\ref{tab:sensitivity-full-varset-lm-top9} is deferred to Appendix~\ref{sec:app-college}.} As $\Gamma$ increases, \texttt{BH} loses power quickly: it retains ten subgroups at $\Gamma=2$ and none beyond that. {Our methods are far more robust, declaring the significance of $25$ data-driven subgroups up to $\Gamma=2.5$ before the selections disappear at $\Gamma=3$.}\footnote{As Table~\ref{tab:sensitivity-full-varset-lm-top9} shows, with smaller values of $\Gamma$, all subgroups are significant. This is because we test the subgroup-level global null: a single matched set with a positive effect suffices to make the entire subgroup nonnull. Since the conditional tree determines subgroups data-adaptively, a coarse partition will naturally yield mostly nonnull subgroups. Table~\ref{tab:sensitivity-full-varset-lm-top9a} presents results under a finer partition for further investigation.} 

Table~\ref{tab:sensitivity-fixed-1to2} only lists the eight subgroups with the highest $W_{\fg}$, the highest estimated benefits of college education.  
We considered a total of $25$ subgroups as shown in Table~\ref{tab:sensitivity-full-varset-lm-top9} of the supplement. All eight subgroups in Table~\ref{tab:sensitivity-fixed-1to2}, the subgroups with the largest estimated benefits of college education, have above-median \texttt{IQ} (\texttt{IQ > 97}). Table~\ref{tab:sensitivity-full-varset-lm-top9} of the supplement shows that although the subgroups with below-median \texttt{IQ} (\texttt{IQ <= 97}) had less estimated benefits of college education, all the subgroups did have positive estimated benefits of college education and were selected as having evidence of positive benefits up to $\Gamma =2.5$.

An interesting feature of the eight subgroups in Table~\ref{tab:sensitivity-fixed-1to2} with the highest estimated benefits of college education is that although they all have above-median \texttt{IQ} (\texttt{IQ > 97}), seven of the eight subgroups consist of graduates who face some structural or resource limitation, e.g., rural residence, limited parental education, or a non-intact family.

{\color{blue}
\renewcommand{\arraystretch}{0.6}
\begin{table}[t]
\centering
\caption{Top 8 subgroups ranked by $W_{\fg}$.
\fullcirc\ = BH + NP + NP-CC; \halfcirc\ = NP/NP-CC only; \bhonly\ = BH only; \notsel\ = not significant.}
\label{tab:sensitivity-fixed-1to2}
\begin{tabular}{@{} >{\raggedright\arraybackslash\baselineskip=10pt}p{9.5cm} r r ccccc @{}}
\toprule
Subgroup & $W_{\fg}$ & $|\fg|$ & $\Gamma{=}1.0$ & $1.5$ & $2.0$ & $2.5$ & $3.0$ \\
\midrule
$\fg_{23}$: \texttt{IQ > 97}, \texttt{num\_sib > 4}, \texttt{father\_edu <= 8}, \texttt{mother\_edu > 10}
  & 617.5 & 51 & \fullcirc & \fullcirc & \fullcirc & \halfcirc & \notsel \\
$\fg_{19}$: \texttt{IQ > 97}, \texttt{num\_sib <= 4}, \texttt{intact = 1}, \texttt{parents\_income > 87.0}, \texttt{friends\_plans = 1}, \texttt{teacher\_encrg <= 1}
  & 384.0 & 69 & \fullcirc & \fullcirc & \halfcirc & \halfcirc & \notsel \\
$\fg_{22}$: \texttt{IQ > 97}, \texttt{num\_sib > 4}, \texttt{father\_edu <= 8}, \texttt{mother\_edu <= 10}
  & 372.4 & 81 & \fullcirc & \fullcirc & \fullcirc & \halfcirc & \notsel \\
$\fg_{8}$: \texttt{IQ > 97}, \texttt{num\_sib <= 4}, \texttt{intact = 0}
  & 369.9 & 84 & \fullcirc & \fullcirc & \halfcirc & \halfcirc & \notsel \\
$\fg_{20}$: \texttt{IQ > 97}, \texttt{num\_sib <= 4}, \texttt{intact = 1}, \texttt{parents\_income > 87.0}, \texttt{friends\_plans = 1}, \texttt{teacher\_encrg > 1}, \texttt{rural\_res = No}
  & 321.4 & 93 & \fullcirc & \fullcirc & \fullcirc & \halfcirc & \notsel \\
$\fg_{9}$: \texttt{IQ > 97}, \texttt{num\_sib <= 4}, \texttt{intact = 1}, \texttt{parents\_income <= 87.0}, \texttt{num\_sib <= 2}, \texttt{rural\_res = No}, \texttt{father\_edu <= 8}
  & 301.2 & 84 & \fullcirc & \fullcirc & \fullcirc & \halfcirc & \notsel \\
$\fg_{16}$: \texttt{IQ > 97}, \texttt{num\_sib <= 4}, \texttt{intact = 1}, \texttt{parents\_income <= 87.0}, \texttt{num\_sib > 2}, \texttt{rural\_res = Yes}, \texttt{father\_edu > 7}
  & 246.7 & 177 & \fullcirc & \fullcirc & \fullcirc & \halfcirc & \notsel \\
$\fg_{21}$: \texttt{IQ > 97}, \texttt{num\_sib <= 4}, \texttt{intact = 1}, \texttt{parents\_income > 87.0}, \texttt{friends\_plans = 1}, \texttt{teacher\_encrg > 1}, \texttt{rural\_res = Yes}
  & 218.5 & 96 & \fullcirc & \halfcirc & \halfcirc & \halfcirc & \notsel \\
\midrule
\# selected by \texttt{BH} & & & 19 & 15 & 10 & 0 & 0 \\
\# selected by \texttt{NP} / \texttt{NP-CC} & & & 25 & 25 & 25 & 25 & 0 \\
\bottomrule
\end{tabular}
\end{table}
}



\section*{Acknowledgement}
The authors would like to thank the Wharton Research Computing team for the computational resources provided and the great support from the staff members.
Z.~Ren is supported by the NSF grant DMS-2413135 and 
the Wharton AI \& Analytics Initiative's AI Research Fund.

\noindent\textbf{Reproducibility.}
Code and data for reproducing the numerical results in this paper are available at \url{https://github.com/yugjerry/ada_subgroup_selection}.

\bibliographystyle{apalike}
\bibliography{ref}

\newpage
\appendix

\section{Additional related work}\label{app:related}

\paragraph{Subgroup partition and selection.}
Subgroup analysis has been of significant interest in a broad range of applications, such as stratified medicine in clinical trials \citep{stallard2014adaptive,rothwell2005subgroup,lipkovich2024modern} and heterogeneous treatment-effect discovery in program and policy evaluation \citep{athey2016recursive}. 
Identifying subgroups from data is the first emerging question.
There is a rich literature on learning subgroups from data, in particular on tree-based methods, including the CART algorithm \citep{su2009subgroup,breiman2017classification}, causal trees \citep{athey2016recursive}, and causal forests \citep{wager2018estimation}.

The data-driven nature of tree-based approaches is a strength, but it can also be a liability: with data-dependent subgroups as inferential targets, how to characterize or relieve the selection bias has become quite challenging, {making valid error control elusive without data splitting.}

A thread of earlier works assumes a fixed class of subgroups \citep{bonetti2004patterns,li2023statistical,chernozhukov2025policy}, with which one can perform simultaneous inference over the entire class, which avoids data-driven partitioning and post-selection complications, at the cost of restricting attention to an analyst-chosen subgroup family and limiting the flexibility of subgroup partition. 
A generic solution that enables valid inference on subgroups is data splitting \citep{cox1975note}, which is assumption-free but may lead to a loss in discovery power due to a reduced sample size.
Recursive/repeated data splitting \citep{chernozhukov2018generic}, on the other hand, can be ambiguous about the inferential targets as inference for subgroups is performed on an aggregation of multiple confirmatory datasets.

\paragraph{Post-selection validity and multiplicity adjustment of subgroup selection.}
Given a collection of data-driven subgroups, 
two key questions arise:
(1) how to identify and select subgroups with nonzero effects using the same dataset, and
(2) how to control the (in-sample) false discoveries, while conditioning on data-driven subgroups?
Several lines of work address these questions in different ways. 
The first type of work focuses on the {\em conditional average treatment effect (CATE)}, i.e., 
the average treatment effect conditional on the observed covariates. 
The goal there is to identify a subset $\cR$ in the covariate space such that 
$\mu(\cR)$---the average treatment effect given the event that $X\in\cR$---exceeds 
some pre-specified threshold $\tau_0$ with probability at least $1-\alpha$.
As demonstrated in the literature, (asymptotically) valid selection of $\cR$ is feasible 
when CATE can be estimated efficiently, allowing for the construction of asymptotically valid 
confidence bands
\citep{armstrong2015inference,guo2021inference,ritzwoller2024simultaneous,wager2018estimation}. 
Recent work further extends these results to classes of CATE functionals under mild shape constraints \citep{reeve2023optimal,muller2025isotonic}. 
More recently,~\citet{cheng2025chiseling} introduces an interactive approach that achieves type-I error 
control in search of $\cR$ without data splitting, although its extension to observational data remains an open question.

The error metric $\PP(\mu(\cR) \leq \tau_0)$ in the aforementioned literature 
is defined at the population level. 
In observational studies with a finite sample, however,  
analysts often seek an {\em in-sample} notion of error control,
where the error rate is defined conditional on the observed sample.
To this end, earlier work by \citet{hsu2013effect,lee2018discovering,fan2024subgroup} 
focuses on controlling the type-I error for testing the global null across subgroups,
while \cite{hsu2015strong} propose a subgroup selection approach that controls
the {\em family-wise error rate (FWER)} among selected subgroups. 
Moving to the FDR \citep{benjamini1995controlling,glickman2014false},
which provides a more informative criterion in large-scale multiple testing problems,
\cite{karmakar2018false} apply the Benjamini-Hochberg (BH) procedure to 
subgroup-level p-values for FDR control with data-driven subgroups. 
More recently, \cite{duan2024interactive} introduced an interactive procedure for subgroup selection 
with FDR control. Their method is the closest to ours, but designed primarily for 
the matched-pair setting without unmeasured confounding. A more detailed comparison 
is described in Section~\ref{sec:baseline}.

\section{Technical proofs}
\subsection{Proof of Lemma~\ref{lem:skewness}}
\label{sec:proof-lemma1}
By definition, $L_{\fg} = 1$ if and only if $\sum_{i\in \cQ_\fg} \ind\{L_i = 1\} > \eta_\fg$.
Thus, we have
\begin{align}
\PP(L_{\fg}  = 1 \mid V_\fg, W_\fg, \cF, \cZ)
&=\PP\left(\sum_{i\in \cQ_\fg} \ind\{L_i = 1\} > \eta_\fg
   \,\Big|\, V_\fg, W_\fg, \cF, \cZ\right)\\
&\le \PP\left(\mathrm{Bin}\bigl(|\cQ_\fg|,\tfrac{\Gamma}{1+\Gamma}\bigr) > \eta_\fg\right),
\label{eq:bscs}
\end{align}
where the inequality follows from the fact that, under $H_{\fg}^{(0)}$, 
the distribution of $\sum_{i\in \cQ_\fg} \ind\{L_i = 1\}$ is stochastically dominated by $\mathrm{Bin}(|\cQ_\fg|,\tfrac{\Gamma}{1+\Gamma})$ conditional on $(V_\fg, W_\fg, \cF, \cZ)$.
\eqref{eq:bscs} implies 
\@\label{eq:bscs2}
\PP(L_{\fg} = -1 \mid V_\fg, W_\fg, \cF, \cZ) \ge 
\PP(\mathrm{Bin}(|\cQ_\fg|,\tfrac{\Gamma}{1+\Gamma}) \le \eta_\fg).
\@
Combining~\eqref{eq:bscs} and~\eqref{eq:bscs2}, we have
\$
\PP(L_{\fg} = 1 \mid V_\fg, W_\fg, \cF, \cZ) & \le \PP\left(\mathrm{Bin}\bigl(|\cQ_\fg|,\tfrac{\Gamma}{1+\Gamma}\bigr) > \eta_\fg\right)\\
& = \frac{\PP\left(\mathrm{Bin}\bigl(|\cQ_\fg|,\tfrac{\Gamma}{1+\Gamma}\bigr) > \eta_\fg\right)}{\PP\left(\mathrm{Bin}\bigl(|\cQ_\fg|,\tfrac{\Gamma}{1+\Gamma}\bigr) \le \eta_\fg\right)} \cdot \PP\left(\mathrm{Bin}\bigl(|\cQ_\fg|,\tfrac{\Gamma}{1+\Gamma}\bigr) \le \eta_\fg\right)\\
& \le \frac{\PP\left(\mathrm{Bin}\bigl(|\cQ_\fg|,\tfrac{\Gamma}{1+\Gamma}\bigr) > \eta_\fg\right)}{\PP\left(\mathrm{Bin}\bigl(|\cQ_\fg|,\tfrac{\Gamma}{1+\Gamma}\bigr) \le \eta_\fg\right)} \cdot \PP(L_{\fg} = -1 \mid V_\fg, W_\fg, \cF, \cZ).
\$
That is, the group-level statistics $(L_{\fg}, V_\fg, W_\fg)$ we construct  
satisfy the bounded skewness condition with $\kappa = \PP(\mathrm{Bin}(|\cQ_\fg|,\tfrac{\Gamma}{1+\Gamma}) > \eta_\fg)/\PP(\mathrm{Bin}(|\cQ_\fg|,\tfrac{\Gamma}{1+\Gamma}) \leq \eta_\fg)$.
Finally, by the choice of $\eta_\fg$ in~\eqref{eq:eta-def}, we have $\PP(\mathrm{Bin}(|\cQ_\fg|,\tfrac{\Gamma}{1+\Gamma}) > \eta_\fg) \le \Gamma/(1+\Gamma)$ and $\PP(\mathrm{Bin}(|\cQ_\fg|,\tfrac{\Gamma}{1+\Gamma}) \le \eta_\fg) \ge 1/(1+\Gamma)$, which implies $\kappa \le \Gamma$.

\subsection{Proof of Theorem~\ref{thm:fdr-general}}
\label{app:proof-fdr-general}
Throughout the proof, we condition on $(\cG,\cF,\cZ)$, and omit 
the conditioning whenever the context is clear. We also define
\$ 
\bL = (L_{\fg})_{\fg\in\cG},~\Xi = (\xi_\fg)_{\fg \in \cG},~\bV = (V_\fg)_{\fg \in \cG},
~\bY = (Y_i)_{ i \in [I]},
\$
and write $\cO_t = \cO_t(\bL,\Xi)$ to emphasize its dependence on $(\bL, \Xi)$.

Recalling the definition of $\hat \cS$, we have 
\$ 
\mathrm{FDR}(\hat\cS;\cG) & 
= \EE\left[\frac{\sum_{\fg\in \cH_{0,\cG}} \ind\{\fg \notin \cO_{\tau}(\bL, \Xi), L_{\fg} = 1 \}}
{1 \vee \sum_{\fg' \in \cG}\ind\{\fg' \notin \cO_{\tau}(\bL, \Xi), L_{\fg'} = 1\} }\right]\\
& = \EE\left[\frac{\sum_{\fg\in \cH_{0,\cG}} \ind\{\fg \notin \cO_{\tau}(\bL, \Xi), L_{\fg} = 1 \}}
{\frac{\kappa}{\gamma}(1+N_{\tau})} \cdot \widehat{\text{FDP}}(\tau) \right]\\
& \le \frac{\gamma \alpha}{\kappa} \cdot 
\EE\left[\frac{\sum_{\fg \in \cH_{0,\cG}} \ind\{\fg \notin \cO_{\tau}(\bL, \Xi), L_{\fg} = 1 \}}
{1 + N_\tau }\right]\\
& = \frac{\gamma \alpha}{\kappa} \cdot \sum_{\fg \in \cH_{0,\cG}} 
\EE\left[\frac{\ind\{\fg \notin \cO_{\tau}(\bL, \Xi), L_{\fg} = 1 \}}
{1 + \sum_{\fg' \in \cH_{0,\cG}, \fg' \neq \fg}\ind\{\fg' \notin \cO_{\tau}(\bL, \Xi), L_{\fg'} = -1\} }\right],
\$
where the inequality follows from the choice of $\tau$.
It suffices to show that
\[
\sum_{\fg \in \cH_{0,\cG}}\;\EE\left[\frac{\ind\{\fg \notin \cO_{\tau}(\bL,\Xi), L_{\fg} = 1 \}}
{1 + \sum_{\fg' \in \cH_{0,\cG}, \fg' \neq \fg}\ind\{\fg' \notin \cO_{\tau}(\bL), L_{\fg'} = -1\} }
\right] \leq \frac{\kappa}{\gamma}.
\]
We carry out the proof in two steps.

\paragraph{Step 1: leave-one-out substitution.}
Fix any $g\in \cH_{0,\cG}$. Define $\tau_\fg$ as the stopping time 
that would result if $L_\fg$ were set to $+1$, while all other sign statistics remain unchanged.
That is, if we write the original stopping time as $\tau = \cT(\bL)$, 
then $\tau_{\fg} = \cT(\bL^{\fg,+})$, where $\bL^{\fg,+}$ denotes the vector obtained 
from $\bL$ by replacing $L_\fg$ with $+1$. 

On the event $\{L_{\fg} = 1\}$, it follows by definition that $\tau = \tau_{\fg}$ and 
$\cO_{\tau}(\bL, \Xi) = \cO_{\tau_{\fg}}(\bL^{\fg,+},\Xi)$. Therefore, 
\@\label{eq:loo} 
& \EE\left[\frac{\ind\{\fg \notin \cO_{\tau}(\bL, \Xi), L_{\fg} = 1 \}}
{1 + \sum_{\fg' \in \cH_{0,\cG}, \fg' \neq \fg}\ind\{\fg' \notin \cO_{\tau}(\bL,\Xi), L_{\fg'} = -1\} }\right] \notag \\
= ~&  
\EE\left[\frac{\ind\{\fg \notin \cO_{\tau_{\fg}}(\bL^{\fg,+},\Xi), L_{\fg} = 1 \}}
{1 + \sum_{\fg' \in \cH_{0,\cG}, \fg' \neq \fg}
\ind\{\fg' \notin \cO_{\tau_{\fg}}(\bL^{\fg,+}, \Xi), L_{\fg'} = -1\} }\right] \notag \\
= ~&  
\EE\left\{\EE\left[\frac{\ind\{\fg \notin \cO_{\tau_{\fg}}(\bL^{\fg,+},\Xi), L_{\fg} = 1 \}}
{1 + \sum_{\fg' \in \cH_{0,\cG}, \fg' \neq \fg}\ind\{\fg' \notin \cO_{\tau_{\fg}}(\bL^{\fg,+},\Xi), L_{\fg'} = -1\} } \bigggiven \bL_{-\fg}, \Xi, \bV, |\bY|\right] \right\} \notag \\
 =~& \EE\left[
\frac{\ind\{\fg \notin \cO_{\tau_{\fg}}(\bL^{\fg,+},\Xi)\} \PP(L_{\fg} = 1 \given \bL_{-\fg},\Xi, \bV, |\bY|)}
{1 + \sum_{\fg' \in \cH_{0,\cG}, \fg' \neq \fg}\ind\{\fg' \notin \cO_{\tau_{\fg}}(\bL^{\fg,+},\Xi), L_{\fg'} = -1\} }\right].
\@
Above, $\bL_{-\fg}$ denotes the vector obtained from removing $L_\fg$ from $\bL$ and 
the last step is because $(\tau_{\fg}, \cO_{\tau_{\fg}}(\bL^{\fg,+},\Xi))$ are fully 
determined given $(\bL_{-\fg}, \Xi, \bV, |\bY|)$.

Since $L_{\fg} \indep (\bL_{-\fg}, \bV, \Xi)$ and by the bounded skewness condition, 
\@\label{eq:bounded-skew}
\PP(L_{\fg} = 1 \given \bL_{-\fg}, \bV, \Xi, |\bY|)
& = \PP(L_{\fg} = 1) \notag \\
& \le \kappa \cdot \PP(L_{\fg} = -1)
= \kappa \cdot \PP(L_{\fg} = -1 \given \bL_{-\fg}, \bV, \Xi,|\bY|). 
\@ 
Plugging~\eqref{eq:bounded-skew} into~\eqref{eq:loo}, we have 
\@\label{eq:loo_negative}
\eqref{eq:loo} & \le  \kappa \cdot \EE\left[
\frac{\ind\{\fg \notin \cO_{\tau_{\fg}}(\bL^{\fg,+},\Xi)\} \cdot 
\PP(L_{\fg} = -1 \given \bL_{-\fg}, \bV, \Xi, |\bY|)}
{1 + \sum_{\fg' \in \cH_{0,\cG}, \fg' \neq \fg}
\ind\{\fg' \notin \cO_{\tau_{\fg}}(\bL^{\fg,+},\Xi), L_{\fg'} = -1\}}\right] \notag\\
& \qquad \qquad = 
\kappa \cdot \EE\left[\frac{\ind\{\fg \notin \cO_{\tau_{\fg}}(\bL^{\fg,+},\Xi), L_{\fg} = -1 \}}
{\sum_{\fg' \in \cH_{0,\cG}}\ind\{\fg' \notin \cO_{\tau_{\fg}}(\bL^{\fg,+},\Xi), L_{\fg'} = -1\} }
\right].
\@
Writing $\Xi^{\fg,+}$ as the vector obtained from $\Xi$ by replacing $\xi_\fg $ with $1$, 
we can check that $\cO_{\tau_g}(\bL^{\fg, +}, \Xi) = \cO_{\tau_g}(\bL^{\fg, +}, \Xi^{\fg,+})$, 
and is therefore independent of $\xi_\fg$. As a result, 
\$ 
\eqref{eq:loo_negative} & = \kappa \cdot \EE\left[\frac{\ind\{\fg \notin \cO_{\tau_{\fg}}(\bL^{\fg,+},\Xi^{\fg,+}), L_{\fg} = -1 \}}
{\sum_{\fg' \in \cH_{0,\cG}}\ind\{\fg' \notin \cO_{\tau_{\fg}}(\bL^{\fg,+},\Xi^{\fg,+}), L_{\fg'} = -1, \xi_{\fg'} = 1\} }
\right]\\
& = \frac{\kappa}{\gamma} \cdot\EE\left[\frac{\ind\{\fg \notin \cO_{\tau_{\fg}}(\bL^{\fg,+},\Xi), L_{\fg} = -1, \xi_\fg = 1\}}
{\sum_{\fg' \in \cH_{0,\cG}}\ind\{\fg' \notin \cO_{\tau_{\fg}}(\bL^{\fg,+},\Xi), L_{\fg'} = -1, \xi_{\fg'} = 1\} }
\right]
\$

\paragraph{Step 2: swapping stopping times and screening sets.}
Next, we will show that for any $\fg, \fg' \in \cH_{0,\cG}$,
\begin{equation}
\label{eq:swap}
\begin{aligned}
& \big\{\fg \notin \cO_{\tau_{\fg}}(\bL^{\fg,+}, \Xi), L_{\fg} = -1, \xi_\fg = 1\big\} \cap 
\big\{\fg' \notin \cO_{\tau_{\fg}}(\bL^{\fg,+}, \Xi), L_{\fg'} = -1, \xi_{\fg'}=1 \big\}\\
& \qquad \qquad \qquad 
\subseteq \{\tau_\fg = \tau_{\fg'}, \cO_t(\bL^{\fg,+},\Xi) = \cO_t(\bL^{\fg',+},\Xi), \forall t \le \tau_\fg \}.
\end{aligned}
\end{equation}
and 
\begin{equation}
\label{eq:swap2}
\begin{aligned}
& \big\{\fg \notin \cO_{\tau_{\fg}}(\bL^{\fg,+}, \Xi), L_{\fg} = -1, \xi_\fg = 1\big\} \cap 
\big\{\fg' \notin \cO_{\tau_{\fg'}}(\bL^{\fg',+}, \Xi), L_{\fg'} = -1, \xi_{\fg'}=1 \big\}\\
& \qquad \qquad \qquad 
\subseteq \{\tau_\fg = \tau_{\fg'}, \cO_t(\bL^{\fg,+},\Xi) = \cO_t(\bL^{\fg',+},\Xi), \forall t \le \tau_\fg \}.
\end{aligned}
\end{equation}

To see~\eqref{eq:swap}, suppose $\{\fg \notin \cO_{\tau_{\fg}}(\bL^{\fg,+},\Xi), L_{\fg} = -1, \xi_\fg = 1\}$ 
and $\{\fg' \notin \cO_{\tau_{\fg}}(\bL^{\fg,+}, \Xi), L_{\fg'} = -1, \xi_{\fg'} = 1\}$ both hold.
We show $\cO_t(\bL^{\fg,+}) = \cO_t(\bL^{\fg',+})$, for $\forall t \le \tau_\fg$, by induction.
\begin{itemize}
\item For the base case $t=0$, 
\$
\cO_0(\bL^{\fg,+},\Xi) = \{\tilde \fg \in \cG: \xi_{\tilde \fg} = 0, 
\bL^{\fg,+}_{\tilde \fg} = -1  \}
= \{\tilde \fg \in \cG: \xi_{\tilde \fg} = 0, \bL^{\fg',+}_{\tilde \fg} = -1  \}
= \cO_0(\bL^{\fg',+},\Xi).
\$ 
\item For the inductive step, suppose 
$\cO_s(\bL^{\fg,+},\Xi) = \cO_s( \bL^{\fg',+},\Xi)$ for all $s \le t < \tau_\fg$. 
Since $\fg, \fg' \in \cO_{\tau_\fg}^c(\bL^{\fg,+},\Xi) \subset \cO_{t}^c(\bL^{\fg,+},\Xi)$, 
there is $\fg,\fg' \in \cO_{t}^c(\bL^{\fg',+},\Xi)$ by the inductive hypothesis. 
As a result, $\bL^{\fg,+}_{\tilde \fg} = \bL^{\fg',+}_{\tilde \fg}$, for 
$\forall \tilde \fg \in \cO_t(\bL^{\fg,+},\Xi)=\cO_t(\bL^{\fg',+},\Xi)$.
Recalling the definition of $\cM_{t}$, we have 
by construction that $\cO_{t+1}(\bL^{\fg,+},\Xi)$  
depends on $\bL^{\fg,+}$ 
only through $\bL^{\fg,+}_{\tilde \fg}$ 
for which $\tilde \fg \in \cO_{t}(\bL^{\fg,+},\Xi)$. 
The same reasoning applies to $\cO_{t+1}(\bL^{\fg',+},\Xi)$.
We then conclude that $\cO_{t+1}(\bL^{\fg,+},\Xi) = \cO_{t+1}(\bL^{\fg',+},\Xi)$.
\end{itemize}
Next, note that the FDP estimate at $\tau_\fg$ with $\bL^{\fg,+}$ is 
\@\label{eq:stop-taug}
\widehat{\text{FDP}}^{\fg,+}(\tau_\fg) & := \frac{\kappa}{\gamma} \cdot 
\frac{1 + \sum_{\tilde \fg \notin \cO_{\tau_{\fg}}(\bL^{\fg,+},\Xi), \tilde \fg \neq \fg, \fg'}\ind\{L_{\tilde g} = -1\} + \ind\{\bL^{\fg,+}_{\fg}=-1\} + \ind\{\bL^{\fg,+}_{\fg'} = -1\}}
{\sum_{\tilde \fg \notin \cO_{\tau_{\fg}}(\bL^{\fg,+},\Xi), \tilde \fg \neq \fg, \fg'} 
\ind\{L_{\tilde \fg}=1 \} + \ind\{\bL_\fg^{\fg,+} = 1\} + \ind\{\bL_{\fg'}^{\fg,+} = 1\}  }\notag \\
& = \frac{\kappa}{\gamma} \cdot \frac{\big(1 + \sum_{\tilde \fg \notin \cO_{\tau_{\fg}}(\bL^{\fg,+},\Xi), \tilde \fg \neq \fg, \fg'}\ind\{L_{\tilde g} = -1\} + 0 + 1\big)}
{\sum_{\tilde \fg \notin \cO_{\tau_{\fg}}(\bL^{\fg,+},\Xi), \tilde \fg \neq \fg, \fg'}\ind\{L_{\tilde \fg}=1\} +1 +0}
 \le {\alpha},
\@
by the definition $\tau_{\fg}$. On the other hand, there is 
\@\label{eq:stop-taugprime}
\widehat{\text{FDP}}^{\fg',+}(\tau_\fg) & := \frac{\kappa}{\gamma} \cdot  
\frac{1 + \sum_{\tilde \fg \notin \cO_{\tau_{\fg}}(\bL^{\fg',+},\Xi), 
\tilde \fg \neq \fg, \fg'}\ind\{L_{\tilde g} = -1\} + \ind\{\bL^{\fg',+}_{\fg}=-1\} + \ind\{\bL^{\fg',+}_{\fg'} = -1\}}
{\sum_{\tilde \fg \notin \cO_{\tau_{\fg}}(\bL^{\fg',+},\Xi), \tilde\fg \neq \fg, \fg'}\ind\{L_{\tilde \fg}=1 \}+ \ind\{\bL^{\fg',+}_{\fg}=1\} + \ind\{\bL^{\fg',+}_{\fg'} = 1\}} \notag \\
& \stepa{=} \frac{\kappa}{\gamma} \cdot 
\frac{1 + \sum_{\tilde \fg \notin \cO_{\tau_{\fg}}(\bL^{\fg,+},\Xi), \tilde \fg \neq \fg, \fg'}\ind\{L_{\tilde g} = -1\} + 1 + 0}
{\sum_{\tilde \fg \notin \cO_{\tau_{\fg}}(\bL^{\fg,+}, \Xi), \tilde \fg \neq \fg, \fg'}
\ind\{L_{\tilde \fg}=1 \} + 0 +1}\notag \\
& \stepb{=}\widehat{\text{FDP}}^{\fg,+}(\tau_\fg) 
\stackrel{\eqref{eq:stop-taug}}{\le} {\alpha},
\@
where step (a) uses $\cO_{\tau_g}(\bL^{\fg',+}, \Xi) = \cO_{\tau_g}(\bL^{\fg,+},\Xi)$,
and step (b) follows from comparing the expression with that of $\widehat{\text{FDP}}^{\fg,+}(\tau_\fg)$.
Combining~\eqref{eq:stop-taugprime} and the definition of $\tau_{\fg'}$, 
we have $\tau_{\fg'} \le \tau_{\fg}$, which further implies
$\fg,\fg' \in \cO^c_{\tau_{\fg'}}(\bL^{\fg,+},\Xi) = \cO^c_{\tau_{\fg'}}(\bL^{\fg',+},\Xi)$. As a result, 
\$
\widehat{\text{FDP}}^{\fg,+}(\tau_{\fg'}) & = \frac{\kappa}{\gamma} \cdot 
\frac{
1 + \sum_{\tilde \fg \notin \cO_{\tau_{\fg'}}(\bL^{\fg,+},\Xi), 
\tilde \fg \neq \fg, \fg'}\ind\{L_{\tilde \fg} =-1\} 
+ \ind\{\bL^{\fg,+}_{\fg} = -1\} + \ind\{\bL^{\fg,+}_{\fg'} = -1\}}
{\sum_{\tilde \fg \notin \cO_{\tau_{\fg'}}(\bL^{\fg,+},\Xi), \tilde \fg \neq \fg, \fg'}\ind\{L_{\tilde \fg}=1\}  
+ \ind\{\bL^{\fg,+}_{\fg} = 1\} + \ind\{\bL^{\fg,+}_{\fg'} = 1\} }\\
& = \frac{\kappa}{\gamma}\cdot \frac{ 
1 + \sum_{\tilde \fg \notin \cO_{\tau_{\fg'}}(\bL^{\fg',+},\Xi), \tilde \fg \neq \fg, \fg'}\ind\{L_{\tilde \fg} =-1\}  
+ 0 + 1 }
{\sum_{\tilde \fg \notin \cO_{\tau_{\fg'}}(\bL^{\fg',+},\Xi), \tilde \fg \neq \fg,\fg'}\ind\{L_{\tilde \fg}=1\}+1+0}\\
& = \widehat{\text{FDP}}^{\fg',+}(\tau_{\fg'}) 
 \le {\alpha},
\$
where the last step follows from the definition of $\tau_{\fg'}$.
Using again the definition of $\tau_{\fg}$, there is $\tau_{\fg} \le \tau_{\fg'}$. 
Collectively, $\tau_{\fg} = \tau_{\fg'}$, which finishes the proof of~\eqref{eq:swap}. 

We now prove~\eqref{eq:swap2}. Suppose 
$\{\fg \notin \cO_{\tau_{\fg}}(\bL^{\fg,+},\Xi), L_{\fg} = -1, \xi_\fg = 1\}$ 
and $\{\fg' \notin \cO_{\tau_{\fg'}}(\bL^{\fg',+}, \Xi), L_{\fg'} = -1, \xi_{\fg'} = 1\}$ both hold. Assume without loss of generality by symmetry that $\tau_\fg \le \tau_{\fg'}$. 
We go through similar steps as the previous case to show that 
$\cO_t(\bL^{\fg,+},\Xi) = \cO_t(\bL^{\fg',+}, \Xi)$, for $\forall t \le \tau_\fg$.
\begin{itemize}
\item For the base case $t = 0$, 
\$
\cO_0(\bL^{\fg,+},\Xi) = \{\tilde \fg \in \cG: \xi_{\tilde \fg} = 0, 
\bL^{\fg,+}_{\tilde \fg} = -1  \}
= \{\tilde \fg \in \cG: \xi_{\tilde \fg} = 0, \bL^{\fg',+}_{\tilde \fg} = -1  \}
= \cO_0(\bL^{\fg',+},\Xi).
\$ 
\item For the inductive step, suppose 
$\cO_s(\bL^{\fg,+},\Xi) = \cO_s( \bL^{\fg',+},\Xi)$ for all $s \le t < \tau_\fg$. 
By assumption, 
\$
& \fg \in \cO_{\tau_\fg}^c(\bL^{\fg,+},\Xi) 
\subset \cO_{t}^c(\bL^{\fg,+},\Xi) = \cO_{t}^c(\bL^{\fg',+},\Xi),\\
& \fg' \in \cO_{\tau_{\fg'}}^c(\bL^{\fg',+},\Xi) 
\subset \cO_{t}^c(\bL^{\fg',+},\Xi) = \cO_{t}^c(\bL^{\fg,+},\Xi).
\$ 
As a result, $\bL^{\fg,+}_{\tilde \fg} = \bL^{\fg',+}_{\tilde \fg}$, for 
$\forall \tilde \fg \in \cO_t(\bL^{\fg,+},\Xi)=\cO_t(\bL^{\fg',+},\Xi)$.
By construction, $\cO_{t+1}(\bL^{\fg,+},\Xi)$  
depends on $\bL^{\fg,+}$ 
only through $\bL^{\fg,+}_{\tilde \fg}$ 
for which $\tilde \fg \in \cO_{t}(\bL^{\fg,+},\Xi)$. 
The same reasoning applies to $\cO_{t+1}(\bL^{\fg',+},\Xi)$.
We then conclude that $\cO_{t+1}(\bL^{\fg,+},\Xi) = \cO_{t+1}(\bL^{\fg',+},\Xi)$.
\end{itemize}
We have shown that $\cO_t(\bL^{\fg,+},\Xi) = \cO_t(\bL^{\fg',+},\Xi)$, 
for $\forall t \le \tau_g$, which further implies that $\fg' \notin \cO_{\tau_\fg}(\bL^{\fg,+},\Xi)$. The result in~\eqref{eq:swap} then leads to $\tau_\fg = \tau_{\fg'}$.

With the results in~\eqref{eq:swap} and~\eqref{eq:swap2} at hand, we 
can see that on the event $\{\fg \notin \cO_{\tau_\fg}, L_\fg = -1, \xi_\fg=1\}$,
\$ 
\{\fg' \in \cO_{\tau_\fg}(\bL^{\fg,+},\Xi), L_{\fg'} = -1, \xi_{\fg'} = 1\}
& = \{\fg' \in \cO_{\tau_\fg}(\bL^{\fg,+},\Xi), \tau_\fg = \tau_{\fg'}, L_{\fg'} = -1, \xi_{\fg'} = 1\}\\
& = \{\fg' \in \cO_{\tau_{\fg'}}(\bL^{\fg',+},\Xi), L_{\fg'} = -1, \xi_{\fg'} = 1\}, 
\quad \forall \fg'\in \cG.
\$
As a result,
\$
& \frac{\kappa}{\gamma}\cdot \EE\left[\frac{\ind\{\fg \notin \cO_{\tau_{\fg}}(\bL^{\fg,+}), L_{\fg} = -1,\xi_{\fg}=1 \}}
{\sum_{\fg' \in \cH_{0,\cG}}\ind\{\fg' \notin \cO_{\tau_{\fg}}(\bL^{\fg,+}), L_{\fg'} = -1,\xi_{\fg'}= -1\} }\right]\\
=~& 
\EE\left[\frac{\ind\{\fg \notin \cO_{\tau_{\fg}}(\bL^{\fg,+}), L_{\fg} = -1,\xi_{\fg}=1 \}}
{\sum_{\fg' \in \cH_{0,\cG}}\ind\{\fg' \notin \cO_{\tau_{\fg'}}(\bL^{\fg',+}), L_{\fg'} = -1,\xi_{\fg}=1\} }\right].
\$
Summing over $\fg \in \cH_{0,\cG}$, we obtain $\fdr \le \alpha$. \qed

\subsection{Proof of Theorem~\ref{thm:cc}}\label{sec:proof-cc}
We start by introducing some notation used in the proof. 
Let $\cC_\fg = (\{L_i\}_{i \notin \fg},\; \{W_{\fg'},\xi_{\fg'}\}_{\fg' \in \cG})$ 
denote the conditioning statistics. Define the budget
\[
B_\fg = \frac{\gamma\alpha}{\kappa} \cdot
\frac{\ind\{\fg \notin \cO_{T^\early},\; L_\fg = 1\}}
     {1 + N_{T^\early(\ell)}}
\]

\paragraph{Step 1: Decomposing the FDR.}
When $\fg \in \hat \cS^\cc$, there is $\hat \cS^\cc = \hat \cS^\cc \cup \{\fg\}$ and 
$|\hat\cS^{\cc}| \ge |\hat\cS \cup \{\fg\}|$. As a result, 
\begin{align}\label{eq:cc-step1}
\mathrm{FDR}(\hat\cS^{\cc};\cG)
= \sum_{\fg \in \cH_{0,\cG}} \EE\bigg[\frac{\ind\{\fg \in \hat\cS^{\cc}\}}{|\hat\cS^{\cc}|\vee 1}\bigg]
\le \sum_{\fg \in \cH_{0,\cG}} \EE\bigg[\frac{\ind\{\fg \in \hat\cS^{\cc}\}}{|\hat\cS\cup\{\fg\}|}\bigg].
\end{align}
For each $\fg$, define the $\cC_\fg$-dependent threshold
\$
\hat c_\fg = \inf\{a \ge 0 : E_\fg(a) \le 0\},
\$
with $\hat c_\fg = \infty$ if no such $a$ exists.
Then $\fg \in \hat\cS^{\cc}$ implies $\ind\{\fg \in \hat\cS\} \vee \ind\{E_\fg(A_\fg) \le 0\} = 1$, so
\begin{align}\label{eq:cc-step1b}
\mathrm{FDR}(\hat\cS^{\cc};\cG)
\le \sum_{\fg \in \cH_{0,\cG}} 
\underbrace{\EE\bigg[\frac{\ind\{\fg \in \hat\cS\}\vee \ind\{E_\fg(A_\fg)\le 0 \}}{|\hat\cS\cup\{\fg\}|}\bigg]}_{\fdr_\fg},
\end{align}
where we use $\fdr_\fg$ to denote the FDR ``contribution'' of $\fg$.

\paragraph{Step 2. Bounding $\fdr_\fg$.}
Fix $\fg \in \cH_{0,\cG}$.
Observe that conditional on $\cC_\fg$,
the only randomness comes from $L_i \overset{\text{ind}}{\sim} \text{Bern}(p_i)$, for $i\in\fg$, 
where $p_i \in [l_\Gamma,u_\Gamma]$ under $H^{(0)}_\fg$. 
We now proceed to show that $\fdr_\fg \le \EE[B_\fg]$.

\begin{itemize}
\item 
If $\hat c_\fg < \infty$ and $E_\fg(\hat c_\fg) \le 0$, then by 
construction 
\$ 
\EE\Big[\frac{\ind\{\fg \in \hat \cS\} \vee \ind\{A_\fg \ge \hat c_g\}}{|\hat \cS \cup \{\fg\}|} 
- B_\fg \Biggiven \cC_\fg\Big] \le 
E_\fg(\hat c_\fg) \le 0.
\$
On the other hand, by the definition of $\hat c_\fg$, 
$E_\fg(A_\fg) \le 0$ implies that $A_{\fg} \ge \hat c_\fg$.
Consequently, 
\$ 
\fdr_\fg \le \EE\Big[\frac{\ind\{\fg \in \hat \cS\} \vee \ind\{A_\fg \ge \hat c_g\}}{|\hat \cS \cup \{\fg\}|} \Big] \le \EE[B_\fg].
\$

\item 
If $\hat c_\fg < \infty$ and $E_\fg(\hat c_\fg) >0$, then by 
construction, there exists a decreasing sequence $c_m \rightarrow \hat c_{\fg}$, 
such that $E_\fg(c_m) \le 0$. For any $m\ge 1$,  
\$ 
\EE\Big[\frac{\ind\{\fg \in \hat \cS\} \vee \ind\{A_\fg \ge c_m\}}{|\hat \cS \cup \{\fg\}|} 
- B_\fg \Biggiven \cC_\fg\Big] \le 
E_\fg(c_m) \le 0.
\$
When $E_\fg(A_\fg) \le 0$, we must have $A_\fg > \hat c_\fg$ and 
there exists $m \ge 1$ such that $c_m \le A_\fg$. Then
\$
\fdr_\fg \le 
\EE\Bigg[\frac{\ind\{\fg \in \hat \cS\} \vee \ind\{A_\fg \ge c_m\}}{|\hat \cS \cup \{\fg\}|}\Bigg] \le \EE[B_\fg]. 
\$

\item 
If $\hat c_{\fg} = \infty$, it holds that
\[
\fdr_\fg = \EE\left[\frac{\ind\{\fg \in \hat\cS\}}{|\hat\cS\cup \{\fg\}|}\right].
\]
It suffices to show that
\@\label{eq:stop-rule}
\frac{\ind\{\fg \in \hat\cS\}}
         {|\hat\cS\cup \{\fg\}|}
    \leq B_\fg,
\@
deterministically.
To see this, if 
$\fg \notin \hat\cS$, then $0 \le B_\fg$ trivially. 
If $\fg \in \hat\cS$, then $\hat\cS \ne \varnothing$ and
$T^\early = \tau$, so
\[
\frac{1}{|\hat\cS \cup \{\fg\}|} = \frac{1}{P_{\tau}}, \qquad
B_\fg = \frac{\gamma\alpha}{\kappa} \cdot \frac{1}{1 + N_{\tau}}.
\]
The stopping condition
$\frac{\kappa}{\gamma} \cdot \frac{1+N_\tau}{P_\tau} \le \alpha$
gives $P_\tau \ge \frac{\kappa}{\gamma\alpha}(1+N_\tau)$, and therefore
\$
\frac{1}{P_\tau}
\le \frac{\gamma\alpha}{\kappa} \frac{1}{1+N_\tau} = B_\fg,
\$ 
which completes the proof of~\eqref{eq:stop-rule}.

\end{itemize}
Summing over $\fg \in \cH_{0,\cG}$, we have 
\begin{align}\label{eq:cc-sum-budget}
\mathrm{FDR}(\hat\cS^{\cc};\cG)
\le \sum_{\fg \in \cH_{0,\cG}} \EE\big[B_\fg\big].
\end{align}

\paragraph{Step 3: Bounding the budget.}
Define
$V_t^+ = \sum_{\fg \in \cH_{0,\cG}}
         \ind\{\fg \notin \cO_t,\, L_\fg = 1\}$,
$V_t^- = \sum_{\fg \in \cH_{0,\cG}}
         \ind\{\fg \notin \cO_t,\, L_\fg = -1\}$,
and $M_t = V_t^+ / (1 + V_t^-)$.
Since $N_t \ge V_t^-$,
\begin{align}\label{eq:cc-mg-bound}
\sum_{\fg \in \cH_{0,\cG}} B_\fg
\le \frac{\gamma \alpha }{\kappa} \cdot
    \frac{V_{T^\early}^+}{1 + V_{T^\early}^-}
= \frac{\alpha \gamma}{\kappa} \cdot M_{T^\early}.
\end{align}
Repeating steps 1 and 2 in the proof
of Theorem~\ref{thm:fdr-general} with $\tau$ replaced by $T^\early$ 
gives the bound
\[
\EE[M_{T^\early}] \le \frac{\kappa}{\gamma}.
\]
Combining the above with \eqref{eq:cc-sum-budget}, we have 
\[
\mathrm{FDR}(\hat\cS^{\cc};\cG)
\le \sum_{\fg \in \cH_{0,\cG}} \EE\big[B_\fg\big] \leq \alpha.
\]

\subsection{Proof of Proposition~\ref{prop:conservative-pval}}
\label{app:proof-conservative-pval}
We first present the concrete definitions of notations in Proposition~\ref{prop:conservative-pval}.
Given an observed Wilcoxon signed-rank statistic $S_{\fg}^\obs$,
the deterministic and randomized one-sided p-values under the
$\Gamma$-sensitivity model are
\@\label{eq:wilcox-pval}
& p_{\fg} = \PP\Bigg(\sum_{i \in \fg}r_i L_i^* \ge S_{\fg}^\obs\Bigg),
\quad p_{\fg}^\rand = \PP\Bigg(\sum_{i \in \fg}r_i L_i^* > S_{\fg}^\obs\Bigg)
+ U \cdot \PP\Bigg(\sum_{i \in \fg}r_i L_i^* = S_{\fg}^\obs\Bigg),
\@
where the probability is over $L_i^* \overset{\text{iid}}{\sim} \text{Bern}(\frac{\Gamma}{1+\Gamma})$
and $U\sim \text{Unif}(0,1)$ is an independent tie-breaking variable.
Letting $p_i = \PP(L_i = 1 \given \cF,\cZ)$ and $p^* = \frac{\Gamma}{1+\Gamma}$,
we define
\$
\Delta_\fg = \frac{\sum_{i \in \fg} r_i(p_i-p^*)}{\big(\sum_{i\in \fg} r_i^2p^*(1-p^*)\big)^{1/2}}.
\$

Throughout the proof, all asymptotics are taken as $|\fg|\rightarrow \infty$. We omit this explicit notation 
when it is clear from the context. 

Note that the $r_i$'s are deterministic given $(\cF,\cZ)$ and the only randomness comes from the $L_i$'s.
Under the $\Gamma$-sensitivity model, we have $ 1 - p^* \le p_i \le p^*$.
We also can check that 
\@ \label{eq:lindberg}
\frac{\sum_{i\in\fg} \EE\big[|r_i L_i^* - r_i p^*|^3\big]}{\big(\sum_{i \in \fg} 
\text{Var}(r_iL_i^*)\big)^{3/2}} \le 
\frac{\sum_{i \in \fg} r_i^3 p^*(1-p^*)}{(\sum_{i\in \fg}r_i^2 p^*(1-p^*))^{3/2}} 
\le \frac{10|\fg|^2(|\fg|-1)^2}{|\fg|^{9/2}\sqrt{p^*(1-p^*)}} 
{\rightarrow} 0 
\@
By the Berry-Esseen Theorem~\citep{feller1971introduction}, the condition in~\eqref{eq:lindberg}
leads to
\$
& \Bigg|\PP\Bigg(\sum_{i \in \fg}r_i L_i^* > S_\fg^\obs\Bigg) 
- \Bigg(1 - \Phi\bigg(\frac{S_\fg^\obs - \sum_{i \in \fg} r_ip^*}{\big(\sum_{i\in \fg} r_i^2p^*(1-p^*)\big)^{1/2}}\bigg)\Bigg)\Bigg|\\
& \qquad \qquad \le \sup_{x \in \RR}
\Bigg|\PP\Bigg(\sum_{i \in \fg}r_i L_i^* > x\Bigg) 
- \Bigg(1 - \Phi\bigg(\frac{x - \sum_{i \in \fg} r_ip^*}{\big(\sum_{i\in \fg} r_i^2p^*(1-p^)\big)^{1/2}}\bigg)\Bigg)\Bigg|
\stackrel{\eqref{eq:lindberg}}{ \rightarrow } 0. 
\$
On the other hand, 
\$ 
\frac{S_\fg^\obs - \sum_{i \in \fg} r_ip^*}{\big(\sum_{i\in \fg} r_i^2p^*(1-p^*)\big)^{1/2}}
= \underbrace{\frac{S_\fg^\obs - \sum_{i \in \fg} r_ip_i}{\big(\sum_{i\in \fg} r_i^2p^*(1-p^*)\big)^{1/2}}}_{\Xi}
+ \underbrace{\frac{\sum_{i \in \fg} r_ip_i - \sum_{i \in \fg}r_i p^*}{\big(\sum_{i\in \fg} r_i^2p^*(1-p^*)\big)^{1/2}}}_{\Delta}.
\$
Since $p_i \le p^*$, $\forall i\in \fg$, there is $\Delta \le 0$.
Since $\Xi = O_P(1)$, 
we have $\Xi + \Delta {\rightarrow} -\infty$ as 
$\Delta \rightarrow -\infty$.
As a result, $p_\fg, p_\fg^\rand \rightarrow 1$. 

\subsection{Counterexample for P-screening validity}\label{app:example}
Recall that the FDR control of P-screening requires 
\@
\PP(L^\pscr_\fg = -1 \given V^\pscr_\fg) \le \PP(L^\pscr_\fg = 1 \given V^\pscr_\fg), 
\text{ for all }\fg \in \cH_{0,\cG}.
\@ 
We show in the following counterexample that, under the sensitivity model, such conditional symmetry is no longer guaranteed.
\begin{counterexample}
Suppose $|\fg|=3$ and $\Gamma = 3$. Let $r_i = i$, for $i=1,2,3$, and $S_\fg = \sum^3_{i=1}r_iL_i$, 
where $L_i \overset{\text{ind}}{\sim} \text{Bern}(p_i)$ with $(p_1,p_2,p_3) = (1/4,3/10,1/4)$. 
Since $1-p^* \le p_i\le p^* := 3/4$, 
the example is compatible with the $\Gamma$-sensitivity model. It can be checked that 
\$
& \PP\big(p_\fg < 0.5 \given \min\{p_\fg,1-p_\fg)\} = \tfrac{27}{64}\big) = 1 
>  \PP\big(p_\fg \ge 0.5 \given \min\{p_\fg,1-p_\fg\} = \tfrac{27}{64}\big),\\
& \PP\big(p_\fg^\rand < 0.5 \given \min\{p_\fg^{\rand},1-p_\fg^{\rand}\} = \tfrac{55}{128}\big)
= \tfrac{2}{5} > \tfrac{14}{45} = \PP\big(p_\fg^\rand \ge 0.5 \given \min\{p_\fg^{\rand},1-p_\fg^{\rand}\} = \tfrac{55}{128}\big).
\$
\end{counterexample}

\subsection{Proof of Lemma~\ref{lem:marg-bdd-skewness}}\label{app:proof-marg-bdd-skewness}
Define $p_r = \PP(r_i = r \mid \{R_{i(j)}:j\in [n_i]\})$, for each $r \in [n_i]$.
Under the $\Gamma$-sensitivity model, there is
\begin{align}\label{eq:bounds-L-pr}
    p_r \in [a, b], \text{ with } a \coloneqq \frac{1}{1 + (n_i-1) \Gamma}, \;b \coloneqq \frac{\Gamma}{n_i-1+\Gamma}.
\end{align}
By the construction of $L_i$, we have 
\begin{align}
    \PP(L_i = 1 \mid \cF, \cZ) = \sum_{r \le \lfloor n_i/2 \rfloor } p_r, \quad  
    \PP(L_i = -1 \mid \cF, \cZ) = \sum_{r \ge \lceil n_i/2 \rceil + 1 } p_r.
\end{align}
Let $\mathbf{p} \coloneqq (p_1,\ldots,p_{n_i})$. Then $\mathbf{p}$ lies in the $n_i$-simplex  
and satisfies the boundedness condition in~\eqref{eq:bounds-L-pr}; equivalently, 
its feasible set is given by
\begin{align}
    \cQ_{a,b} = \left\{\mathbf{p} \in \RR^{n_i}_+:\;\sum_{r \in [n_i]} p_r=1,\;p_r \in [a,b]\right\}.
\end{align}
We then claim that 
\begin{align}
    & \max_{\mathbf{p} \in \cQ_{a,b}}\;\sum_{r \le \lfloor n_i/2 \rfloor} p_r \le 
    \min\big\{\lfloor n_i/2 \rfloor \cdot b,\;1 - (n_i - \lfloor n_i/2 \rfloor)\cdot a\big\}
    \eqqcolon \overline{q}^* ,\label{eq:max-L}\\
    & \min_{\mathbf{p} \in \cQ_{a,b}}\;\sum_{r \ge \lceil n_i/2 \rceil} p_r 
    \ge \max\left\{\lfloor n_i/2 \rfloor \cdot a,\;1 - (n_i - \lfloor n_i / 2\rfloor)\cdot b\right\} 
    \eqqcolon \underline{q}^*. \label{eq:min-L}
\end{align}
To see why~\eqref{eq:max-L} and~\eqref{eq:min-L} are true, we consider
any $\mathbf{p}' = (p_1',\ldots,p_{n_i}') \in \cQ_{a,b}$, and there is
\$ 
& \sum_{r \le \lfloor n_i/2 \rfloor} p_r' \le b \lfloor n_i/2\rfloor,\\
& \sum_{r \le \lfloor n_i/2 \rfloor} p_r' = 1 - \sum_{r > \lfloor n_i/2 \rfloor}p_r'
\le 1 - a(n_i - \lfloor n_i/2 \rfloor ).
\$
The above implies that $\sum_{r \le \lfloor n_i/2 \rfloor} p_r' \le \overline{q}^*$.
and proves~\eqref{eq:max-L}.
Similarly, we have 
\$ 
& \sum_{r \ge \lceil n_i/2 \rceil + 1} p_r' \ge a \cdot (n_i - \lceil n_i/2\rceil) = 
a \lfloor n_i/2 \rfloor,\\
& \sum_{r \ge \lceil n_i/2 \rceil + 1} p_r' = 1 - 
\sum_{r \le \lceil n_i/2 \rceil } p_r' \ge 1 - b\lceil n_i/2 \rceil 
= 1 - b(n_i - \lceil n_i/2 \rceil),
\$
which implies that $\sum_{r \ge \lceil n_i/2 \rceil + 1}p'_r \ge \underline{q}^*$
and proves~\eqref{eq:min-L}.

Plugging in the value of $a$ and $b$, we can also explicitly write 
\$ 
\overline{q}^* = \frac{(n_i-1)(\Gamma-1) + \lfloor n_i/2\rfloor}{1+(n_i-1)\Gamma}, \quad 
\underline{q}^* = \frac{\lfloor n_i/2 \rfloor}{1+(n_i-1)\Gamma}.
\$
As a result, there is
\$ 
\PP(L_i = 1 \given \cF,\cZ) & \le  \overline{q}^* \\
& = \frac{(n_i-1)(\Gamma-1) + \lfloor n_i/2 \rfloor }{\lfloor n_i/2\rfloor} \cdot \underline{q}^*\\
& \le \frac{(n_i-1)(\Gamma-1) + \lfloor n_i/2 \rfloor }{\lfloor n_i/2\rfloor}
\cdot \PP(L_i = -1 \given \cF,\cZ).
\$

\subsection{Proof of Lemma~\ref{lem:multic-L}}\label{sec:proof-skewness}

When $n_i$ is odd and $g(r_i) = \ceil{n_i/2}$, 
\$
\PP(L_i = 1 \given g(r_i) = \ceil{n_i/2}, \cF,\cZ) = 
\PP(L_i = - 1 \given g(r_i) = \ceil{n_i/2}, \cF,\cZ)  =0,
\$
and the result trivially holds.

When $n_i$ is even or $g(n_i/2) \neq \ceil{n_i/2}$,
the pre-image of $g(r_i)$ has cardinality $2$, 
and we can write $g^{-1}(g(r_i)) = \{a_1,a_2\}$ with $a_1 \le \floor{n_i/2}< a_2$. Then,
\begin{align}
    \frac{\PP\left(L_i = 1 \mid r_i \in \{a_1, a_2\}, \cF, \cZ\right)}{\PP\left(L_i = -1 \mid r_i \in \{a_1, a_2\}, \cF, \cZ\right)} & = \frac{\PP\left(r_i \leq \floor{n_i/2} \mid r_i \in \{a_1, a_2\}, \cF, \cZ\right)}{\PP\left(r_i > \floor{n_i/2} \mid r_i \in \{a_1, a_2\}, \cF, \cZ\right)}\\
    & = \frac{\PP\left(r_i = a_1 \mid  \cF, \cZ\right)}{\PP\left(r_i = a_2 \mid  \cF, \cZ\right)} = \frac{p_{a_1}}{p_{a_2}}.
\end{align}
By the sensitivity model in \eqref{eq:sens}, for any $a_1,a_2 \in [n_i]$, it holds that $\Gamma^{-1} \leq p_{a_1} / p_{a_2} \leq \Gamma$.
Hence, conditioning on $g^{-1}(g(r_i))$, $L_i$ satisfies $\Gamma$-bounded skewness.

\subsection{Proof of Theorem~\ref{thm:optim}}\label{sec:proof-optim}
Throughout the proof, we omit the index $i$ for the matched set to simplify notation.
\paragraph{Proof of (a).}
The proof essentially follows from the proof of the Neyman-Pearson lemma. 
For any prediction $\tilde L \in \{\pm 1\}$ that is a  function of $(\cJ, \tilde R,X)$,  
we can decompose the classification risk as:
\@\label{eq:err-decomp} 
& \PP(L = 1, \tilde L = -1) + \PP(L = -1, \tilde L = 1) \notag \\
& = \PP(L = 1) - \PP(L = 1, \tilde L = 1) +  \PP(L=-1, \tilde L = 1) \notag \\
& = \PP(L = 1) + \EE\big[\ind\{\tilde L = 1\}\cdot (\ind\{L=-1\}-\ind\{L=1\})\big] \notag\\
& = \PP(L = 1) + \EE\big[\ind\{\tilde L = 1\}\cdot (\PP(L=-1\given \cJ,\tilde R,X)-
\PP(L=1\given \cJ, \tilde R,X))\big],
\@
where the last step follows from the fact that $\tilde L$ is a function of $\cJ,\tilde R,X$
and the tower property. Define the events
\$ 
\cE_+ = \big\{\beta(\cJ,\tilde R,X)>0\big\} \text{ and }\cE_- = \big\{\beta(\cJ,\tilde R,X)<0\big\}.
\$
We then have 
\$ 
\eqref{eq:err-decomp} = \PP(L = 1) & - \EE\Big[\ind\{\tilde L = 1, \cE_+\} \cdot 
\big |\PP(L=1\given \cJ,\tilde R,X)- \PP(L=-1\given \cJ, \tilde R,X)\big| \Big] \\
&\quad  + \EE\Big[\ind\{\tilde L = 1, \cE_-\} \cdot \big |\PP(L=1\given \cJ,\tilde R,X)-
\PP(L=-1\given \cJ, \tilde R,X)\big| \Big]\\
\ge \PP(L = 1) & - \EE\Big[\big |\PP(L=1\given \cJ,\tilde R,X)-
\PP(L=-1\given \cJ, \tilde R,X)\big| \Big],
\$
where the lower bound is exactly the classification error 
corresponding to $\hat L = \sgn(\beta(\cJ, \tilde R,X))$.
We have therefore established the optimality of $\hat L$.

\paragraph{Proof of (b).}
We consider the following decomposition of the order statistics $\tilde R$: 
\begin{align}\label{eq:reparam}
\zeta_{1} = R_{q} - R_{q'}, \qquad \zeta_{2} = \frac{R_{q} + R_{q'}}{2}, \qquad U = R_{-\cJ},
\end{align}
which are mutually independent conditional on $X$ under the Gaussian model~\eqref{eq:gaussian}.

For any $z > 0$, any $a \in \RR$, any $u \in \RR^{n_i-2}$, and $x\in \RR^d$, it holds that
\begin{align}
\PP(L=1 \mid |\zeta_1| = z,\;\zeta_2 = a,\;U = u,X = x) = \PP(\zeta_1 = z \mid |\zeta_1| = z,\;\zeta_2 = a,\;U = u,\; X = x).
\end{align}
Since $\zeta_1 \indep (\zeta_2, U)$,
\begin{align}
\PP(\zeta_1 = z \mid |\zeta_1| = z,\;\zeta_2 = a,\;U = u) = \PP(\zeta_1 = z \mid |\zeta_1| = z) = \frac{g_{\tau}(z)}{g_{\tau}(z) + g_{\tau}(-z)}, 
\end{align}
where $g_{\tau}$ denotes the density of $\cN(\tau,2)$. 
Then, 
\begin{align}
\PP(L = 1 \mid \cJ, \tilde R, X) & \stackrel{\text{tower prop.}}{=} 
\EE\big[\PP(L = 1 \mid |\zeta_1|,\zeta_2,U, X) \given \cJ,\tilde R,X\big] \\
& = \frac{g_{\tau}(|\zeta_1|)}{g_{\tau}(|\zeta_1|) + g_{\tau}(-|\zeta_1|)}
= \frac{g_{\tau}(W^{\text{NP}})}{g_{\tau}(W^{\text{NP}}) + g_{\tau}(-W^{\text{NP}})}.
\end{align}
Since $\zeta_1 \sim \cN(\tau, 2\sigma^2)$, 
\begin{align}
\beta(\cJ,\tilde R, X) & = \log \frac{g_{\tau}(W^{\text{NP}})}{g_{\tau}(-W^{\text{NP}})} 
= \log \frac{\exp(-(W^{\text{NP}}-\tau)^2/(4\sigma^2))}{\exp(-(-W^{\text{NP}}-\tau)^2/(4\sigma^2))} \\
& = \frac{(W^{\text{NP}}+\tau)^2 - (W^{\text{NP}}-\tau)^2}{4\sigma^2} = \frac{W^{\text{NP}} \cdot \tau}{\sigma^2}.
\end{align}
Since $\tau > 0$, the log-likelihood ratio is strictly increasing in $z \in \RR_+$, which completes the proof.

\subsection{Proof of Lemma~\ref{lem:multiouts} (multi-outcome extension)}\label{app:proof-multiouts}
To ease the notation, we define $g_m = g(r_i^{(m)})$, for $m\in[M]$, 
where $r_i^{(m)}$ is the rank of the treated unit with respect to outcome 
$m$. Observe that knowing
$\{L_i^{(m^*(i))}, g_1,\cdots,g_m\}$
is equivalent to knowing the index of treated unit.
We denote the index of the treated unit of the $i$-th set by $T_i$
and write $\cT$ as a function that maps 
$(L_i^{(m^*(i))},g_1,\cdots,g_m)$ to the index. 

Consider any realized values of $(g_1,\cdots,g_m)$ with positive probability, denoted by 
$(\bar g_1,\cdots, \bar g_m)$. We have 
\$ 
\PP(L_i = 1, g_1 = \bar g_1,\cdots,g_m = \bar g_m \given \cF,\cZ) 
& = \PP(T_i = \cT(1, \bar{g}_1,\cdots,\bar{g}_m) \given \cF,\cZ) \\
& \le \Gamma \cdot \PP(T_i = n_i+1-\cT(1, \bar{g}_1,\cdots,\bar{g}_m) \given \cF,\cZ) \\
& = \Gamma \cdot \PP(L_i = -1, g_1 = \bar g_1,\cdots,g_m = \bar g_m \given \cF,\cZ).
\$
Dividing both hand-sides by $\PP(g_1 = \bar g_1,\cdots,g_m = \bar g_m \given \cF,\cZ)$, 
we arrive at 
\$ 
\PP(L_i = 1 \given g_1 = \bar g_1,\cdots, g_m =\bar g_m, \cF,\cZ )
\le \Gamma\cdot \PP(L_i = -1 \given g_1 = \bar g_1,\cdots, g_m =\bar g_m, \cF,\cZ).
\$
Applying the tower property completes the proof.

\section{Extensions and variants of our approach}\label{sec:variant}
This section provides extensions and variants of our method, 
including its application to scenarios with multiple outcomes, variants that incorporate other considerations (e.g., subgroup sizes), and extensions to two-sided effects. 
Implementations of these extensions are presented in Section~\ref{sec:add-simu}.

\subsection{Adaptive discovery of effect modification with multiple outcomes}\label{sec:multiouts}
In many practical scenarios, multiple outcomes can be affected by the exposure $Z_{ij}$ \citep{bekerman2024planning,small2024protocols}. 
For each outcome $m \in [M]$, we assume
\begin{align}
    R^{(m)}_{ij} = Z_{ij} \cdot t^{(m)}_{ij} + (1-Z_{ij}) \cdot c^{(m)}_{ij}.
\end{align}
When the goal is to detect effect modification, we should pay attention to subgroups in which 
{\em any} outcome is impacted by the treatment. In this sense, leveraging information across 
outcomes can improve the power of subgroup selection.

Concretely, at the unit level, for each $i \in [I]$ and each $j \in [n_i]$, we consider the hypothesis:
\begin{align}
    H^{(0)}_{ij}: t^{(m)}_{ij} = c^{(m)}_{ij},\;\forall\;m \in [M],\;\;\text{versus}\;\; H^{(1)}_{ij}:\;\exists\;m \in [M],\;t^{(m)}_{ij} \neq c^{(m)}_{ij}.
\end{align}
Given a (possibly data-drive) partition $\cG = \{\fg_1,\cdots, \fg_K\}$, 
we define the group level null as 
\@\label{eq:grp-null-mul-otc}
H^{(0)}_{\fg} = \bigcap_{i \in \fg} H_i^{(0)}, \text{ where } 
H_i^{(0)} = \bigcap_{j \in [n_i]} H^{(0)}_{ij}.
\@

\paragraph{Adaptation of our method to multiple outcomes.}
Suppose that for each matched set $i \in [I]$ and each outcome $m \in [M]$,
we have constructed a pair $(L^{(m)}_{i}, W^{(m)}_{i})$.
We now combine them into a single magnitude-sign pair $(L_i,W_i)$.
The cross-outcome aggregation must ensure that 
the sign statistic $L_i$ satisfies the bounded skewness condition 
under the null $H^{(0)}_\fg$ defined in~\eqref{eq:grp-null-mul-otc}, and that the magnitude statistic is sensitive to signals in any outcome.
To this end, we define the magnitude statistic as
\begin{align}
    W_i = \max_{m \in [M]}\;W^{(m)}_{i}.
\end{align}
and the sign statistic as
\[
L_i = L^{(m^*(i))}_{i} \qquad \text{where}\;m^*(i) \in \argmax{m \in [M]}\;W^{(m)}_i.
\]
We have the following result on $(W_i,L_i)$, whose proof is in Appendix~\ref{app:proof-multiouts}.
\begin{lemma}\label{lem:multiouts}
For any $i \in [I]$, under $H^{(0)}_{i}$, there is 
\begin{align}
    \PP(L_i = 1 \given W_i, \cF,\cG) 
    \le \Gamma \cdot \PP(L_i = -1 \given W_i, \cF,\cG).
\end{align}
\end{lemma}

We then obtain the subgroup-level statistics $(W_{\fg}, L_{\fg})_{\fg \in \cG}$ 
using the aggregation rule in Section~\ref{sec:warmup}.
By Theorem~\ref{thm:fdr-general}, the bounded skewness stated in Lemma~\ref{lem:multiouts} 
carries over to $L_{\fg}$; the finite-sample FDR control with multiple outcomes follows.

\paragraph{P-value based baseline.}
As we can see, when the goal is to detect whether {\em any} of these outcomes 
is affected by the treatment, the problem can be cast in a form similar to the single-outcome case,
but it additionally requires valid and efficient summary statistics that 
capture the signals among outcomes. Concretely, suppose we can get access to p-values $p^{(m)}_{\fg}$ for each outcome $m \in [M]$. Testing the cross-outcome global null then
requires a valid combination of these p-values at the same nominal level, e.g., Bonferroni's combination $M\cdot \min_{m \in [M]}\;p^{(m)}_{\fg}$ or Fisher's combination 
\begin{align}
    p_{\fg,{\rm Fisher}} = 1 - F_{\chi^2_{2M}}\left(-2\sum_{m \in [M]}\log p^{(m)}_{\fg}\right),
\end{align}
where $F_{\chi^2_{2M}}$ is the c.d.f. of the distribution $\chi^2_{2M}$ (the latter may be invalid 
under dependence across the outcomes).
With combined p-values, the BH adjustment can be applied \citep{karmakar2018false} for subgroup selection.

\subsection{Screening by incorporating additional information}\label{sec:side-screen}
In our framework, the sign-magnitude pair is used in a disentangled way: 
the sign is used for FDP estimation, while the magnitude determines the screening ordering.
For FDR control, the magnitude statistic is subject to essentially no restriction 
beyond a near-independence condition with the sign.
This separation offers substantial flexibility that allows additional considerations 
or side information to be incorporated in the design of $W_{\fg}$---provided  
they are independent of the treatment assignment.

One practically important consideration is subgroup size. 
In many applications, very small subgroups may raise concerns about reproducibility 
and/or exacerbate multiplicity issues for classical p-value-based 
methods~\citep{burke2015three,wang2007statistics}. 
On the other hand, larger subgroups are also often more relevant in clinical settings, 
as they indicate broader applicability of a discovery.


This motivates us to incorporate $|\fg|$, which is independent of $L_{\fg}$,
into the magnitude statistic without affecting the bounded skewness and hence the FDR control.
One straightforward variant is to define $\tilde W_{\fg} = |\fg|^b \cdot W_g$,
where $b > 0$ is up to practitioners' choice. 
Since $W^{\rm NP}_{\fg}$ is shown to be optimal under Gaussian models, replacing it with its counterpart $\tilde W^{\rm NP}_{\fg}$ tends to exhibit the loss in power (in terms of the 
number of selected groups).

\subsection{Testing two-sided effects}\label{sec:two-sided}
In practical scenarios, however, the signs of underlying effects are unknown without prior knowledge. Then, to identify affected subgroups, we expect our inferential procedure to be aware of both positive and negative effects, in which the alternative hypothesis takes the form $H^{(0)}_i:\;t_{ij} \neq c_{ij}$. 

Recalling the procedure targeted at non-negative effects, both the validity and the power of discoveries rely on the desired decomposition $(W_i, L_i)$. Concretely, the statistic $W_i \cdot L_i = Y_i = (R_{i1} - R_{i2})(Z_{i1} - Z_{i2})$ tends to be positive when the one-sided alternative $t_{ij} > c_{ij}$ is true for all $j \in [n_i]$ and is centered around zero otherwise. 
Then, to identify two-sided effects, one solution is to define a statistic $Y_i$ as a function of $(R_{ij}, Z_{ij}, X_i)$ such that it is nearly symmetric around zero under the null hypothesis and antisymmetric under the alternative, with which, the decomposition $(W_i = |Y_i|, L_i = \sgn(Y_i))$ still satisfies the ideal properties in Section~\ref{sec:warmup}. 

To start with, it suffices to define a variant $\tilde R_{ij}$ of $R_{ij}$ that can inform the treatment effect such that, under the alternative, with a high probability, $\tilde R_{ij} > \tilde R_{i j'}$ if $Z_{ij}=1$ regardless of the sign of the effect, and in the meantime, $\tilde R_{ij} \approx \tilde R_{i j'}$ both of which are centered around zero under the null.
To achieve this, we need to center the original outcome $R_{ij}$ and flip its sign when the effect is negative, which requires knowledge of the null distribution of outcomes.
Suppose we have a predictive model $R_0(X_i)$ for $R_{ij}$ under the null (i.e., $R_0(X_i) \overset{d}{\approx} (R_{ij} \mid Z_{ij}=0)$, which is only a function of covariates.
Then, we can define $\tilde R_{ij} = (R_{ij} - R_0(X_i))^2$, with which, $Y_i$ can be defined by
\begin{align}\label{eq:stat-two-side}
    Y_i = (\tilde R_{i1} - \tilde R_{i2})(Z_{i1} - Z_{i2}) = (R_{i1} - R_{i2})(R_{i1} + R_{i2} - 2R_0(X_i))(Z_{i1} - Z_{i2}).
\end{align}
Accordingly, we can follow the one-sided case to define subgroup-level statistics $(W_{\fg},L_{\fg})$ for all $\fg \in \cG$. 
We note that, in \eqref{eq:stat-two-side}, the power of our procedure may rely on the quality of the predictive model $R_0$. To avoid double-dipping, the predictive model can be updated at each step $t$ in Algorithm~\ref{alg:SS} using screened data from $\cO_t$.

\section{More implementation details}

In this section, we begin by summarizing our procedure in Algorithm~\ref{alg:SS}, followed by implementation details for conditional calibration in Section~\ref{sec:app-cc} and additional discussion of the P-screening method in Section~\ref{app:split-method}.

\begin{algorithm}[ht]
\DontPrintSemicolon
\caption{Data-driven subgroup selection}
\label{alg:SS}
\textbf{Input:} subgroups $\cG=\cA_G(\cF)$; test statistics 
$(L_{\fg}, V_{\fg}, W_\fg)_{\fg\in\cG}$;
target FDR level $\alpha$; skewness parameter $\kappa$; 
sample-splitting ratio $\gamma$; 
semi-supervised learning algorithm $\cA_L$.\;
\textbf{Initialization:} set $t\gets 0$.
Generate $\xi_\fg \overset{\text{iid}}{\sim} \text{Bern}(\gamma)$, $\forall \fg \in \cG$.\;
Define $\cI^-_1$, $\cI^-_2$, and $\cI^+$ according to Equation~\eqref{eq:split} and set $\cO_t \gets \cI^-_1$.\;
\While{$t\le |\cI^-_2 \cup \cI^+|$}{
     $P_t \gets \sum_{\fg\notin\cO_t}\ind\{L_{\fg}=1\}$,
           $N_t \gets \sum_{\fg\notin\cO_t}\ind\{L_{\fg}=-1\}$.\;
     $\widehat{\text{FDP}}(t) \gets \frac{\kappa}{\gamma} \cdot \frac{1+N_t}{1\vee P_t}$.\;
    \If{$\widehat{\textnormal{FDP}}(t)\le \alpha$}{
        \textbf{break}   
        }
     Update $\hat \mu_{t} \gets \cA_L(\cO_t; \cG\backslash \cO_t)$.
     \tcp*{$\cA_L$ treats $\cO_t$ as labeled and $\cG\backslash \cO_t$ unlabeled.} 
     Update $\hat L_\fg \gets \hat \mu_t((x_i)_{i\in\fg}, V_\fg, W_\fg)$, for $\forall \fg \in \cG \backslash \cO_t$.\;
     Choose $\pi(t+1)\in\arg\min_{\fg\notin\cO_t} \hat L_\fg$, with ties 
    broken arbitrarily.\; 
     Set $\cO_{t+1}\gets \cO_t\cup\{\pi(t+1)\}$. 
     $t\gets t+1$.
     }
     \eIf{$\widehat{\textnormal{FDP}}(t) \le \alpha$}{
     $\hat{\cS}\gets\{\fg\in\cG: \fg\notin\cO_{t}, L_{\fg}>0\}$.
     }{
     $\hat \cS \gets \varnothing$. \tcp*{No threshold meets the stopping criterion; return the empty set.}
     }
\textbf{Output} $\hat{\cS}$.
\end{algorithm}

\subsection{Implementation details of conditional calibration}\label{sec:app-cc}
This section describes the implementation details of the conditional calibration 
(\texttt{CC}) step introduced in Section~\ref{sec:cc}.
Throughout, we take $h(L_{\fg}, V_{\fg}) = -\log p_{\fg,\Gamma}$, 
where $p_{\fg,\Gamma}$ is the sensitivity p-value for group 
$\fg$ resulting from the group-global signed-rank test.

It remains to specify how to compute $E_\fg(A_\fg)$.
Recall that we condition on $\cC_{\fg} = (\{L_i\}_{i \notin \fg},\; \{W_{\fg'},\xi_{\fg'}\}_{\fg'\in \cG})$, which can be viewed as fixed in the computation. 
To emphasize the dependence on $(L_\fg, V_\fg)$, we write the budget as 
\[
B_\fg(\ell,v) = \frac{\gamma\alpha}{\kappa} \cdot
\frac{\ind\{\fg \notin \cO_{T^\early(\ell, v)}(\ell,v),\; \ell = 1\}}
     {1 + N_{T^\early(\ell,v)}(\ell,v)}
\]
and 
\begin{align}
    \zeta_a(\ell,v) = \frac{\ind\{\fg \in \hat\cS(\ell,v)\}
          \vee \ind\{h(\ell, v) \ge a\}}
         {|\hat\cS(\ell,v) \cup \{\fg\}|}
    - B_\fg(\ell,v),
\end{align}
where the respective quantities computed with $(L_\fg, V_\fg)$ 
set to $(\ell,v)$.
Then, we can rewrite $E_{\fg}(a)$ as 
\begin{align}
E_\fg(a) & = \sup_{p_i \in [\ell_{\Gamma}, u_{\Gamma}]}
  \EE_{L_i \sim \mathrm{Bern}(p_i), i\in \fg}\big[\zeta_a(L'_{\fg}, V'_{\fg})\big]\\
  & = \sup_{p_i \in [\ell_{\Gamma}, u_{\Gamma}]}
  \sum_{\mathbf{L} \in \{-1,1\}^{|\fg|}} \zeta_a(\mathbf{L}) \cdot \prod_{i \in [|\fg|]} p_i^{L_i} (1-p_i)^{1-L_i}.
\end{align}
This is a multi-linear function of $\mathbf{p} = (p_1,\cdots, p_{|\fg|})$, thus obtaining the maximum at a certain vertex.
The conditional calibrated selection is then defined by $\hat \cS^{\cc,{\rm full}} = \hat \cS \cup \{\fg \in \cG:\;E_\fg(h(L_{\fg},V_{\fg})) \leq 0\}$.

\paragraph{A computationally light variant.}
As an alternative, we condition on $\tilde \cC_\fg = (\{L_i\}_{i \notin \fg},\;V_{\fg},\; \{W_{\fg'},\xi_{\fg'}\}_{\fg'\in \cG})$ when defining $E_\fg(a)$. 
Here, $\tilde \cC$ treats $V_{\fg}$ as fixed and is therefore a finer conditioning than $\cC_{\fg}$. 
Concretely, we define 
\begin{align}
\tilde E_\fg(a)
= \sup_{p \in [\ell_{\Gamma}, u_{\Gamma}]}
  \EE_{L'_\fg \sim \mathrm{Bern}(p)}
  \bigg[
    \frac{\ind\{\fg \in \hat\cS(L'_\fg, V_\fg)\}
          \vee \ind\{h(L'_\fg, V_\fg) \ge a\}}
         {|\hat\cS(L'_\fg, V_\fg) \cup \{\fg\}|}
    - B_\fg(L'_\fg, V_\fg)
  \;\bigg|\; \tilde \cC_\fg
  \bigg],
\end{align}
where the expectation is {\em only} over the randomness of $L_\fg'$.
To calculate $\tilde E_{\fg}(a)$, we similarly define  
\begin{align}
    \tilde \zeta(\ell) = \frac{\ind\{\fg \in \hat\cS(\ell,V_\fg)\}
          \vee \ind\{h(\ell, V_{\fg}) \ge a\}}
         {|\hat\cS(\ell,V_\fg) \cup \{\fg\}|}
    - B_\fg(\ell,V_\fg),
\end{align}
with which we have
\begin{align}
\tilde E_\fg(a) & = \sup_{p \in [\ell_{\Gamma}, u_{\Gamma}]}
  \EE_{L'_\fg \sim \mathrm{Bern}(p)}\big[\tilde \zeta(L'_{\fg})\big]\\
  & = \sup_{p \in [\ell_{\Gamma}, u_{\Gamma}]} \left\{p \cdot \tilde \zeta(1) + (1 - p) \cdot \tilde \zeta(-1)\right\}\\
  & = \max\left\{\tilde \zeta(-1) + \ell_{\Gamma} (\tilde \zeta(1) - \tilde \zeta(-1)), \;\tilde \zeta(-1) + u_{\Gamma} (\tilde \zeta(1) - \tilde \zeta(-1)) \right\}.
\end{align}
Then, we define $\hat \cS^{\cc} = \hat \cS \cup \{\fg \in \cG:\;\tilde E_\fg(h(L_{\fg},V_{\fg})) \leq 0\}$.

We note that evaluating $\tilde E_\fg(a)$ has constant computational cost, 
which is much more lightweight compared with the evaluation of $E_\fg(a)$, 
whose cost is exponential in $|\fg|$. The trade-off is that the finer 
conditioning reduces the remaining randomness and may lead to some 
loss in statistical efficiency.
In the main paper, we report the results conditional on fixed $V_{\fg}$; 
as discussed below, these two versions are comparable empirically.

\paragraph{Comparison between two versions of conditional calibration.}
In this section, we compare the performance of two versions of conditional calibration: \texttt{-cc} that treats $V_{\fg}$ as fixed and \texttt{-ccfull} that randomizes both $L_{\fg}$ and $V_{\fg}$.
Following the same data-generating process as in Section~\ref{sec:simu}, we consider the one-sided effect and tree-based subgroup partition with $I=200$ matched sets, in which we set \texttt{minsplit} and \texttt{minbucket} to be $1$. We vary $\Gamma \in \{1, 1.5\}$ and other settings remain the same with Section~\ref{sec:simu-positive}. 
\begin{figure}[ht]
    \centering
    \begin{subfigure}[t]{0.49\textwidth}
        \centering
        \includegraphics[height=0.65\textwidth]{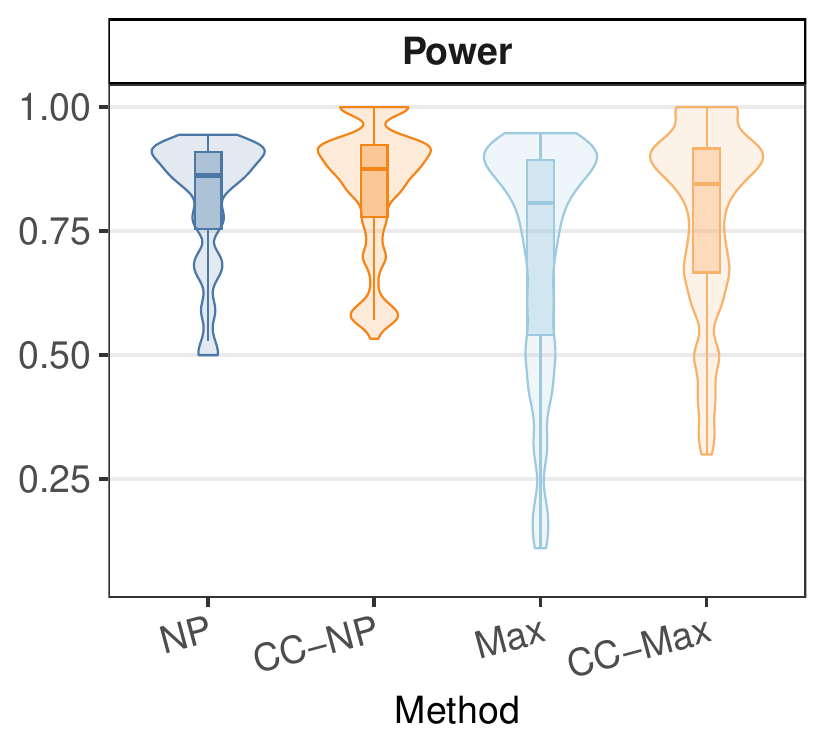}
        \caption{$\Gamma=1$.}
        \label{fig:cc_compare_regression_gamma1}
    \end{subfigure}\hfill
    \begin{subfigure}[t]{0.49\textwidth}
        \centering
        \includegraphics[height=0.65\textwidth]{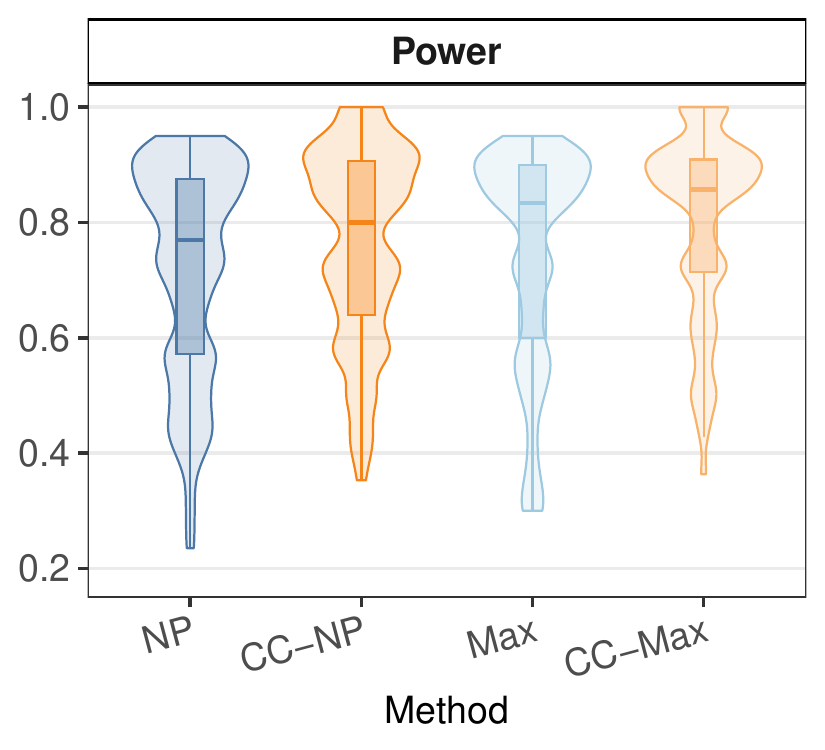}
        \caption{$\Gamma=1.5$.}
        \label{fig:cc_compare_regression_gamma15}
    \end{subfigure}
    \caption{Boxplots of FDR and power for comparison with and without conditional calibration.}
    \label{fig:cc_compare_regression}
\end{figure}

From Figure~\ref{fig:cc_compare_regression}, we can see that both \texttt{Ours-cc-NP} and \texttt{Ours-ccfull-NP} improve upon the power of \texttt{Ours-NP} in these two settings. In the first setting, without sensitivity adjustment, conditional calibration is less conservative; thus, its advantage is more pronounced.

\subsection{Alternative splitting methods in P-screening}\label{app:split-method}
\cite{chao2021adapt} generalizes the p-value splitting method in
~Section~\ref{sec:baseline}
to account for conservative (i.e., super-uniform) p-values.
Rather than estimating the number of null p-values in $[0,1/2)$ 
by counting those in $[1/2,1]$, they propose 
using p-values in an interval $[\lambda,\nu] \subseteq [0,1]$ to infer those in $[0,\underline{\alpha})$,
where $\underline{\alpha} \leq \lambda$ accounts for stretching factors.
Concretely, with $0 < \underline{\alpha} \leq \lambda < \nu \leq 1$, their proposed splitting is 
\begin{align}\label{eq:mask}
    L_{\fg}^\pscr = 
    \begin{cases}
    1, & p_\fg \in [0,\underline{\alpha})\\
    -1, & p_\fg \in [\lambda, \nu],\\
    0, & \text{otherwise,}
    \end{cases}
    \quad 
    V_{\fg}^\pscr = 
    \begin{cases}
        (\nu - p_{\fg})/\zeta, & p_{\fg} \in [\lambda, \nu],\\
        p_{\fg}, &\text{otherwise},
    \end{cases}
\end{align}
where $\zeta = (\nu - \lambda)/\underline{\alpha}$.
Note that, with $\underline{\alpha} = \lambda = 0.5$ and $\nu = 1$, \eqref{eq:mask} covers 
the splitting method in
~Section~\ref{sec:baseline}
as a special example.
Adapting the result of~\citet[Theorem 2.1]{chao2021adapt} to our setting, we 
obtain the following implication: if the subgroup-level p-value $p_\fg$ 
has a density that is {\em non-decreasing} under the null, then---by their theory---the 
$L^\pscr_\fg$'s satisfy the bounded skewness condition with $\kappa = \zeta$.
However, with the presence of unmeasured confounding ($\Gamma>1$), the group-level
p-values are not guaranteed to have non-decreasing density, as illustrated by the following example.

\begin{counterexample}
Suppose $|\fg|=3$ and $\Gamma = 3$. Let $r_i= i$, for $i=1,2,3$, and define 
\$
S_\fg = \sum^3_{i=1}r_i \ind\{L_i=1\},
\$ 
where $\ind\{L_i = 1\} \overset{\text{ind}}{\sim} \text{Bern}(p_i)$ with $(p_1,p_2,p_3) = (\tfrac{3}{5}, \tfrac{1}{2}, \tfrac{3}{4})$. 
Since $\tfrac{1}{1+\Gamma} \le p_i \le \frac{\Gamma}{1+\Gamma}$, for $\forall i =1,2,3$, the example is compatible 
with the sensitivity model. It can be verified that the density $f$ of the 
randomized p-values resulting from the Wilcoxon signed-rank test defined in~\eqref{eq:wilcox-pval}
satisfies the following: 
\$
f(40) = \frac{8}{5} > \frac{6}{5} = f(50).
\$
\end{counterexample}
The above counterexample demonstrates that the density of the signed-rank p-values  
need not be non-decreasing under the null; in particular, the bounded skewness is not 
guaranteed to hold. Similar to the case in Section~\ref{sec:baseline}, P-screening 
does not guarantee the FDR control unless the screening order is independent of 
$V^\pscr_\fg$. 

That said, the overestimation issue can be mitigated by carefully choosing 
$\lambda$ and $\nu$, so that the probability mass in $[0,\underline{\alpha})$ 
is comparable to that in $[\lambda,\nu]$.  

\renewcommand{\arraystretch}{0.8}
\begin{table}[h]
\centering
\small
\caption{Mapping between covariate names and original dataset columns.}
\label{tab:covariate_name_map}
\renewcommand{\arraystretch}{1.3}
\setlength{\tabcolsep}{6pt}
\begin{tabular}{lll}
\toprule
\textbf{Covariate name} & \textbf{Original column} & \textbf{Description} \\
\midrule
\texttt{parents\_income}  & \texttt{bmpin1}      & Parents' income\\
\texttt{mother\_edu}      & \texttt{bmmaedu}     & Mother's education\\
\texttt{father\_edu}      & \texttt{bmfaedu}     & Father's education\\
\texttt{intact}           & \texttt{z\_bklvpr}    &\makecell[l]{Indicator of family intactness\\(Live with both parents up to 16)}\\
\texttt{num\_sib}         & \texttt{sibstt}      & Number of siblings\\
\texttt{rural\_res}       & \texttt{res57}       &
\makecell[l]{Residential area of graduate. Degree of urbanization\\ (Counties with no city or with a city of less than 50,000)}\\
\texttt{prox\_college}    & \texttt{avcl57}      & \makecell[l]{Geographic availability of college\\(High school in community $\leq$ 15 miles from any college)}\\
\texttt{class\_rank}      & \texttt{hsrscorq}    & High school grades percentile rank-normalized\\
\texttt{IQ}               & \texttt{gwiiq\_j}    &
\makecell[l]{IQ score mapped from raw junior-year\\Henmon--Nelson test score}\\
\texttt{college\_prep}    & \texttt{hsprog}      &
\makecell[l]{Did graduate's high school classes fulfill\\the R01 entrance requirements for UW--Madison?}\\
\texttt{friends\_plans}   & \texttt{zfrplc}      & Did graduate's friends plan to go to college?\\
\texttt{parents\_encrg}   & \texttt{z\_iv205rer} &
\makecell[l]{Up until you were 18, how often did your parents\\encourage you to go to college?\\(1-5: never, rarely, sometimes, often, very often)}\\
\texttt{teacher\_encrg}   & \texttt{tchencq}     & \makecell[l]{Have your teachers encouraged you to attend college?\\(1-3: encouraged, discouraged, no effect)}\\
$Y$ & \texttt{z\_rg021jjd} & hourly wages at the age of 53 in 1993\\
\bottomrule
\end{tabular}
\end{table}

\section{Additional details and results in real data application}\label{sec:app-college}
In this section, we present additional details on the preprocessing and matching, as well as additional results on the college education dataset.
We use the Wisconsin Longitudinal Study (WLS) dataset to examine the heterogeneous effect of college education on earnings. 
The WLS dataset consists of $10,317$ men and women who graduated from Wisconsin high schools in 1957 and includes seven rounds of follow-up studies of this cohort to assess their educational and occupational attainment, well-being, and other aspects of the life course.

\renewcommand{\arraystretch}{0.8}
\begin{table}[p]
\centering
\caption{Summary statistics of baseline covariates.}
\label{tab:data_summary_updated}
\setlength{\tabcolsep}{3.0pt}     
\begin{tabular}{lrrrrrr}
\toprule
\textbf{Covariate} & \textbf{Min} & \textbf{1st Qu.} & \textbf{Median} & \textbf{Mean} & \textbf{3rd Qu.} & \textbf{Max} \\
\midrule
\multicolumn{7}{l}{\textbf{Continuous / ordinal covariates}} \\
\texttt{parents\_income} &  1.00 & 37.00 & 55.00 & 63.31 & 74.00 & 940.00 \\
$\log(1+\texttt{parents\_income})$ &  0.69 &  3.64 &  4.03 &  3.96 &  4.32 &   6.85 \\
\texttt{mother\_edu}     &  0.00 &  8.00 & 12.00 & 10.49 & 12.00 &  20.00 \\
\texttt{father\_edu}     &  0.00 &  8.00 &  8.00 &  9.75 & 12.00 &  26.00 \\
\texttt{num\_sib}        &  0.00 &  1.00 &  3.00 &  3.18 &  4.00 &  26.00 \\
\texttt{class\_rank}     & 61.00 & 93.00 &103.00 &102.75 &113.00 & 139.00 \\
\texttt{IQ}              & 61.00 & 92.00 &102.00 &102.30 &112.00 & 145.00 \\
\midrule
\multicolumn{7}{l}{\textbf{Factor covariates (counts per level)}} \\
\addlinespace
& \multicolumn{1}{c}{\textbf{1}} & \multicolumn{1}{c}{\textbf{2}} 
& \multicolumn{1}{c}{\textbf{3}} & \multicolumn{1}{c}{\textbf{4}} 
& \multicolumn{1}{c}{\textbf{5}} & \\
\cmidrule(lr){2-6}
\texttt{parents\_encrg}  & 1440 & 634 & 755 & 602 & 732 & \\
\texttt{teacher\_encrg}  & 2020 &  48 & 2095 &  &  & \\
\midrule
\multicolumn{7}{l}{\textbf{Binary covariates (counts)}} \\
\addlinespace
\multicolumn{1}{l}{\textbf{Covariate}} &
\multicolumn{3}{c}{\textbf{Level 0}} &
\multicolumn{3}{c}{\textbf{Level 1}} \\
\cmidrule(lr){2-4}\cmidrule(lr){5-7}
\texttt{intact}         & \multicolumn{3}{c}{331}  & \multicolumn{3}{c}{4375} \\
\texttt{rural\_res}     & \multicolumn{3}{c}{1535} & \multicolumn{3}{c}{3171} \\
\texttt{prox\_college}  & \multicolumn{3}{c}{1781} & \multicolumn{3}{c}{2925} \\
\texttt{college\_prep}  & \multicolumn{3}{c}{1844} & \multicolumn{3}{c}{2862} \\
\texttt{friends\_plans} & \multicolumn{3}{c}{2965} & \multicolumn{3}{c}{1741} \\
\texttt{college\_edu}   & \multicolumn{3}{c}{3391} & \multicolumn{3}{c}{1315} \\
\bottomrule
\end{tabular}
\end{table}

\subsection{Preprocessing}

With the focus on participants' earnings, we follow the setup in \cite{brand2010benefits} and consider the outcome $Y$--hourly wages at the age of $53$ in 1993.
The mapping between the analysis covariate names and the original WLS variable names is provided in Table~\ref{tab:covariate_name_map}.

We preprocess the WLS data prior to matching and inference: observations with incomplete treatment coding (treatment value $-2$) are recoded as untreated. We exclude individuals with extreme parental income values by removing observations with $\texttt{parents\_income} \ge 998$, and recode \texttt{intact} as a binary indicator of family intactness.
Parents' income is log-transformed to reduce skewness prior to matching and analysis.
We retain only observations with nonnegative values across all covariates, treatment, and outcomes.

Several covariates are then recoded into binary factors to align with substantive definitions: \texttt{rural\_res} indicates residence in a less urbanized area; \texttt{prox\_college} indicates geographic proximity to a college; \texttt{college\_prep} indicates completion of a college preparatory curriculum; and \texttt{friends\_plans} indicates whether peers planned to attend college. The treatment indicator is converted to a binary factor denoting college attendance. After removing missing data, with $Y$, we obtain a dataset with $4706$ effective entries for later matching.

\subsection{Additional details of matching}
\begin{figure}
    \centering
    \includegraphics[width=0.5\linewidth]{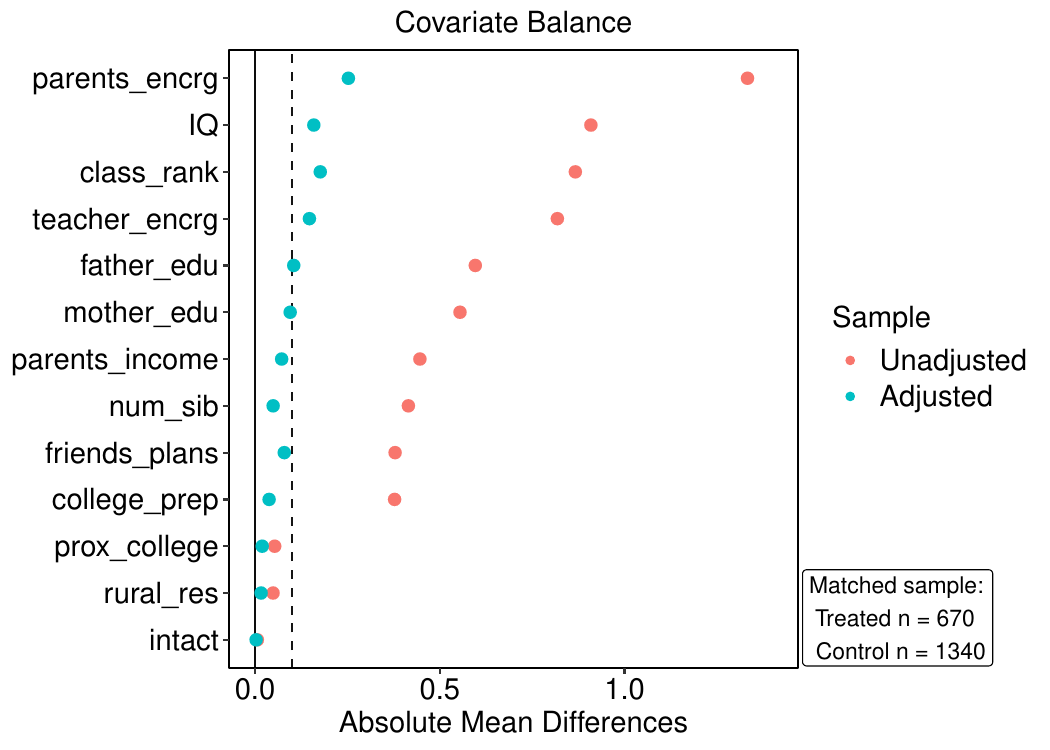}
    \caption{Covariate balance with propensity score matching: $\text{caliper}=0.4$.}
    \label{fig:love-plot}
\end{figure}

We adopt propensity score nearest-neighbor matching to construct covariate-balanced matched sets contrasting college and non-college graduates. Propensity scores are estimated by logistic regression via \texttt{distance = "glm"} in the R package \texttt{MatchIt} \citep{ho2011matchit}. To avoid poor-quality matches, we impose a standardized caliper of $0.4$ on the logit of the propensity score (\texttt{caliper = 0.4, std.caliper = TRUE}), so that any treated--control pair whose propensity-score difference exceeds $0.4$ standard deviations is excluded. Matching is performed without replacement (\texttt{replace = FALSE}), and the match ratio controls the number of controls per treated unit. With the cleaned dataset using $Y$ (hourly wages at age~$53$) as the outcome, a $1{:}2$ matching ratio yields $670$ matched sets, each containing one treated and two control units.

\renewcommand{\arraystretch}{1.0}
\begin{longtable}{@{} >{\raggedright\arraybackslash\baselineskip=10pt}p{9cm} r r ccccc @{}}
\caption{Subgroups from the conditional inference tree under $1{:}2$ matching.
\fullcirc\ = BH + NP + NP-CC; \halfcirc\ = NP/NP-CC only; \bhonly\ = BH only; \notsel\ = not selected.}
\label{tab:sensitivity-full-varset-lm-top9} \\

\toprule
Subgroup & $L_{\fg} \cdot W_{\fg}$ & $|\fg|$ & $\Gamma{=}1.0$ & $1.5$ & $2.0$ & $2.5$ & $3.0$ \\
\midrule
\endfirsthead

\toprule
Subgroup & $L_{\fg} \cdot W_{\fg}$ & $|\fg|$ & $\Gamma{=}1.0$ & $1.5$ & $2.0$ & $2.5$ & $3.0$ \\
\midrule
\endhead

\midrule
\multicolumn{8}{r}{\textit{continued on next page}} \\
\endfoot

\endlastfoot
$\fg_{1}$: \texttt{IQ <= 97}, \texttt{friends\_plans = 0}, \texttt{teacher\_encrg <= 1}, \texttt{parents\_income <= 47.0}
  & 107.7 & 57 & \fullcirc & \fullcirc & \halfcirc & \halfcirc & \notsel \\
$\fg_{2}$: \texttt{IQ <= 97}, \texttt{friends\_plans = 0}, \texttt{teacher\_encrg <= 1}, \texttt{parents\_income > 47.0}
  & 129.1 & 60 & \halfcirc & \halfcirc & \halfcirc & \halfcirc & \notsel \\
$\fg_{3}$: \texttt{IQ <= 97}, \texttt{friends\_plans = 0}, \texttt{teacher\_encrg > 1}, \texttt{num\_sib <= 3}
  & 162.2 & 75 & \fullcirc & \fullcirc & \halfcirc & \halfcirc & \notsel \\
$\fg_{4}$: \texttt{IQ <= 97}, \texttt{friends\_plans = 0}, \texttt{teacher\_encrg > 1}, \texttt{num\_sib > 3}
  & 154.9 & 54 & \halfcirc & \halfcirc & \halfcirc & \halfcirc & \notsel \\
$\fg_{5}$: \texttt{IQ <= 97}, \texttt{friends\_plans = 1}, \texttt{father\_edu <= 11}, \texttt{parents\_income <= 37.0}
  & 121.9 & 48 & \fullcirc & \halfcirc & \halfcirc & \halfcirc & \notsel \\
$\fg_{6}$: \texttt{IQ <= 97}, \texttt{friends\_plans = 1}, \texttt{father\_edu <= 11}, \texttt{parents\_income > 37.0}
  & 98.0 & 78 & \fullcirc & \fullcirc & \halfcirc & \halfcirc & \notsel \\
$\fg_{7}$: \texttt{IQ <= 97}, \texttt{friends\_plans = 1}, \texttt{father\_edu > 11}
  & 108.6 & 66 & \fullcirc & \halfcirc & \halfcirc & \halfcirc & \notsel \\
$\fg_{8}$: \texttt{IQ > 97}, \texttt{num\_sib <= 4}, \texttt{intact = 0}
  & 369.9 & 84 & \fullcirc & \fullcirc & \halfcirc & \halfcirc & \notsel \\
$\fg_{9}$: \texttt{IQ > 97}, \texttt{num\_sib <= 4}, \texttt{intact = 1}, \texttt{parents\_income <= 87.0}, \texttt{num\_sib <= 2}, \texttt{rural\_res = No}, \texttt{father\_edu <= 8}
  & 301.2 & 84 & \fullcirc & \fullcirc & \fullcirc & \halfcirc & \notsel \\
$\fg_{10}$: \texttt{IQ > 97}, \texttt{num\_sib <= 4}, \texttt{intact = 1}, \texttt{parents\_income <= 87.0}, \texttt{num\_sib <= 2}, \texttt{rural\_res = No}, \texttt{father\_edu > 8}
  & 194.7 & 90 & \fullcirc & \fullcirc & \fullcirc & \halfcirc & \notsel \\
$\fg_{11}$: \texttt{IQ > 97}, \texttt{num\_sib <= 4}, \texttt{intact = 1}, \texttt{parents\_income <= 87.0}, \texttt{num\_sib <= 2}, \texttt{rural\_res = Yes}, \texttt{mother\_edu <= 13}
  & 175.0 & 321 & \fullcirc & \fullcirc & \fullcirc & \halfcirc & \notsel \\
$\fg_{12}$: \texttt{IQ > 97}, \texttt{num\_sib <= 4}, \texttt{intact = 1}, \texttt{parents\_income <= 87.0}, \texttt{num\_sib <= 2}, \texttt{rural\_res = Yes}, \texttt{mother\_edu > 13}
  & 97.9 & 78 & \fullcirc & \halfcirc & \halfcirc & \halfcirc & \notsel \\
$\fg_{13}$: \texttt{IQ > 97}, \texttt{num\_sib <= 4}, \texttt{intact = 1}, \texttt{parents\_income <= 87.0}, \texttt{num\_sib > 2}, \texttt{rural\_res = No}, \texttt{father\_edu <= 10}
  & 215.7 & 39 & \halfcirc & \halfcirc & \halfcirc & \halfcirc & \notsel \\
$\fg_{14}$: \texttt{IQ > 97}, \texttt{num\_sib <= 4}, \texttt{intact = 1}, \texttt{parents\_income <= 87.0}, \texttt{num\_sib > 2}, \texttt{rural\_res = No}, \texttt{father\_edu > 10}
  & 106.0 & 51 & \halfcirc & \halfcirc & \halfcirc & \halfcirc & \notsel \\
$\fg_{15}$: \texttt{IQ > 97}, \texttt{num\_sib <= 4}, \texttt{intact = 1}, \texttt{parents\_income <= 87.0}, \texttt{num\_sib > 2}, \texttt{rural\_res = Yes}, \texttt{father\_edu <= 7}
  & 138.8 & 42 & \fullcirc & \fullcirc & \fullcirc & \halfcirc & \notsel \\
$\fg_{16}$: \texttt{IQ > 97}, \texttt{num\_sib <= 4}, \texttt{intact = 1}, \texttt{parents\_income <= 87.0}, \texttt{num\_sib > 2}, \texttt{rural\_res = Yes}, \texttt{father\_edu > 7}
  & 246.7 & 177 & \fullcirc & \fullcirc & \fullcirc & \halfcirc & \notsel \\
$\fg_{17}$: \texttt{IQ > 97}, \texttt{num\_sib <= 4}, \texttt{intact = 1}, \texttt{parents\_income > 87.0}, \texttt{friends\_plans = 0}, \texttt{IQ <= 107}
  & 211.9 & 39 & \halfcirc & \halfcirc & \halfcirc & \halfcirc & \notsel \\
$\fg_{18}$: \texttt{IQ > 97}, \texttt{num\_sib <= 4}, \texttt{intact = 1}, \texttt{parents\_income > 87.0}, \texttt{friends\_plans = 0}, \texttt{IQ > 107}
  & 170.2 & 69 & \fullcirc & \fullcirc & \fullcirc & \halfcirc & \notsel \\
$\fg_{19}$: \texttt{IQ > 97}, \texttt{num\_sib <= 4}, \texttt{intact = 1}, \texttt{parents\_income > 87.0}, \texttt{friends\_plans = 1}, \texttt{teacher\_encrg <= 1}
  & 384.0 & 69 & \fullcirc & \fullcirc & \halfcirc & \halfcirc & \notsel \\
$\fg_{20}$: \texttt{IQ > 97}, \texttt{num\_sib <= 4}, \texttt{intact = 1}, \texttt{parents\_income > 87.0}, \texttt{friends\_plans = 1}, \texttt{teacher\_encrg > 1}, \texttt{rural\_res = No}
  & 321.4 & 93 & \fullcirc & \fullcirc & \fullcirc & \halfcirc & \notsel \\
$\fg_{21}$: \texttt{IQ > 97}, \texttt{num\_sib <= 4}, \texttt{intact = 1}, \texttt{parents\_income > 87.0}, \texttt{friends\_plans = 1}, \texttt{teacher\_encrg > 1}, \texttt{rural\_res = Yes}
  & 218.5 & 96 & \fullcirc & \halfcirc & \halfcirc & \halfcirc & \notsel \\
$\fg_{22}$: \texttt{IQ > 97}, \texttt{num\_sib > 4}, \texttt{father\_edu <= 8}, \texttt{mother\_edu <= 10}
  & 372.4 & 81 & \fullcirc & \fullcirc & \fullcirc & \halfcirc & \notsel \\
$\fg_{23}$: \texttt{IQ > 97}, \texttt{num\_sib > 4}, \texttt{father\_edu <= 8}, \texttt{mother\_edu > 10}
  & 617.5 & 51 & \fullcirc & \fullcirc & \fullcirc & \halfcirc & \notsel \\
$\fg_{24}$: \texttt{IQ > 97}, \texttt{num\_sib > 4}, \texttt{father\_edu > 8}, \texttt{father\_edu <= 12}
  & 183.3 & 66 & \fullcirc & \fullcirc & \fullcirc & \halfcirc & \notsel \\
$\fg_{25}$: \texttt{IQ > 97}, \texttt{num\_sib > 4}, \texttt{father\_edu > 8}, \texttt{father\_edu > 12}
  & 200.9 & 42 & \halfcirc & \halfcirc & \halfcirc & \halfcirc & \notsel \\
\midrule
\# selected by \texttt{BH} & & & 19 & 15 & 10 & 0 & 0 \\
\# selected by \texttt{NP} / \texttt{NP-CC} & & & 25 & 25 & 25 & 25 & 0 \\
\bottomrule
\end{longtable}

\subsection{Additional details on subgroup partition}
\label{sec:app-subgroup}

With the matched dataset, to select covariates for tree fitting, we regress the magnitude statistics on all thirteen base covariates listed in Table~\ref{tab:covariate_name_map}. At the screening stage with a threshold of $0.6$ on p-values to encourage more covariates, nine covariates are retained: \texttt{num\_sib} ($p = 0.025$), \texttt{IQ} ($p = 0.084$), \texttt{intact} ($p = 0.097$), \texttt{parents\_income} ($p = 0.118$), \texttt{mother\_edu} ($0.244$), \texttt{father\_edu} ($0.275$), \texttt{friends\_plans} ($p=0.430$), \texttt{rural\_res} ($p=0.477$), \texttt{teacher\_encrg} ($p=0.546$). 
Because this screening uses only the magnitude statistics, it does not affect the validity of subsequent sign-based inference, giving us considerable flexibility in the choice of threshold. 
A conditional inference tree \citep{hothorn2006unbiased} is then fitted to the magnitude statistics using these discretized covariates with maximum depth $7$, and minimum leaf size $13$, but with a nominal level for subgroups splitting of $0.005$ to encourage more subgroups to be identified. The $25$ terminal nodes of this tree define the subgroups reported in Table~\ref{tab:sensitivity-fixed-1to2} and Table~\ref{tab:sensitivity-full-varset-lm-top9}.

\subsection{Results with a finer subgroup partition}
\renewcommand{\arraystretch}{1.0}
\begin{longtable}{@{} >{\raggedright\arraybackslash\baselineskip=10pt}p{9cm} r r ccccc @{}}
\caption{Finer subgroups from the conditional inference tree under $1{:}2$ matching.
\fullcirc\ = BH + NP + NP-CC; \halfcirc\ = NP/NP-CC only; \bhonly\ = BH only; \notsel\ = not selected.}
\label{tab:sensitivity-full-varset-lm-top9a} \\

\toprule
Subgroup & $L_{\fg} \cdot W_{\fg}$ & $|\fg|$ & $\Gamma{=}1.0$ & $1.5$ & $2.0$ & $2.5$ & $3.0$ \\
\midrule
\endfirsthead

\toprule
Subgroup & $L_{\fg} \cdot W_{\fg}$ & $|\fg|$ & $\Gamma{=}1.0$ & $1.5$ & $2.0$ & $2.5$ & $3.0$ \\
\midrule
\endhead

\midrule
\multicolumn{8}{r}{\textit{continued on next page}} \\
\endfoot

\endlastfoot
$\fg_{1}$: \texttt{IQ = low}, \texttt{intact = 0}, \texttt{rural\_res = No}
  & 131.9 & 30 & \fullcirc & \bhonly & \notsel & \notsel & \notsel \\
$\fg_{2}$: \texttt{IQ = low}, \texttt{intact = 0}, \texttt{rural\_res = Yes}, \texttt{teacher\_encrg = low}
  & 147.8 & 24 & \halfcirc & \notsel & \notsel & \notsel & \notsel \\
$\fg_{3}$: \texttt{IQ = low}, \texttt{intact = 0}, \texttt{rural\_res = Yes}, \texttt{teacher\_encrg = high}
  & 617.9 & 24 & \halfcirc & \halfcirc & \halfcirc & \notsel & \notsel \\
$\fg_{4}$: \texttt{IQ = low}, \texttt{intact = 1}, \texttt{friends\_plans = 0}, \texttt{teacher\_encrg = low}, \texttt{rural\_res = No}, \texttt{num\_sib = low}
  & 201.2 & 33 & \fullcirc & \halfcirc & \halfcirc & \notsel & \notsel \\
$\fg_{5}$: \texttt{IQ = low}, \texttt{intact = 1}, \texttt{friends\_plans = 0}, \texttt{teacher\_encrg = low}, \texttt{rural\_res = No}, \texttt{num\_sib = high}
  & 188.6 & 27 & \fullcirc & \notsel & \notsel & \notsel & \notsel \\
$\fg_{6}$: \texttt{IQ = low}, \texttt{intact = 1}, \texttt{friends\_plans = 0}, \texttt{teacher\_encrg = low}, \texttt{rural\_res = Yes}, \texttt{parents\_income = low}, \texttt{num\_sib = low}
  & 132.2 & 30 & \fullcirc & \bhonly & \notsel & \notsel & \notsel \\
$\fg_{7}$: \texttt{IQ = low}, \texttt{intact = 1}, \texttt{friends\_plans = 0}, \texttt{teacher\_encrg = low}, \texttt{rural\_res = Yes}, \texttt{parents\_income = low}, \texttt{num\_sib = high}
  & 132.3 & 63 & \fullcirc & \notsel & \notsel & \notsel & \notsel \\
$\fg_{8}$: \texttt{IQ = low}, \texttt{intact = 1}, \texttt{friends\_plans = 0}, \texttt{teacher\_encrg = low}, \texttt{rural\_res = Yes}, \texttt{parents\_income = high}, \texttt{father\_edu = low}
  & -134.2 & 15 & \notsel & \notsel & \notsel & \notsel & \notsel \\
$\fg_{9}$: \texttt{IQ = low}, \texttt{intact = 1}, \texttt{friends\_plans = 0}, \texttt{teacher\_encrg = low}, \texttt{rural\_res = Yes}, \texttt{parents\_income = high}, \texttt{father\_edu = high}
  & 164.9 & 45 & \fullcirc & \bhonly & \notsel & \notsel & \notsel \\
$\fg_{10}$: \texttt{IQ = low}, \texttt{intact = 1}, \texttt{friends\_plans = 0}, \texttt{teacher\_encrg = high}, \texttt{num\_sib = low}, \texttt{parents\_income = low}, \texttt{mother\_edu = low}, \texttt{father\_edu = low}
  & -146.9 & 48 & \bhonly & \notsel & \notsel & \notsel & \notsel \\
$\fg_{11}$: \texttt{IQ = low}, \texttt{intact = 1}, \texttt{friends\_plans = 0}, \texttt{teacher\_encrg = high}, \texttt{num\_sib = low}, \texttt{parents\_income = low}, \texttt{mother\_edu = low}, \texttt{father\_edu = high}
  & -170.0 & 24 & \notsel & \notsel & \notsel & \notsel & \notsel \\
$\fg_{12}$: \texttt{IQ = low}, \texttt{intact = 1}, \texttt{friends\_plans = 0}, \texttt{teacher\_encrg = high}, \texttt{num\_sib = low}, \texttt{parents\_income = low}, \texttt{mother\_edu = high}
  & -51.2 & 21 & \notsel & \notsel & \notsel & \notsel & \notsel \\
$\fg_{13}$: \texttt{IQ = low}, \texttt{intact = 1}, \texttt{friends\_plans = 0}, \texttt{teacher\_encrg = high}, \texttt{num\_sib = low}, \texttt{parents\_income = high}, \texttt{father\_edu = low}, \texttt{rural\_res = No}
  & 179.5 & 15 & \fullcirc & \bhonly & \notsel & \notsel & \notsel \\
$\fg_{14}$: \texttt{IQ = low}, \texttt{intact = 1}, \texttt{friends\_plans = 0}, \texttt{teacher\_encrg = high}, \texttt{num\_sib = low}, \texttt{parents\_income = high}, \texttt{father\_edu = low}, \texttt{rural\_res = Yes}
  & 164.6 & 24 & \fullcirc & \notsel & \notsel & \notsel & \notsel \\
$\fg_{15}$: \texttt{IQ = low}, \texttt{intact = 1}, \texttt{friends\_plans = 0}, \texttt{teacher\_encrg = high}, \texttt{num\_sib = low}, \texttt{parents\_income = high}, \texttt{father\_edu = high}
  & 330.7 & 30 & \fullcirc & \halfcirc & \halfcirc & \notsel & \notsel \\
$\fg_{16}$: \texttt{IQ = low}, \texttt{intact = 1}, \texttt{friends\_plans = 0}, \texttt{teacher\_encrg = high}, \texttt{num\_sib = high}, \texttt{father\_edu = low}, \texttt{rural\_res = No}
  & 132.0 & 18 & \halfcirc & \notsel & \notsel & \notsel & \notsel \\
$\fg_{17}$: \texttt{IQ = low}, \texttt{intact = 1}, \texttt{friends\_plans = 0}, \texttt{teacher\_encrg = high}, \texttt{num\_sib = high}, \texttt{father\_edu = low}, \texttt{rural\_res = Yes}, \texttt{parents\_income = low}
  & 376.2 & 75 & \fullcirc & \halfcirc & \halfcirc & \notsel & \notsel \\
$\fg_{18}$: \texttt{IQ = low}, \texttt{intact = 1}, \texttt{friends\_plans = 0}, \texttt{teacher\_encrg = high}, \texttt{num\_sib = high}, \texttt{father\_edu = low}, \texttt{rural\_res = Yes}, \texttt{parents\_income = high}
  & 572.5 & 24 & \fullcirc & \halfcirc & \halfcirc & \notsel & \notsel \\
$\fg_{19}$: \texttt{IQ = low}, \texttt{intact = 1}, \texttt{friends\_plans = 0}, \texttt{teacher\_encrg = high}, \texttt{num\_sib = high}, \texttt{father\_edu = high}, \texttt{rural\_res = No}
  & -195.7 & 18 & \notsel & \notsel & \notsel & \notsel & \notsel \\
$\fg_{20}$: \texttt{IQ = low}, \texttt{intact = 1}, \texttt{friends\_plans = 0}, \texttt{teacher\_encrg = high}, \texttt{num\_sib = high}, \texttt{father\_edu = high}, \texttt{rural\_res = Yes}
  & 170.5 & 33 & \halfcirc & \notsel & \notsel & \notsel & \notsel \\
$\fg_{21}$: \texttt{IQ = low}, \texttt{intact = 1}, \texttt{friends\_plans = 1}, \texttt{parents\_income = low}, \texttt{rural\_res = No}, \texttt{teacher\_encrg = low}, \texttt{num\_sib = low}
  & 149.8 & 21 & \fullcirc & \notsel & \notsel & \notsel & \notsel \\
$\fg_{22}$: \texttt{IQ = low}, \texttt{intact = 1}, \texttt{friends\_plans = 1}, \texttt{parents\_income = low}, \texttt{rural\_res = No}, \texttt{teacher\_encrg = low}, \texttt{num\_sib = high}
  & 157.3 & 18 & \halfcirc & \notsel & \notsel & \notsel & \notsel \\
$\fg_{23}$: \texttt{IQ = low}, \texttt{intact = 1}, \texttt{friends\_plans = 1}, \texttt{parents\_income = low}, \texttt{rural\_res = No}, \texttt{teacher\_encrg = high}
  & 124.5 & 27 & \halfcirc & \notsel & \notsel & \notsel & \notsel \\
$\fg_{24}$: \texttt{IQ = low}, \texttt{intact = 1}, \texttt{friends\_plans = 1}, \texttt{parents\_income = low}, \texttt{rural\_res = Yes}, \texttt{num\_sib = low}, \texttt{teacher\_encrg = low}, \texttt{father\_edu = low}
  & 87.7 & 18 & \fullcirc & \notsel & \notsel & \notsel & \notsel \\
$\fg_{25}$: \texttt{IQ = low}, \texttt{intact = 1}, \texttt{friends\_plans = 1}, \texttt{parents\_income = low}, \texttt{rural\_res = Yes}, \texttt{num\_sib = low}, \texttt{teacher\_encrg = low}, \texttt{father\_edu = high}
  & -70.2 & 18 & \notsel & \notsel & \notsel & \notsel & \notsel \\
$\fg_{26}$: \texttt{IQ = low}, \texttt{intact = 1}, \texttt{friends\_plans = 1}, \texttt{parents\_income = low}, \texttt{rural\_res = Yes}, \texttt{num\_sib = low}, \texttt{teacher\_encrg = high}, \texttt{father\_edu = low}
  & 101.2 & 33 & \halfcirc & \notsel & \notsel & \notsel & \notsel \\
$\fg_{27}$: \texttt{IQ = low}, \texttt{intact = 1}, \texttt{friends\_plans = 1}, \texttt{parents\_income = low}, \texttt{rural\_res = Yes}, \texttt{num\_sib = low}, \texttt{teacher\_encrg = high}, \texttt{father\_edu = high}
  & 119.8 & 21 & \fullcirc & \bhonly & \notsel & \notsel & \notsel \\
$\fg_{28}$: \texttt{IQ = low}, \texttt{intact = 1}, \texttt{friends\_plans = 1}, \texttt{parents\_income = low}, \texttt{rural\_res = Yes}, \texttt{num\_sib = high}, \texttt{father\_edu = low}, \texttt{teacher\_encrg = low}
  & 88.6 & 30 & \halfcirc & \notsel & \notsel & \notsel & \notsel \\
$\fg_{29}$: \texttt{IQ = low}, \texttt{intact = 1}, \texttt{friends\_plans = 1}, \texttt{parents\_income = low}, \texttt{rural\_res = Yes}, \texttt{num\_sib = high}, \texttt{father\_edu = low}, \texttt{teacher\_encrg = high}
  & 181.0 & 45 & \fullcirc & \notsel & \notsel & \notsel & \notsel \\
$\fg_{30}$: \texttt{IQ = low}, \texttt{intact = 1}, \texttt{friends\_plans = 1}, \texttt{parents\_income = low}, \texttt{rural\_res = Yes}, \texttt{num\_sib = high}, \texttt{father\_edu = high}
  & 200.7 & 33 & \fullcirc & \fullcirc & \halfcirc & \notsel & \notsel \\
$\fg_{31}$: \texttt{IQ = low}, \texttt{intact = 1}, \texttt{friends\_plans = 1}, \texttt{parents\_income = high}, \texttt{num\_sib = low}, \texttt{rural\_res = No}, \texttt{mother\_edu = low}, \texttt{father\_edu = low}, \texttt{teacher\_encrg = low}
  & 550.8 & 18 & \fullcirc & \halfcirc & \halfcirc & \notsel & \notsel \\
$\fg_{32}$: \texttt{IQ = low}, \texttt{intact = 1}, \texttt{friends\_plans = 1}, \texttt{parents\_income = high}, \texttt{num\_sib = low}, \texttt{rural\_res = No}, \texttt{mother\_edu = low}, \texttt{father\_edu = low}, \texttt{teacher\_encrg = high}
  & 158.3 & 24 & \fullcirc & \bhonly & \notsel & \notsel & \notsel \\
$\fg_{33}$: \texttt{IQ = low}, \texttt{intact = 1}, \texttt{friends\_plans = 1}, \texttt{parents\_income = high}, \texttt{num\_sib = low}, \texttt{rural\_res = No}, \texttt{mother\_edu = low}, \texttt{father\_edu = high}
  & 155.4 & 24 & \fullcirc & \notsel & \notsel & \notsel & \notsel \\
$\fg_{34}$: \texttt{IQ = low}, \texttt{intact = 1}, \texttt{friends\_plans = 1}, \texttt{parents\_income = high}, \texttt{num\_sib = low}, \texttt{rural\_res = No}, \texttt{mother\_edu = high}
  & 166.8 & 18 & \halfcirc & \notsel & \notsel & \notsel & \notsel \\
$\fg_{35}$: \texttt{IQ = low}, \texttt{intact = 1}, \texttt{friends\_plans = 1}, \texttt{parents\_income = high}, \texttt{num\_sib = low}, \texttt{rural\_res = Yes}, \texttt{mother\_edu = low}, \texttt{teacher\_encrg = low}
  & 72.9 & 24 & \halfcirc & \notsel & \notsel & \notsel & \notsel \\
$\fg_{36}$: \texttt{IQ = low}, \texttt{intact = 1}, \texttt{friends\_plans = 1}, \texttt{parents\_income = high}, \texttt{num\_sib = low}, \texttt{rural\_res = Yes}, \texttt{mother\_edu = low}, \texttt{teacher\_encrg = high}, \texttt{father\_edu = low}
  & 299.1 & 15 & \fullcirc & \halfcirc & \halfcirc & \notsel & \notsel \\
$\fg_{37}$: \texttt{IQ = low}, \texttt{intact = 1}, \texttt{friends\_plans = 1}, \texttt{parents\_income = high}, \texttt{num\_sib = low}, \texttt{rural\_res = Yes}, \texttt{mother\_edu = low}, \texttt{teacher\_encrg = high}, \texttt{father\_edu = high}
  & 134.7 & 42 & \halfcirc & \notsel & \notsel & \notsel & \notsel \\
$\fg_{38}$: \texttt{IQ = low}, \texttt{intact = 1}, \texttt{friends\_plans = 1}, \texttt{parents\_income = high}, \texttt{num\_sib = low}, \texttt{rural\_res = Yes}, \texttt{mother\_edu = high}
  & 91.6 & 30 & \halfcirc & \notsel & \notsel & \notsel & \notsel \\
$\fg_{39}$: \texttt{IQ = low}, \texttt{intact = 1}, \texttt{friends\_plans = 1}, \texttt{parents\_income = high}, \texttt{num\_sib = high}, \texttt{rural\_res = No}, \texttt{teacher\_encrg = low}
  & 185.1 & 15 & \fullcirc & \notsel & \notsel & \notsel & \notsel \\
$\fg_{40}$: \texttt{IQ = low}, \texttt{intact = 1}, \texttt{friends\_plans = 1}, \texttt{parents\_income = high}, \texttt{num\_sib = high}, \texttt{rural\_res = No}, \texttt{teacher\_encrg = high}
  & 85.9 & 30 & \fullcirc & \notsel & \notsel & \notsel & \notsel \\
$\fg_{41}$: \texttt{IQ = low}, \texttt{intact = 1}, \texttt{friends\_plans = 1}, \texttt{parents\_income = high}, \texttt{num\_sib = high}, \texttt{rural\_res = Yes}, \texttt{father\_edu = low}
  & 129.5 & 24 & \fullcirc & \notsel & \notsel & \notsel & \notsel \\
$\fg_{42}$: \texttt{IQ = low}, \texttt{intact = 1}, \texttt{friends\_plans = 1}, \texttt{parents\_income = high}, \texttt{num\_sib = high}, \texttt{rural\_res = Yes}, \texttt{father\_edu = high}
  & 157.3 & 18 & \halfcirc & \notsel & \notsel & \notsel & \notsel \\
$\fg_{43}$: \texttt{IQ = high}, \texttt{rural\_res = No}, \texttt{num\_sib = low}, \texttt{friends\_plans = 0}, \texttt{parents\_income = low}
  & 180.8 & 15 & \fullcirc & \notsel & \notsel & \notsel & \notsel \\
$\fg_{44}$: \texttt{IQ = high}, \texttt{rural\_res = No}, \texttt{num\_sib = low}, \texttt{friends\_plans = 0}, \texttt{parents\_income = high}, \texttt{father\_edu = low}, \texttt{teacher\_encrg = low}
  & 151.4 & 18 & \fullcirc & \notsel & \notsel & \notsel & \notsel \\
$\fg_{45}$: \texttt{IQ = high}, \texttt{rural\_res = No}, \texttt{num\_sib = low}, \texttt{friends\_plans = 0}, \texttt{parents\_income = high}, \texttt{father\_edu = low}, \texttt{teacher\_encrg = high}
  & 175.4 & 18 & \fullcirc & \bhonly & \notsel & \notsel & \notsel \\
$\fg_{46}$: \texttt{IQ = high}, \texttt{rural\_res = No}, \texttt{num\_sib = low}, \texttt{friends\_plans = 0}, \texttt{parents\_income = high}, \texttt{father\_edu = high}
  & 167.6 & 33 & \fullcirc & \notsel & \notsel & \notsel & \notsel \\
$\fg_{47}$: \texttt{IQ = high}, \texttt{rural\_res = No}, \texttt{num\_sib = low}, \texttt{friends\_plans = 1}, \texttt{parents\_income = low}
  & 156.3 & 21 & \fullcirc & \bhonly & \notsel & \notsel & \notsel \\
$\fg_{48}$: \texttt{IQ = high}, \texttt{rural\_res = No}, \texttt{num\_sib = low}, \texttt{friends\_plans = 1}, \texttt{parents\_income = high}, \texttt{mother\_edu = low}
  & 371.9 & 66 & \fullcirc & \fullcirc & \halfcirc & \notsel & \notsel \\
$\fg_{49}$: \texttt{IQ = high}, \texttt{rural\_res = No}, \texttt{num\_sib = low}, \texttt{friends\_plans = 1}, \texttt{parents\_income = high}, \texttt{mother\_edu = high}
  & 213.1 & 30 & \fullcirc & \halfcirc & \halfcirc & \notsel & \notsel \\
$\fg_{50}$: \texttt{IQ = high}, \texttt{rural\_res = No}, \texttt{num\_sib = high}, \texttt{father\_edu = low}, \texttt{friends\_plans = 0}
  & 228.9 & 18 & \fullcirc & \halfcirc & \halfcirc & \notsel & \notsel \\
$\fg_{51}$: \texttt{IQ = high}, \texttt{rural\_res = No}, \texttt{num\_sib = high}, \texttt{father\_edu = low}, \texttt{friends\_plans = 1}
  & 738.8 & 18 & \fullcirc & \halfcirc & \halfcirc & \notsel & \notsel \\
$\fg_{52}$: \texttt{IQ = high}, \texttt{rural\_res = No}, \texttt{num\_sib = high}, \texttt{father\_edu = high}
  & 270.8 & 60 & \fullcirc & \halfcirc & \halfcirc & \notsel & \notsel \\
$\fg_{53}$: \texttt{IQ = high}, \texttt{rural\_res = Yes}, \texttt{teacher\_encrg = low}, \texttt{mother\_edu = low}, \texttt{father\_edu = low}, \texttt{num\_sib = low}
  & 258.9 & 24 & \fullcirc & \halfcirc & \halfcirc & \notsel & \notsel \\
$\fg_{54}$: \texttt{IQ = high}, \texttt{rural\_res = Yes}, \texttt{teacher\_encrg = low}, \texttt{mother\_edu = low}, \texttt{father\_edu = low}, \texttt{num\_sib = high}
  & 198.2 & 27 & \fullcirc & \fullcirc & \halfcirc & \notsel & \notsel \\
$\fg_{55}$: \texttt{IQ = high}, \texttt{rural\_res = Yes}, \texttt{teacher\_encrg = low}, \texttt{mother\_edu = low}, \texttt{father\_edu = high}, \texttt{parents\_income = low}
  & 202.9 & 27 & \halfcirc & \halfcirc & \halfcirc & \notsel & \notsel \\
$\fg_{56}$: \texttt{IQ = high}, \texttt{rural\_res = Yes}, \texttt{teacher\_encrg = low}, \texttt{mother\_edu = low}, \texttt{father\_edu = high}, \texttt{parents\_income = high}
  & 140.3 & 21 & \halfcirc & \notsel & \notsel & \notsel & \notsel \\
$\fg_{57}$: \texttt{IQ = high}, \texttt{rural\_res = Yes}, \texttt{teacher\_encrg = low}, \texttt{mother\_edu = high}
  & 582.8 & 18 & \halfcirc & \halfcirc & \halfcirc & \notsel & \notsel \\
$\fg_{58}$: \texttt{IQ = high}, \texttt{rural\_res = Yes}, \texttt{teacher\_encrg = high}, \texttt{intact = 0}, \texttt{father\_edu = low}
  & 180.1 & 18 & \fullcirc & \notsel & \notsel & \notsel & \notsel \\
$\fg_{59}$: \texttt{IQ = high}, \texttt{rural\_res = Yes}, \texttt{teacher\_encrg = high}, \texttt{intact = 0}, \texttt{father\_edu = high}
  & 134.5 & 15 & \fullcirc & \bhonly & \notsel & \notsel & \notsel \\
$\fg_{60}$: \texttt{IQ = high}, \texttt{rural\_res = Yes}, \texttt{teacher\_encrg = high}, \texttt{intact = 1}, \texttt{father\_edu = low}, \texttt{mother\_edu = low}, \texttt{friends\_plans = 0}, \texttt{parents\_income = low}, \texttt{num\_sib = low}
  & 165.4 & 24 & \fullcirc & \notsel & \notsel & \notsel & \notsel \\
$\fg_{61}$: \texttt{IQ = high}, \texttt{rural\_res = Yes}, \texttt{teacher\_encrg = high}, \texttt{intact = 1}, \texttt{father\_edu = low}, \texttt{mother\_edu = low}, \texttt{friends\_plans = 0}, \texttt{parents\_income = low}, \texttt{num\_sib = high}
  & 208.6 & 27 & \fullcirc & \fullcirc & \halfcirc & \notsel & \notsel \\
$\fg_{62}$: \texttt{IQ = high}, \texttt{rural\_res = Yes}, \texttt{teacher\_encrg = high}, \texttt{intact = 1}, \texttt{father\_edu = low}, \texttt{mother\_edu = low}, \texttt{friends\_plans = 0}, \texttt{parents\_income = high}
  & 233.2 & 21 & \fullcirc & \halfcirc & \halfcirc & \notsel & \notsel \\
$\fg_{63}$: \texttt{IQ = high}, \texttt{rural\_res = Yes}, \texttt{teacher\_encrg = high}, \texttt{intact = 1}, \texttt{father\_edu = low}, \texttt{mother\_edu = low}, \texttt{friends\_plans = 1}, \texttt{parents\_income = low}, \texttt{num\_sib = low}
  & 146.9 & 18 & \fullcirc & \bhonly & \notsel & \notsel & \notsel \\
$\fg_{64}$: \texttt{IQ = high}, \texttt{rural\_res = Yes}, \texttt{teacher\_encrg = high}, \texttt{intact = 1}, \texttt{father\_edu = low}, \texttt{mother\_edu = low}, \texttt{friends\_plans = 1}, \texttt{parents\_income = low}, \texttt{num\_sib = high}
  & 162.5 & 24 & \fullcirc & \bhonly & \notsel & \notsel & \notsel \\
$\fg_{65}$: \texttt{IQ = high}, \texttt{rural\_res = Yes}, \texttt{teacher\_encrg = high}, \texttt{intact = 1}, \texttt{father\_edu = low}, \texttt{mother\_edu = low}, \texttt{friends\_plans = 1}, \texttt{parents\_income = high}
  & 124.6 & 33 & \fullcirc & \notsel & \notsel & \notsel & \notsel \\
$\fg_{66}$: \texttt{IQ = high}, \texttt{rural\_res = Yes}, \texttt{teacher\_encrg = high}, \texttt{intact = 1}, \texttt{father\_edu = low}, \texttt{mother\_edu = high}, \texttt{num\_sib = low}
  & 115.7 & 18 & \fullcirc & \notsel & \notsel & \notsel & \notsel \\
$\fg_{67}$: \texttt{IQ = high}, \texttt{rural\_res = Yes}, \texttt{teacher\_encrg = high}, \texttt{intact = 1}, \texttt{father\_edu = low}, \texttt{mother\_edu = high}, \texttt{num\_sib = high}
  & 157.7 & 18 & \fullcirc & \notsel & \notsel & \notsel & \notsel \\
$\fg_{68}$: \texttt{IQ = high}, \texttt{rural\_res = Yes}, \texttt{teacher\_encrg = high}, \texttt{intact = 1}, \texttt{father\_edu = high}
  & 250.1 & 213 & \fullcirc & \fullcirc & \fullcirc & \notsel & \notsel \\
\midrule
\# selected by \texttt{BH} & & & 47 & 16 & 1 & 0 & 0 \\
\# selected by \texttt{NP} / \texttt{NP-CC} & & & 62 & 20 & 20 & 0 & 0 \\
\bottomrule
\end{longtable}

We can see that subgroups shown in Table~\ref{tab:sensitivity-full-varset-lm-top9} are all selected with small values of $\Gamma$, which is due to the fact that we are interested in the subgroup-level global null, so any matched set with a positive effect will make the entire subgroup a nonnull. Since the subgroups are determined data-drivenly using conditional trees, when the partition is not fine enough, most of the identified subgroups are likely to be nonnull.
Based on this observation, we also consider a refinement of the subgroups partition, where we allow deeper tree structures (the depth of $15$) and a smaller lower bound ($5$) on the sample size in each leaf. Here, to enhance interpretability and avoid scenarios with too small subgroups, we add the constraint that each variable can be split only at the median during tree fitting. 
Identified subgroups and their significance under different levels of $\Gamma$ are presented in Table~\ref{tab:sensitivity-full-varset-lm-top9a}.

\section{Additional simulation details and results}\label{sec:add-simu}

\subsection{A formal introduction of simulation setup}\label{app:simu_setup}
We begin by describing the data-generating process used throughout this section.
Given the number of matched sets \(I\) and \(n_i\ge 2\) for \(i=1,\dots,I\), 
we generate for each matched set \(i\in\{1,\dots,I\}\) 
a $d$-dimensional covariate vector shared by all units in the set
from a multivariate Gaussian distribution: $X_i \sim \mathcal{N}(0, I_d)$.

To define the ``important'' sets, we assign each set $i\in [I]$ an importance score 
\(s_i = X_{i1} + 0.5 X_{i2}\).
We then let \(t\) be the \((1-p_{\mathrm{imp}})\)-quantile 
of \(\{s_i\}_{i=1}^I\) and define the group label $S_i \;=\; \mathbf{1}\{s_i \ge t\}\in\{0,1\}$.
Thus, $p_{\mathrm{imp}}$ controls the proportion of ``important'' units, and we fix $p_{\mathrm{imp}} = 0.3$ throughout this section.
The potential outcomes are generated by $R_{ij} \;=\; \beta^\top X_i \;+\; \alpha_u\, u_{ij} \;+\;\tau(X_i) \cdot S_i \cdot \ind\{Z_{ij}=1\} \;+\; \varepsilon_{ij}$,
where \(\varepsilon_{ij}\overset{\mathrm{i.i.d.}}{\sim}\mathcal{N}(0,1)\),
$\alpha_u = 0.2$, and $u_{ij} \sim {\rm Unif}(0,1)$ is the unit-level latent confounder.
The vector $\beta \in \RR^d$ is sampled once with independent entries
$\beta_j \sim {\rm Unif}(0,1)$, and is 
held fixed across simulation runs.  
The heterogeneous treatment effect $\tau(X_i)$ takes the form $\tau(X_i) = \tau^* \cdot \Phi( s^{-1}_Z(Z_i - \bar{Z}) )$, where $Z_i = X_i^\top \eta - 1$.
Above, $\tau^* = 3.0$ for random subgroup partition and $\tau^* = 4.0$ for tree-based partition, $\eta \in \RR^d$ is 
sampled once with independent entries $\eta_j\sim {\rm Unif}(0,1)$,
and $\bar Z$ and $s_Z$ are the sample mean and standard deviation of $\{Z_i\}_{i\in[I]}$.

To model treatment assignment under unmeasured confounding, we adopt Rosenbaum's sensitivity model \citep{rosenbaum2020design}. Specifically, we define
\[
\pi_{ij} \;=\;\frac{w_{ij}}{\sum_{\ell=1}^{n_i} w_{i\ell}}, \text{ where }
w_{ij} \;=\; \Gamma^{\,u_{ij}} \;=\; \exp\bigl((\log\Gamma)\,u_{ij}\bigr), \forall j\in[n_i].
\]
We then sample the index of the treated unit $T_i$ from a multinomial distribution 
with parameters $(\pi_{i1},\cdots,\pi_{in_i})$ and let $Z_{iT_i} = 1 \text{ and } Z_{ij} =0$, $\forall j \neq T_i$.
By construction,
\(
\pi_{ij}/\pi_{ik} = \Gamma^{\,u_{ij}-u_{ik}} \in [\Gamma^{-1},\,\Gamma]
\)
for any \(j,k\), i.e., treatment probabilities within a set are bounded by \(\Gamma\).
In tree-based subgroup partitions, since the final FDR control is free of configurations of the tree, we set the nominal level to be $0.05$ to produce a sufficiently large number of subgroups and will later vary the sample size restriction \texttt{minsplit} to control the size of subgroups.

\subsection{Effect modifications with varying group sizes}\label{sec:add-simu-group}
\begin{figure}[ht]
  \centering
  \begin{subfigure}[b]{0.49\textwidth}
    \centering
    \includegraphics[height=0.55\textwidth]{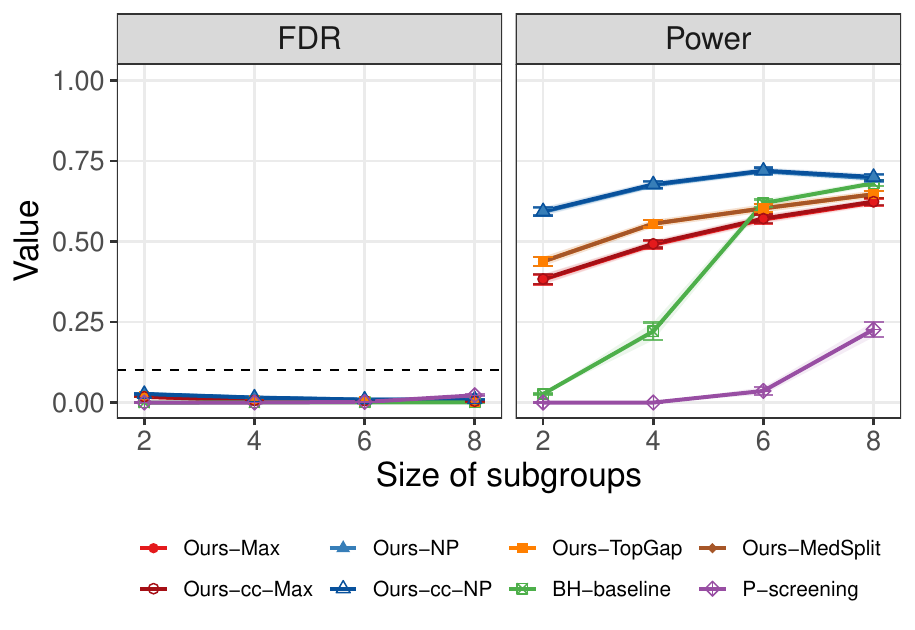}
    \caption{One-sided effects.}
    \label{fig:1side-1out-tree}
  \end{subfigure}
  \hfill
  \begin{subfigure}[b]{0.49\textwidth}
    \centering
    \includegraphics[height=0.55\textwidth]{figs_paper/MI_2sided_random_nout1.pdf}
    \caption{Two-sided effects.}
    \label{fig:2side-1out-tree}
  \end{subfigure}
  \caption{FDR and Power comparison versus varying group sizes with tree-based subgroup partition.}
  \label{fig:1out-tree}
\end{figure}

We follow the same data-generating process as in Section~\ref{sec:simu-positive} and present results with tree-based partitions in Figure~\ref{fig:1out-tree}.
Here, group size refers to the \texttt{minsplit} parameter in \texttt{partykit::ctree}, which controls the minimum sample size in each subgroup.
Unlike random partitions, the tree-based partition is fully data-driven, and increasing \texttt{minsplit} can cause matched sets with nonzero effects to be spread across different subgroups, which may reduce the subgroup-level signal-to-noise ratio.  As a result, power may decrease as group size increases.

This phenomenon is visible in Figure~\ref{fig:1out-tree}: under one-sided effects (Figure~\ref{fig:1side-1out-tree}), \texttt{Ours-NP} outperforms not only two baselines but also variants of our method across all subgroup sizes.  The \texttt{BH-baseline}, starting near zero at small group sizes, gradually improves but does not surpass $\texttt{Ours-NP}$ even at the largest group size.  Under two-sided effects (Figure~\ref{fig:2side-1out-tree}), the baselines tend to have a lower power, while our methods continue to outperform.

\subsection{Effect modifications with multiple controls}\label{sec:add-simu-ctrl}

\begin{figure}[ht]
    \centering
    \begin{subfigure}[b]{0.49\textwidth}
        \centering
        \includegraphics[height=0.55\textwidth]{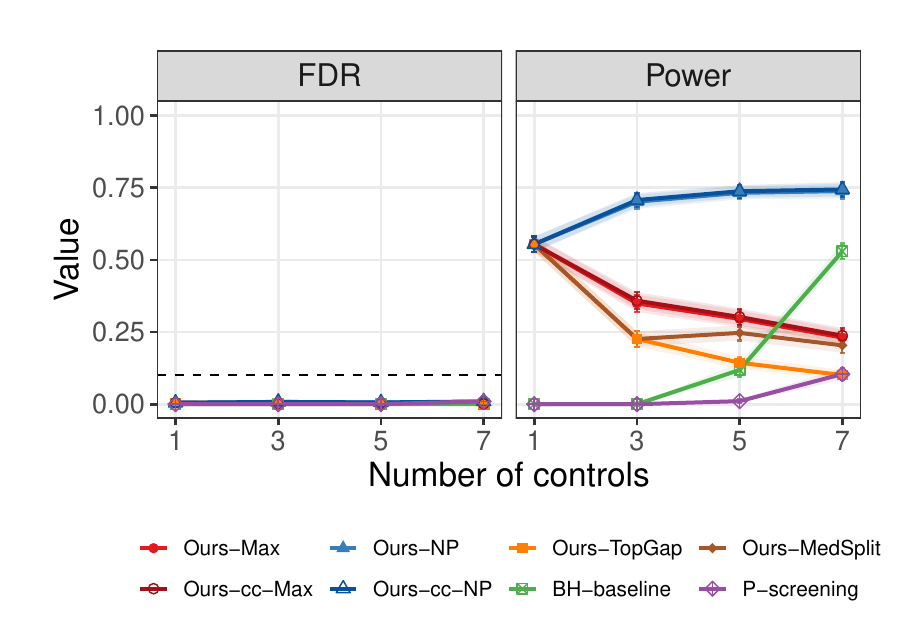}
        \caption{Group size $=6$.}
        \label{fig:2side-1out-random-ctrl6}
    \end{subfigure}
    \hfill
    \begin{subfigure}[b]{0.49\textwidth}
        \centering
        \includegraphics[height=0.55\textwidth]{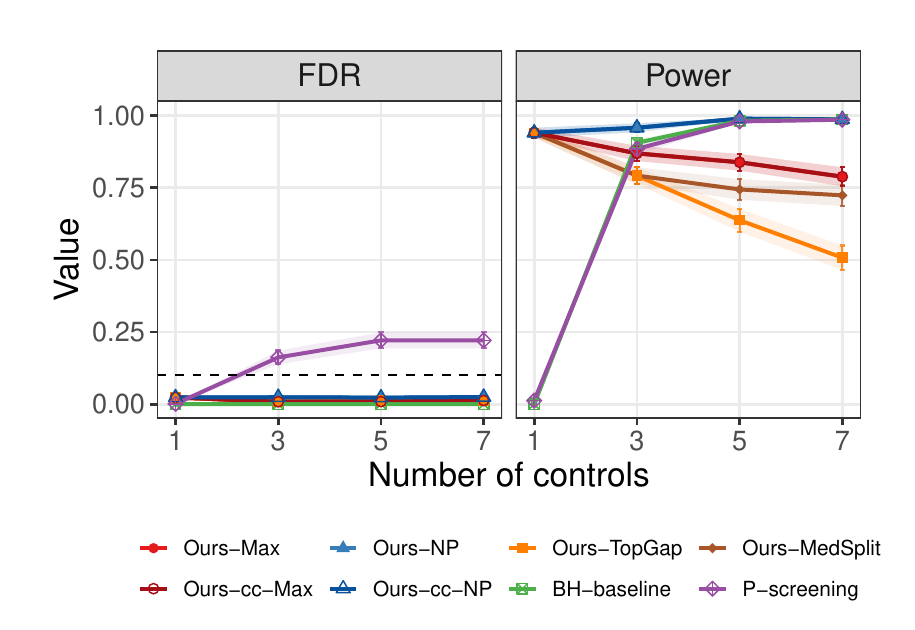}
        \caption{Group size $=18$.}
        \label{fig:2side-1out-random-ctrl18}
    \end{subfigure}
    \caption{FDR and Power comparison with random subgroup partition: two-sided effects and varying number of controls.}
    \label{fig:2side-1out-random-ctrl}
\end{figure}

We adopt the same data-generating process as in Section~\ref{sec:simu-ctrl}, now under two-sided effects.  We vary $n_i$ in $\{2,4,6,8\}$ and present results with random partitions (Figure~\ref{fig:2side-1out-random-ctrl}).

For random partitions (Figure~\ref{fig:2side-1out-random-ctrl}), the patterns are consistent with the one-sided case but with even starker contrasts.
With small subgroups of size $6$ (Figure~\ref{fig:2side-1out-random-ctrl6}), both baseline methods, the \texttt{BH-baseline} and \texttt{P-screening}, have substantially lower power compared to \texttt{Ours-NP}, regardless of the number of controls, whereas our method \texttt{Ours-NP} achieves power close to $0.7$ uniformly across all settings.
Moreover, variants \texttt{Ours-Max}, \texttt{Ours-TopGap}, and \texttt{Ours-MedSplit} undergo a loss in power when the number of control units increases, as discussed in Section~\ref{sec:multictrls}. 
With larger subgroups of size $18$ (Figure~\ref{fig:2side-1out-random-ctrl18}), the \texttt{BH-baseline} improves as controls are added but remains below our methods.


\subsection{Effect modifications with multiple outcomes}\label{sec:add-simu-outs}

In this setting, for each outcome $m \in [M]$, we generate $R^{(m)}_{ij}$ independently following Section~\ref{sec:simu_setup}, in which the random vector $\beta_m$ varies across outcomes.
To design potential weak signals across outcomes, for each $(i,j)$, we mask the underlying effects of $q \times 100\%$ outcomes by zero, and in this simulation, we choose $q = 0.6$. Each method is implemented following procedures in Section~\ref{sec:multiouts}.

\begin{figure}[t]
  \centering
  \begin{subfigure}[b]{0.49\textwidth}
    \centering
    \includegraphics[height=0.55\textwidth]{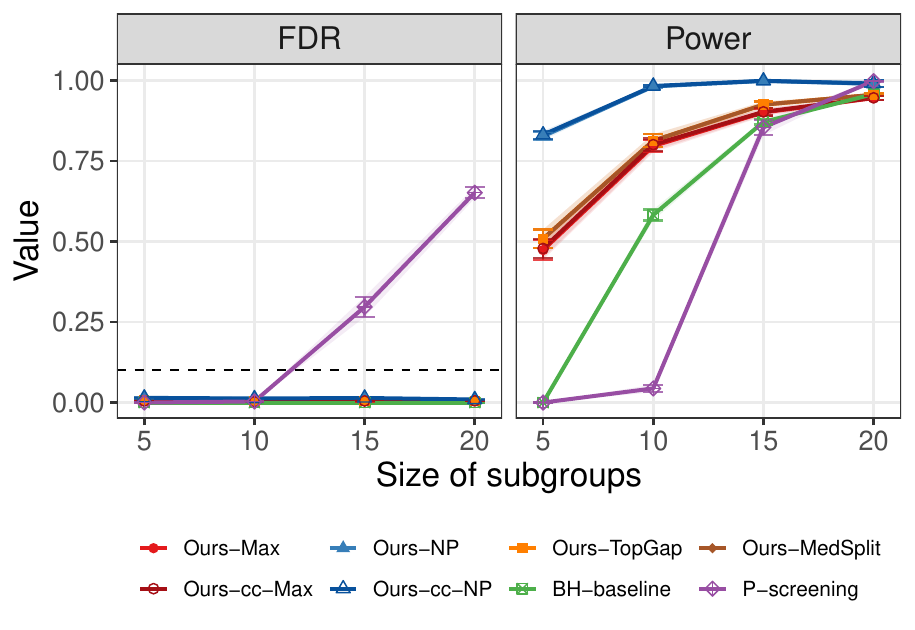}
    \caption{One-sided effects.}
    \label{fig:1side-5out-random}
  \end{subfigure}
  \hfill
  \begin{subfigure}[b]{0.49\textwidth}
    \centering
    \includegraphics[height=0.55\textwidth]{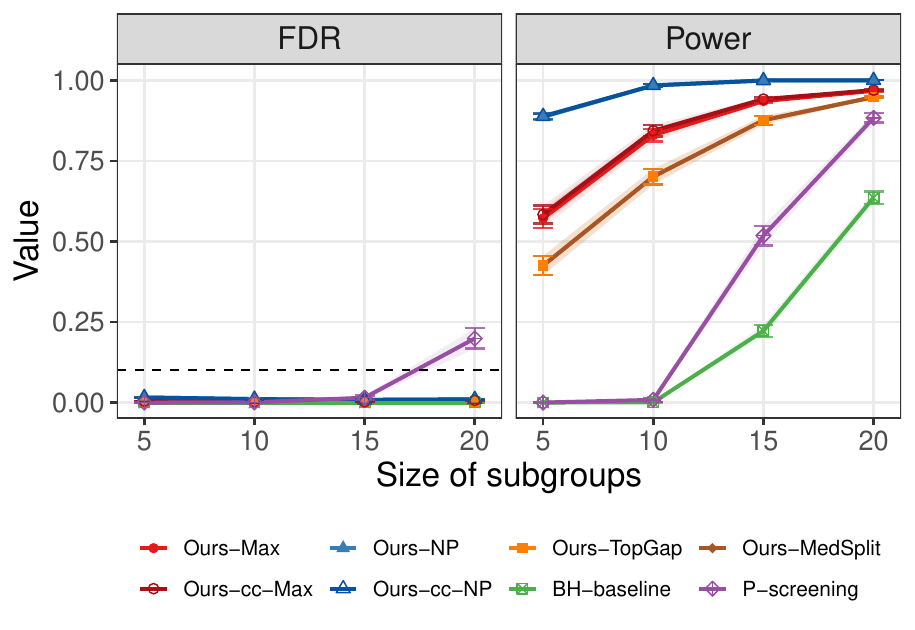}
    \caption{Two-sided effects.}
    \label{fig:2side-5out-random}
  \end{subfigure}
  \caption{FDR and power versus group size: multiple outcomes and varying subgroup sizes.}
  \label{fig:5out-random}
\end{figure}

\paragraph{Random subgroup partition.}
We start with simulations with the random subgroup partition.
In Figure~\ref{fig:5out-random}, we present results in this setting with varying sizes of subgroups in $\{5,10,15,20\}$. Similar to the results with a single outcome, the performance of two baselines is sensitive to subgroup sizes, and they tend to lose power when subgroups are not sufficiently large. Our method with NP-based $W_{\fg}$ outperforms other approaches in discovery power, demonstrating the benefit of leveraging the masked rank of the treated outcomes.

\begin{figure}[t]
    \centering
    \begin{subfigure}[t]{0.49\linewidth}
        \centering
        \includegraphics[height=0.55\textwidth]{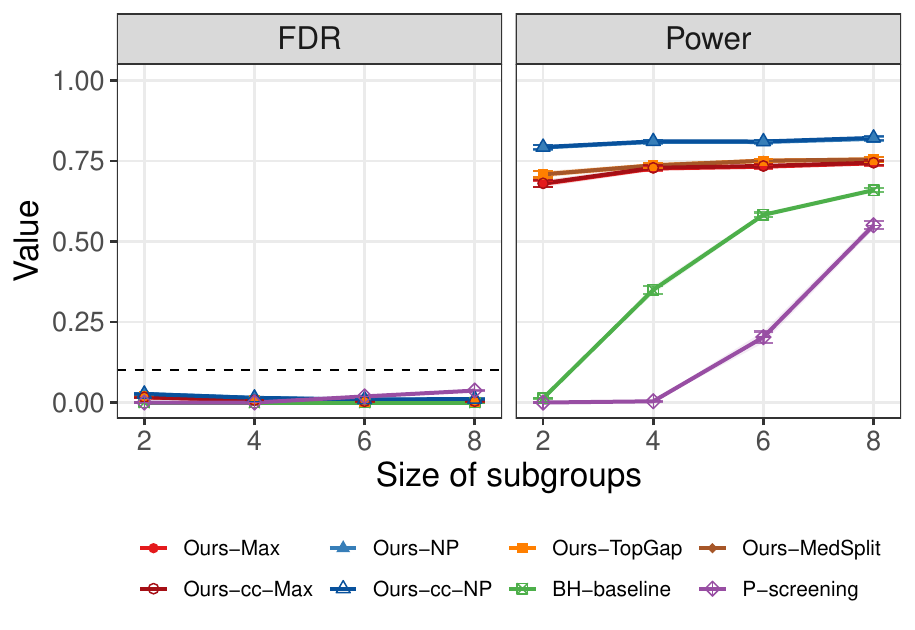}
        \caption{One-sided effects.}
        \label{fig:1side-5out-tree}
    \end{subfigure}
    \hfill
    \begin{subfigure}[t]{0.49\linewidth}
        \centering
        \includegraphics[height=0.55\textwidth]{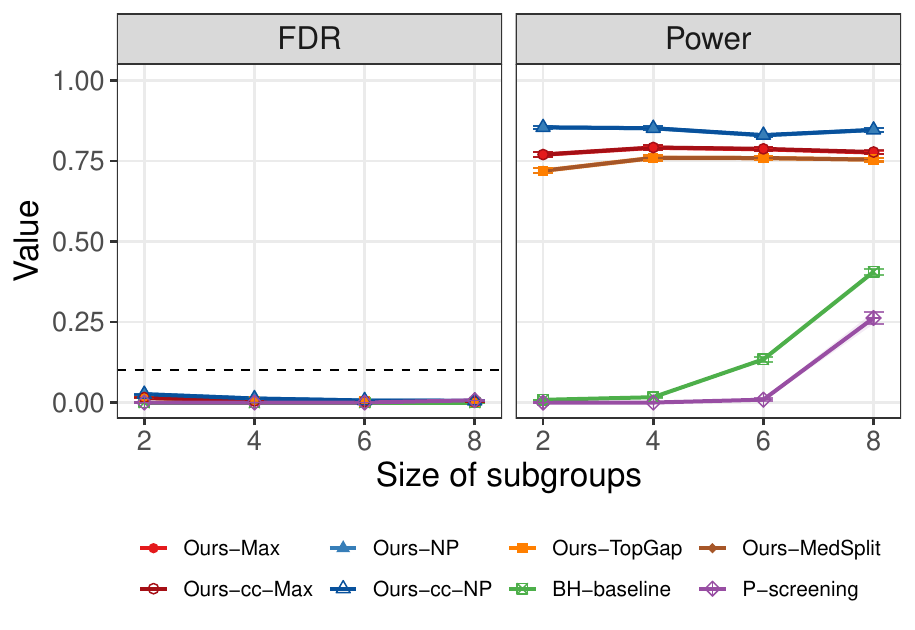}
        \caption{Two-sided effects.}
        \label{fig:2side-5out-tree}
    \end{subfigure}
    \caption{FDR and Power comparison versus varying group sizes with tree-based subgroup partition: multiple outcomes and varying subgroup sizes.}
    \label{fig:5out-tree}
\end{figure}

\paragraph{Tree-based subgroup partition.}
We complement the random-partition results with tree-based partitions in Figure~\ref{fig:5out-tree}.
Under one-sided effects (Figure~\ref{fig:1side-5out-tree}), \texttt{Ours-NP} achieves power above $0.8$ at small group sizes, and the \texttt{BH-baseline} improves with group size but plateaus well below our methods.
Under two-sided effects (Figure~\ref{fig:2side-5out-tree}), the baselines, the \texttt{BH-baseline} and \texttt{P-screening}, are substantially weaker, while our methods maintain high discovery power.

\subsection{Additional simulations with varying aggregation rules}\label{sec:ablate-agg}
\begin{figure}[!ht]
    \centering
    \includegraphics[width=0.5\linewidth]{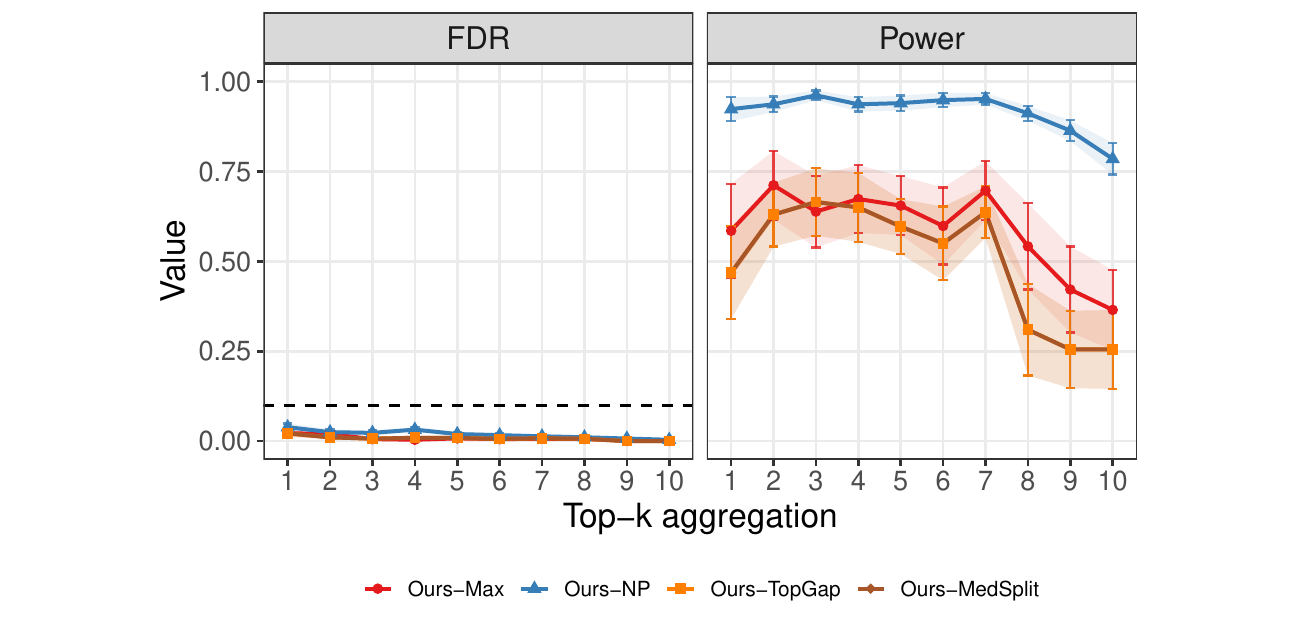}
    \caption{Performance with different values of $|\cQ_{\fg}|$.}
    \label{fig:agg-k}
\end{figure}

\begin{figure}[!ht]
    \centering
    \includegraphics[width=0.5\linewidth]{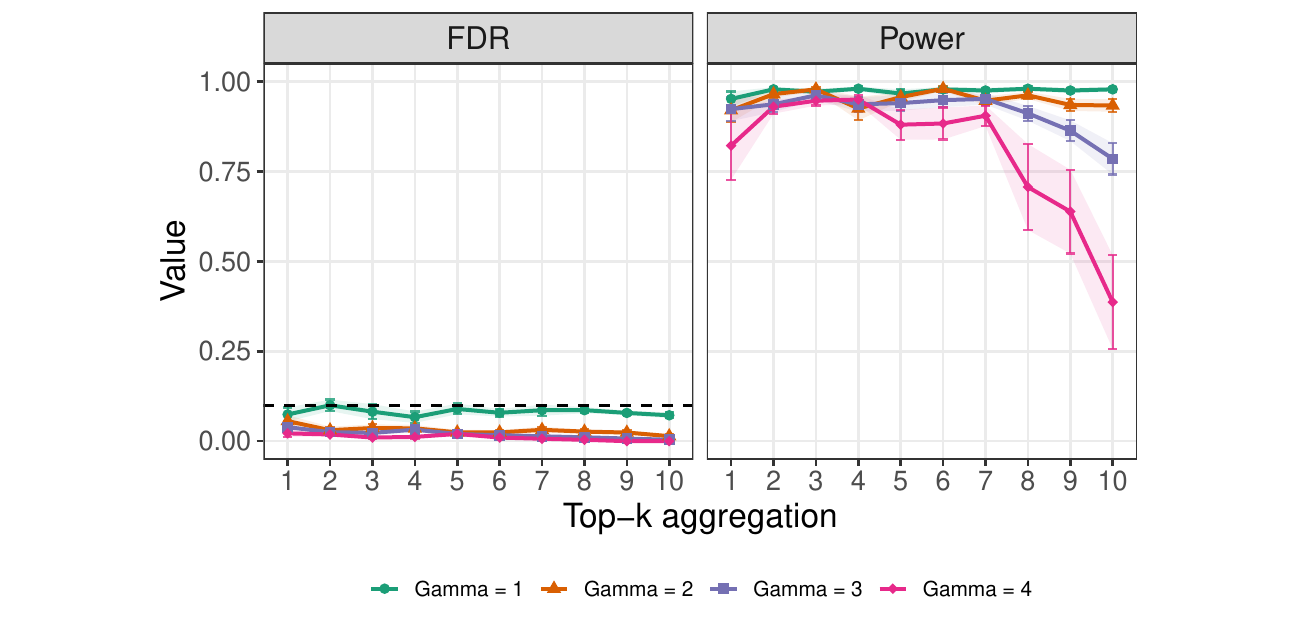}
    \caption{Performance of \texttt{Ours-NP} with different values of $|\cQ_{\fg}|$ and $\Gamma$.}
    \label{fig:agg-k-g}
\end{figure}

Recall the sign aggregation in Section~\ref{sec:L-agg}:
\[
L_\fg \;=\; 2 \cdot \ind\!\left\{\,\sum_{i \in \cQ_{\fg}}\, L_i \;>\; 2\,\eta_\fg - |\cQ_\fg|\,\right\} - 1,
\qquad
V_\fg \;=\; \bigl(L_{j}\bigr)_{j\in \fg,\; j \notin \cQ_{\fg}},
\]
where $\cQ_{\fg} = \{i \in \fg: W_i \geq W_{(k)}\}$, which consists of indices whose $W_i$'s are among the top-$k$ largest values in $\fg$, and the threshold $\eta_\fg$ is the upper $\Gamma/(1+\Gamma)$-quantile of
the worst-case binomial distribution,
\begin{equation}
\eta_\fg \;:=\; \min\!\left\{\, t \in \ZZ_{\ge 0} \;:\;
  \PP\!\left(\mathrm{Bin}\bigl(|\cQ_\fg|,\, \tfrac{\Gamma}{1+\Gamma}\bigr) > t\right)
  \;\le\; \tfrac{\Gamma}{1+\Gamma} \,\right\}.
\end{equation}
In this section, we use a simple simulation (same setting as Figure~\ref{fig:1side-1out-random}) to demonstrate the effect of $|\cQ_{\fg}|$ on the power of \texttt{Ours-NP} in Figures~\ref{fig:agg-k} and \ref{fig:agg-k-g}, where the subgroups have a size of $10$. Figures~\ref{fig:agg-k} and \ref{fig:agg-k-g} suggest that, with moderate values of $\Gamma$, the power is robust to the choice of $|\cQ_{\fg}|$, and with larger $\Gamma$'s, the optimal power is given by $|\cQ_{\fg}| \approx 4$. Hence, the choice of $|\cQ_{\fg}| = |\fg|/2$ turns out to be reasonable for a wide range of $\Gamma$ values.

\end{document}